\documentclass[12pt]{article}
\usepackage{natbib}
\usepackage{a4wide}
\usepackage{subfigure}
\usepackage{lineno}

\usepackage{graphicx}
\usepackage{subfigure}
\usepackage{color}
\usepackage[pdftex,colorlinks,
                citecolor=darkgreen,linkcolor=darkblue,urlcolor=darkblue]{hyperref}
\definecolor{darkgreen}{rgb}{0.1,0.4,0.1}
\definecolor{darkblue}{rgb}{0.1,0.1,0.3}

\newcommand{\beq}  { \begin{linenomath*}\begin{eqnarray} }
\newcommand{\eeq}  { \end{eqnarray}\end{linenomath*}}
\newcommand{\ol}   { \overline}
\newcommand{\rvec} [1]{ \raisebox{-1.5ex}{$\stackrel{\textstyle{#1}}{\neg}$} }
\newcommand{\rnabla} { \rvec{\nabla} }

\title{Higher-order approximations of the residual-mean eddy streamfunction and the quasi-Stokes streamfunction}

\author{\textsc{Jan Viebahn}
				\thanks{\textit{Corresponding author address:} 
				Jan Viebahn, KlimaCampus, Institute of Oceanography, University of Hamburg,
                                  Bundesstra\ss e 53, 20146 Hamburg, Germany.
				\newline{E-mail: jan.viebahn@zmaw.de. Tel/fax: +49 40 42838 -8215/-2995.}}
				\quad\textsc{and Carsten Eden}\\
\centerline{\textit{\footnotesize{KlimaCampus, University Hamburg, Hamburg, Germany}}}
}

\date{\today}

\begin{document}
\maketitle
\begin{abstract}
The series expansion of the residual-mean eddy streamfunction and
the quasi-Stokes streamfunction are compared up to third order in buoyancy perturbation,
both formally and by using several idealised eddy-permitting zonal channel model experiments.
In model configurations with flat bottom, both streamfunctions may be well
approximated by the first one or two leading order terms in the ocean interior,
although terms up to  third  order still significantly impact the implied interior circulations.
Further, differences in both series expansions up to third order remain small here.
Near surface and bottom boundaries, on the other hand,
the leading order terms differ and are initially of alternating sign and of increasing magnitude
such that the low order approximate expressions break down there.
In more realistic model configurations with significant topographic features,
physically inconsistent recirculation cells also appear in the ocean interior and are not effectively
reduced by the next higher order terms. A measure indicating an initially increasing
or decreasing series expansion is proposed.
%
\end{abstract}

\newpage
\section{Introduction}\label{introduction}
Eulerian averaging of velocities and tracers is usually considered as the simplest way of averaging:
Time and ensemble averages are performed at fixed position
and space averages are solely defined by the geometrical framework (i.e. by the coordinate lines of the geometrically natural coordinate system).
In other words, Eulerian averaging appears to be the most straightforward averaging procedure,
because no physical properties of the fluid are taken into account in the integration conditions \citep{Andrews:78b}.
Accordingly, the Eulerian meridional transport streamfunction $\Lambda$,
i.e. the zonally integrated meridional transport of fluid across a given latitude $y$ and below a constant height surface $z$,
is defined in regard to the space coordinates $y$ and $z$.

However, $\Lambda$ gives rise to spurious diabatic circulations such as the Deacon cell \citep{Doos:94}.
In an isopycnal averaging framework \citep{Nurser:04a}, the Deacon cell is reduced and therefore
the isopycnal meridional transport streamfunction,
i.e. the zonally integrated meridional transport of fluid across a given $y$ and denser than a given density,
is considered as a more appropriate description of the  meridional overturning circulation (MOC).
In order to obtain a physically meaningful MOC in the Eulerian framework,
the initial simplification (i.e. the insensitivity of the transport integration to the physical state of the fluid) has to be revised
and a more complicated redefinition of the total overturning streamfunction has to be introduced.
More precisely, two different approaches of constructing physically more satisfying overturning streamfunctions in the Eulerian framework
have been put forward:
the residual-mean theory \citep{Andrews:76,Eden:07} and
the quasi-Stokes streamfunction 
\citep{McDougall:01,Nurser:04b}.

The residual streamfunction $\psi_{res}$ is defined as the streamfunction which advects the Eulerian-mean buoyancy and
it is constituted as the residual of two parts:
On the one hand, the advection is due to the Eulerian-mean velocities (given by $\Lambda$),
on the other hand, there is an eddy-induced streamfunction $\psi^*$ due to the advective part of the eddy buoyancy flux.
Physically, it is desired that,
if there is no instantaneous diabatic buoyancy forcing, there should be also no diabatic effects in the Eulerian-mean buoyancy budget,
i.e. the eddy-induced diabatic forcing should vanish too.
\cite{Eden:07} (extending ideas of \cite{McDougall:96,Medvedev:04}) demonstrate, by explicitly incorporating rotational eddy fluxes,
that this physical criterion uniquely sets $\psi^*$ and with it $\psi_{res}$.
However, $\psi^*$ is then given by a series involving fluxes of eddy buoyancy moments.

The quasi-Stokes streamfunction $\Psi^*$ is the eddy-induced component of the total isopycnal streamfunction transformed into Eulerian space \citep{McDougall:01,Nurser:04b}.
That is, the isopycnal streamfunction mapped into Eulerian space
may be given by the sum of $\Lambda$ and $\Psi^*$ and advects the isopycnally averaged buoyancy\footnote{
The isopycnally averaged buoyancy is defined as inverse function of the mean height of isopycnals \citep{deSzoeke:93,McDougall:01,Nurser:04a}.
}.
\cite{McDougall:01} (see also \cite{Nurser:04b}) apply a Taylor series analysis centred around the mean height of isopycnals in order to express $\Psi^*$ by Eulerian-mean quantities.
Consequently, $\Psi^*$ is given in two ways: On the one hand, it may be computed out of an isopycnal averaging framework.
On the other hand, $\Psi^*$ may be given directly in Eulerian space (i.e. in height coordinates) and is then expressed by a series expansion. Of course, this series expansion is different to the one of the residual-mean eddy streamfunction $\psi^*$, however, both are intimately connected as we discuss in this study.

Hence, if physically meaningful streamfunctions of the MOC are sought directly in the Eulerian framework,
it seems that the appearance of series expansions generally represents a necessary and severe complication.
Most problematic is that, practically, it is inevitable to cut off the series expansions at a certain order
and hence one is left with approximate formulas.
Typically, in a zonal-mean framework the first order terms of both series expansions are considered as
good approximations in the nearly adiabatic ocean interior,
but near horizontal boundaries (surface, bottom) the approximate formulas are found to break down,
i.e. unphysical nonzero (and relatively large) values appear at the horizontal boundaries \citep{Killworth:01,McDougall:01,Nurser:04b}.
A physically satisfactory solution of this serious problem is outstanding.

In this study, this problem will be further explored.
The subjects of this study are the following:
We formally compare for the first time (at least to our knowledge)
the first three orders of the series expansions of $\psi^*$ and $\Psi^*$
in order to specify the essential differences between these two intimately linked streamfunctions at low orders.
Furthermore, we consider all terms up to the third order of both series expansions
in different idealised models of the Southern Ocean (SO)
in order to investigate the behaviour of both series expansions in different concrete model setups
which have been previously used to analyse eddy streamfunctions. 
It will turn out that, in a zonal-mean framework, the problems due to the convergence behaviour of both series expansions are more severe
and hence the limitations of both approaches are stronger than discussed so far.
Finally, we propose a measure to diagnose regions in the ocean where approximations of the series expansions break down.

The study is structured as follows:
In section \ref{experiments} we present our different model setups and experiments.
In section \ref{RME} we consider approximations of the series expansion of the residual-mean eddy streamfunction $\psi^*$,
while in section \ref{STOKES} we additionally examine approximations of the Taylor series of the quasi-Stokes streamfunction $\Psi^*$.
Finally, in section \ref{seriesN} we propose the series number as a measure to diagnose the break down of the approximations and section \ref{Summary} provides a summary and discussion.

\section{Models and experiments}\label{experiments}
In this study, we use the code of CPFLAME\footnote{\it http://www.ifm.zmaw.de/$\sim$cpflame} in two different configurations
which both have been previously used to analyse eddy streamfunctions. 
The first configuration is essentially a reproduction of the model setup considered in the study of \cite{Nurser:04a,Nurser:04b},
who used an idealised eddy-permitting zonal channel model in order to compare the isopycnal transport streamfunction with
the first order approximation of the quasi-Stokes streamfunction given by Eulerian-mean quantities.
In our version, the primitive equations are formulated in Cartesian coordinates.
The zonally reentrant channel extends over $L_x=600$km in zonal direction and $L_y=1000$km in meridional direction
with $10$km horizontal resolution. It is $1000$m deep with $20$m vertical resolution.
The beta-plane approximation is used with a reference latitude situated at $49.31^\circ$S, such that at the centre of the channel
the Coriolis parameter becomes the one at $45^\circ$S for spherical coordinates.
We simulate only buoyancy $b$ in the model, which might be thought as proportional to temperature.
The model is not forced with winds, but the circulation in the model is driven by three buoyancy restoring regions:
At the surface, $b$ is relaxed towards a target buoyancy varying linearly between $21.764\times10^{-3}$ms$^{-2}$ at $y=95$km and
$45.269\times10^{-3}$ms$^{-2}$ at $y=905$km with a restoring time scale of $7$ days.
Within the southernmost $95$km and the northernmost $95$km, model buoyancies are relaxed throughout the water column to specified values:
linearly varying with depth from $19.346\times10^{-3}$ms$^{-2}$ at the surface to zero at the bottom in the southern zone
and from $48.365\times10^{-3}$ms$^{-2}$ to zero in the northern zone.
The relaxation rate varies linearly between $1/(2$ days) at the boundaries, and zero at the inner edges of the relaxation zone.
Vertical viscosity is $5\times10^{-4}$m$^2$s$^{-1}$ and we use a horizontal biharmonic viscosity of $1.25\times10^{11}$m$^4$s$^{-1}$. The linear bottom friction parameter is $2\times10^{-5}$s$^{-1}$.
Vertical diffusivity is $5\times10^{-5}$m$^2$s$^{-1}$, but we use no explicit lateral diffusion.
The Quicker advection scheme is used as the advection scheme of buoyancy.
The model was run for a total of $30$ years.
The diagnostics below are presented as temporal averages over the last $10$ years of the run.
We refer to this experiment as the \emph{NL case}.

The second configuration is the idealised SO model setup introduced and discussed by \cite{Viebahn:10,Viebahn:12},
i.e. an eddy-permitting primitive equation model consisting of a zonally reentrant channel, which is connected to a northern ocean basin enclosed by
land. The circulation in the model is driven by a sinusoidal westerly wind stress over the channel with a magnitude of $\tau=1\times10^{-4} m^2s^{-2}$, and a surface restoring boundary condition for buoyancy $b$ (again, there is only buoyancy in the model).
The corresponding target buoyancy increases northward over the channel, remains constant over the southern half of the northern ocean basin
and decreases while approaching the northern end of the domain.
Boundary conditions on the northern and southern edges of the domain are simply given by no-flux conditions.
Hence, the water mass distribution is solely determined by the surface boundary conditions.
The domain of the idealised model extends over $L = 2520$km in the zonal and meridional direction,
with $20$km horizontal resolution and $40$ vertical levels with $50$m thickness ($1900$m maximal water depth).
The channel (i.e. the SO) extends from the southern boundary ($x=0$km) to $x=L/2$.
Further details may be found in \cite{Viebahn:10,Viebahn:12}.
In particular, we consider the same two experiments already discussed by \cite{Viebahn:12} in an isopycnal averaging framework.
In the \emph{flat case} experiment, the bottom is completely flat.
In the \emph{hill case} experiment, a simple hill-like topographic feature is imposed in the channel:
The top of the hill is located at $z=-950$m and $x=0$ (and $x=2520$km respectively).
According to an exponential map the height of the hill decreases eastward (westward), such that at the longitudes of the northern ocean basin (from $x=850$km to $x=1690$km) the channel has a flat bottom.
In both experiments, the model has been run for $240$ years.
Additionally, we introduced harmonic viscosities (which act to damp EKE) in both experiments,
in order to discuss the convergence behaviour of the series expansions of both $\psi^*$ and $\Psi^*$
subject to the ``strength'' of the eddy field.
In the flat case, we introduced $A_h=2000\mathrm{m}^2\mathrm{s}^{-1}$ and the model has been run for another $50$ years, i.e. $290$ years in total.
In the hill case, we introduced $A_h=2000\mathrm{m}^2\mathrm{s}^{-1}$, $A_h=5000\mathrm{m}^2\mathrm{s}^{-1}$ and $A_h=10000\mathrm{m}^2\mathrm{s}^{-1}$ respectively after $200$ years of the initial model run and the model has been run for another $60$ years, i.e. $260$ years in total.
In each experiment the time-mean is performed over the last 10 years\footnote{
Note that in this study, we perform the analysis in a time-zonal-mean context.
That is, each quantity $q$ generally may be decomposed into its temporal and zonal average $\ol{q}$
and its temporal and zonal deviation $q'\equiv q-\ol{q}$, i.e. $q=\ol{q}+q'$.
However, we denote the time-zonal-mean by an overbar only regarding the basic physical quantities, i.e. the velocities and buoyancy.
The meridional streamfunctions considered in this study are zonally integrated quantities from the outset
and we do not explicitly denote the additional time average by another symbol, but implicitly assume it from now on.
}.

In a way, the NL case represents the simplest model configuration in this study:
Due to the lack of both zonal wind stress and the connection to a northern ocean basin, the time-zonal-mean meridional velocity $\ol{v}$ disappears almost completely in the NL case and hence it holds $\Lambda\approx0$, in contrast to the wind-driven flat case and the hill case experiments.
Since $\Lambda$ generally opposes the eddy-induced streamfunctions in the SO,
the wind-driven model configuration shows eddy-induced streamfunctions of higher magnitudes
at equal magnitudes of the overall overturning circulation.

\section{Approximations of the residual-mean streamfunctions}\label{RME}
The time-zonal-mean residual streamfunction $\psi_{res}(y,z)$ is defined as the meridional streamfunction,
which advects the Eulerian time-zonal-mean buoyancy $\ol{b}$.
As outlined in appendix A, it is given by the sum of the time-zonal-mean Eulerian streamfunction ${\Lambda}$ (defined by Eq. (\ref{EulerPsi})) and the eddy streamfunction $\psi^*$,
\beq
\psi_{res}={\Lambda}+\psi^*\ .
\eeq
A physically consistent determination of $\psi^*$ (and with it $\psi_{res}$) was given by \cite{Eden:07} by explicitly incorporating rotational eddy fluxes (see appendix A for a synopsis).
$\psi^*$ is then given by a series involving fluxes of eddy buoyancy moments (see Eq. (\ref{ApsiEdFULL})),
\beq\label{psiEdFULL}
\psi^*|\nabla\ol{b}|=
-J_1+\partial_mJ_2-\frac{1}{2}\partial^2_mJ_3+O(b'^4)\ ,
\eeq
where $\partial_m()\equiv|\nabla\ol{b}|^{-1}\nabla\ol{b}\cdot\nabla|\nabla\ol{b}|^{-1}()$
and the $J_n\equiv\mathbf{F}_n\cdot\rnabla\ol{b}|\nabla\ol{b}|^{-1}$ represent the along-isopycnal fluxes of the eddy buoyancy moments\footnote{
The eddy buoyancy moments are defined as $\phi_n=b'^n/n$ and the fluxes of the eddy buoyancy moments are given by
$\mathbf{F}_n=L_x(\ol{v\phi_n},\ol{w\phi_n})$, where $n$ represents the order.
The operator $\rnabla$ is defined as $\rnabla\equiv(-\partial_z,\partial_y)$.
See appendix A for more details.
}.
The terminology $O(b'^4)$ indicates additional terms that are of fourth or higher order in buoyancy perturbations.
The orders of the series expansion (\ref{psiEdFULL}) are defined solely by the fluxes of $b'^n$, i.e. via the order of the eddy buoyancy moment.
The first order term in the expansion for $\psi^*$ is identical to an eddy streamfunction of the transformed Eulerian mean (TEM) framework \citep{Andrews:76,Andrews:78}.
The remainder of the expansion is due to the introduction of the rotational flux potential $\theta$ (see Eq. (\ref{AthetaFULL})).
In the interior ocean, it typically holds $|\partial_y\ol{b}|\ll|\partial_z\ol{b}|$ and $|w|\ll|v|$ and we obtain
\beq
\psi^*&\approx&\label{psiEdAPPROX}
\frac{L_x\ol{v'b'}}{\partial_z\ol{b}}
-\frac{1}{\partial_z\ol{b}}\partial_z\biggl(\frac{L_x\ol{\phi_2v}}{\partial_z\ol{b}}\biggr)
+\frac{1}{2}\frac{1}{\partial_z\ol{b}}\partial_z\biggl(\frac{1}{\partial_z\ol{b}}\partial_z\biggl(\frac{L_x\ol{\phi_3v}}{\partial_z\ol{b}}\biggr)\biggr)+O(b'^4)
\eeq
Now we consider the first three orders of the series expansion of the residual-mean eddy streamfunction $\psi^*$ in our different model experiments.
Notice that we calculated the terms as given by Eq. (\ref{psiEdFULL}), but that differences to the terms as given by Eq. (\ref{psiEdAPPROX})
are small in the entire model domain (including the diabatic boundary regions).
In the following, we index the terms of the different orders of the series expansion of $\psi^*$ by Roman numerals, i.e. $\psi^*=\psi^*_{I}+\psi^*_{II}+\psi^*_{III}+\psi^*_{IV}+...$.

\subsection{NL case}\label{NLcase}
\begin{figure}[t]
  \centering
  \subfigure{
    \includegraphics[scale=0.269]{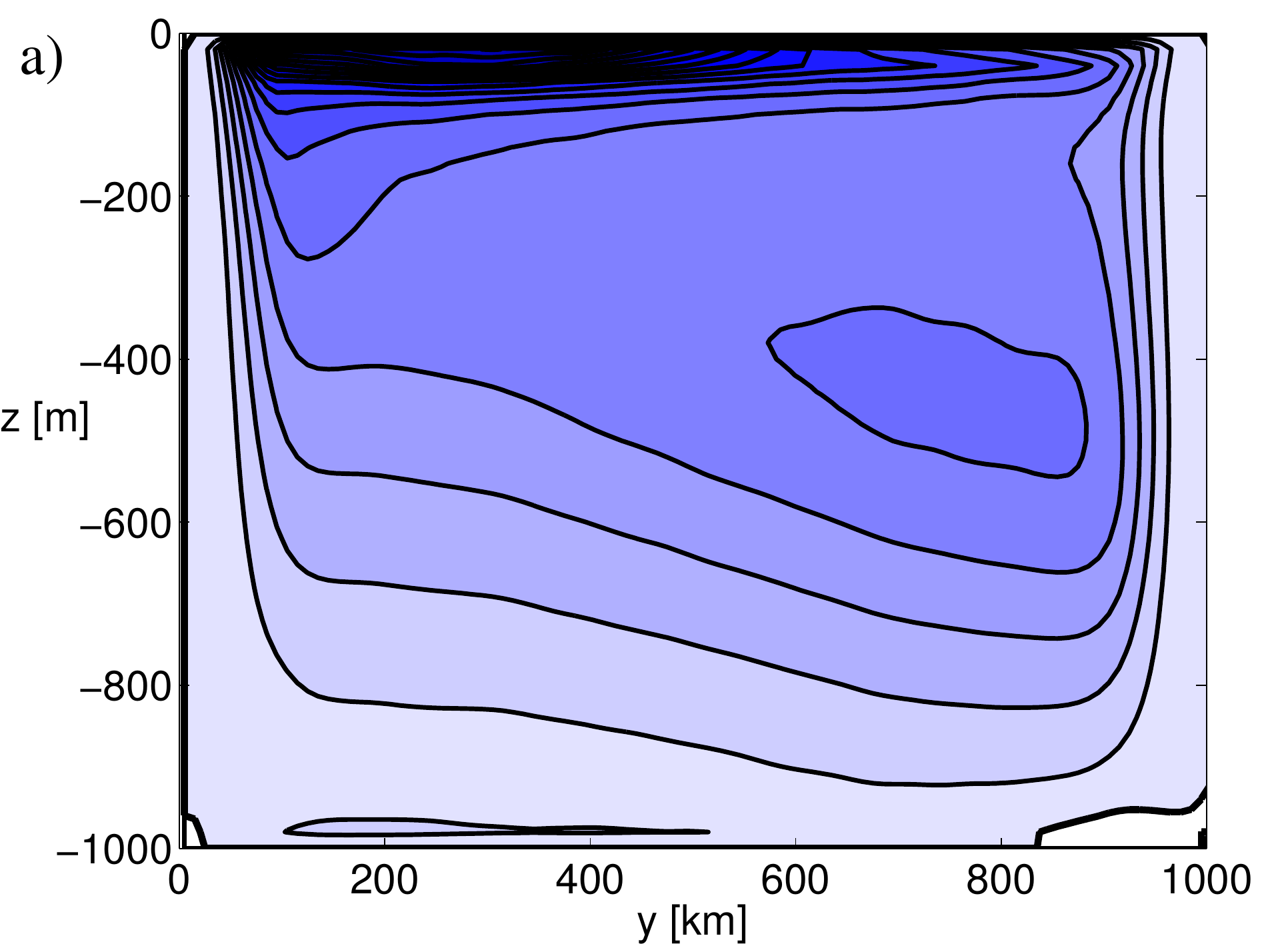}
  }
  \subfigure{
    \includegraphics[scale=0.269]{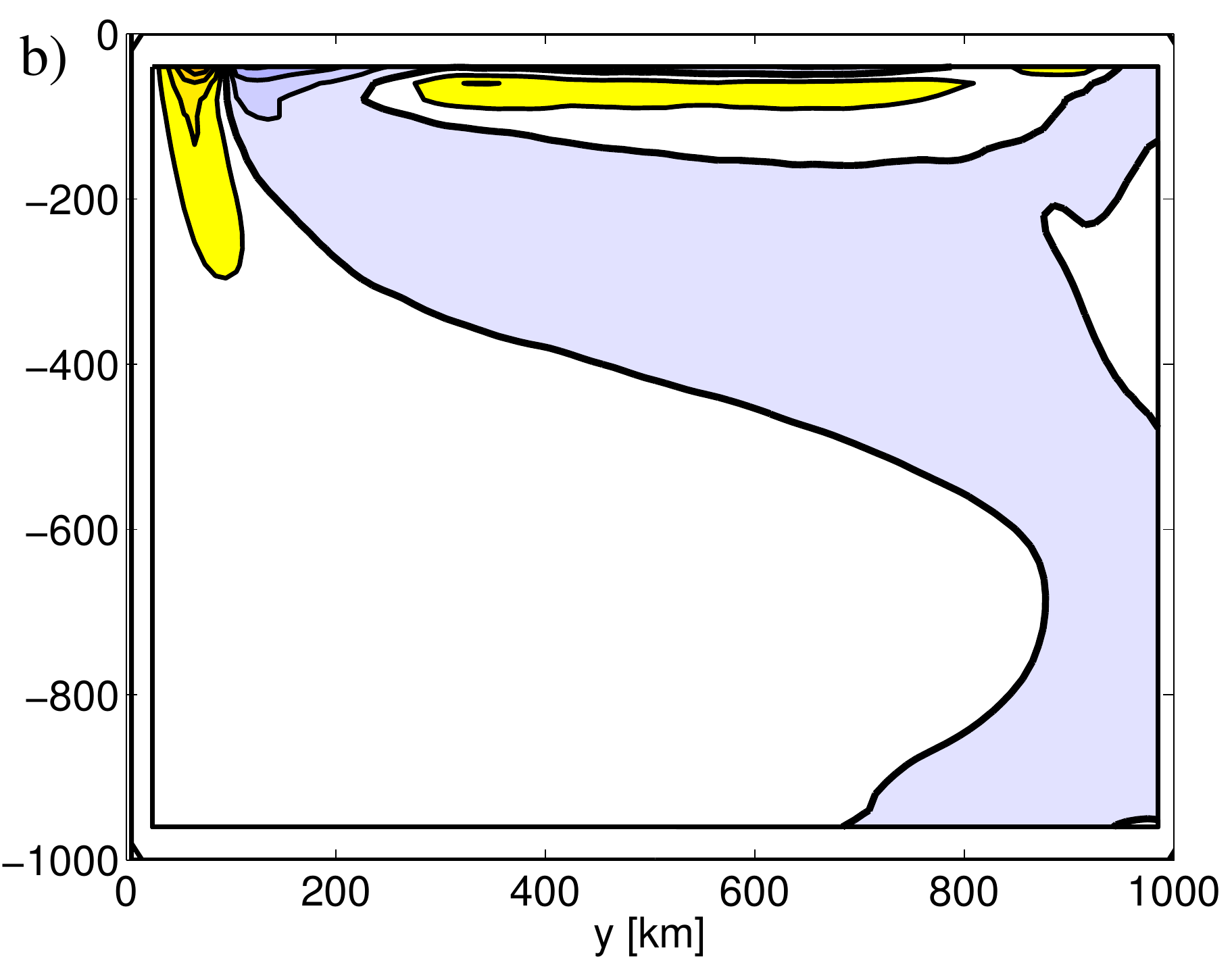}
  }
  \subfigure{
    \includegraphics[scale=0.269]{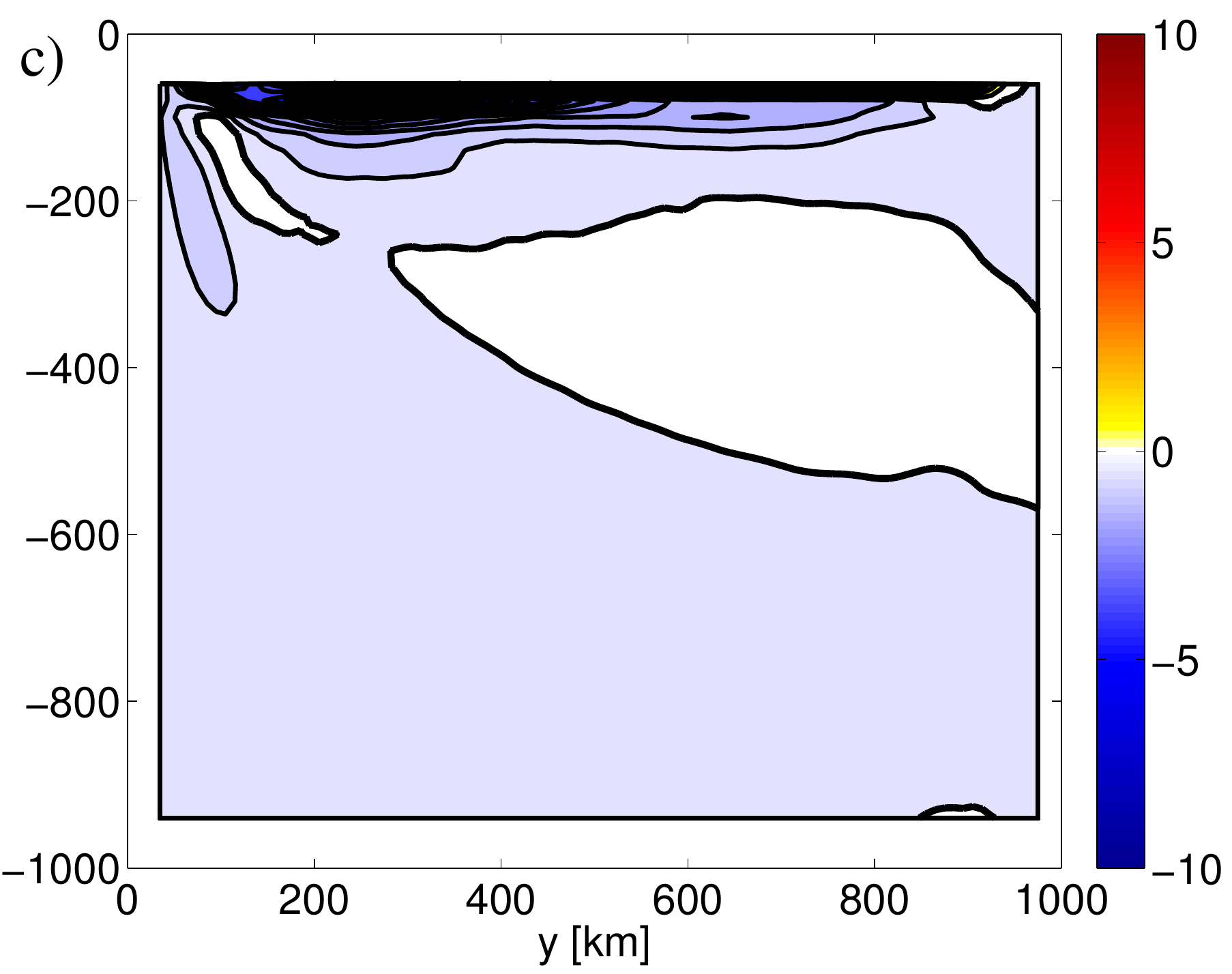}
  }
  \subfigure{
    \includegraphics[scale=0.269]{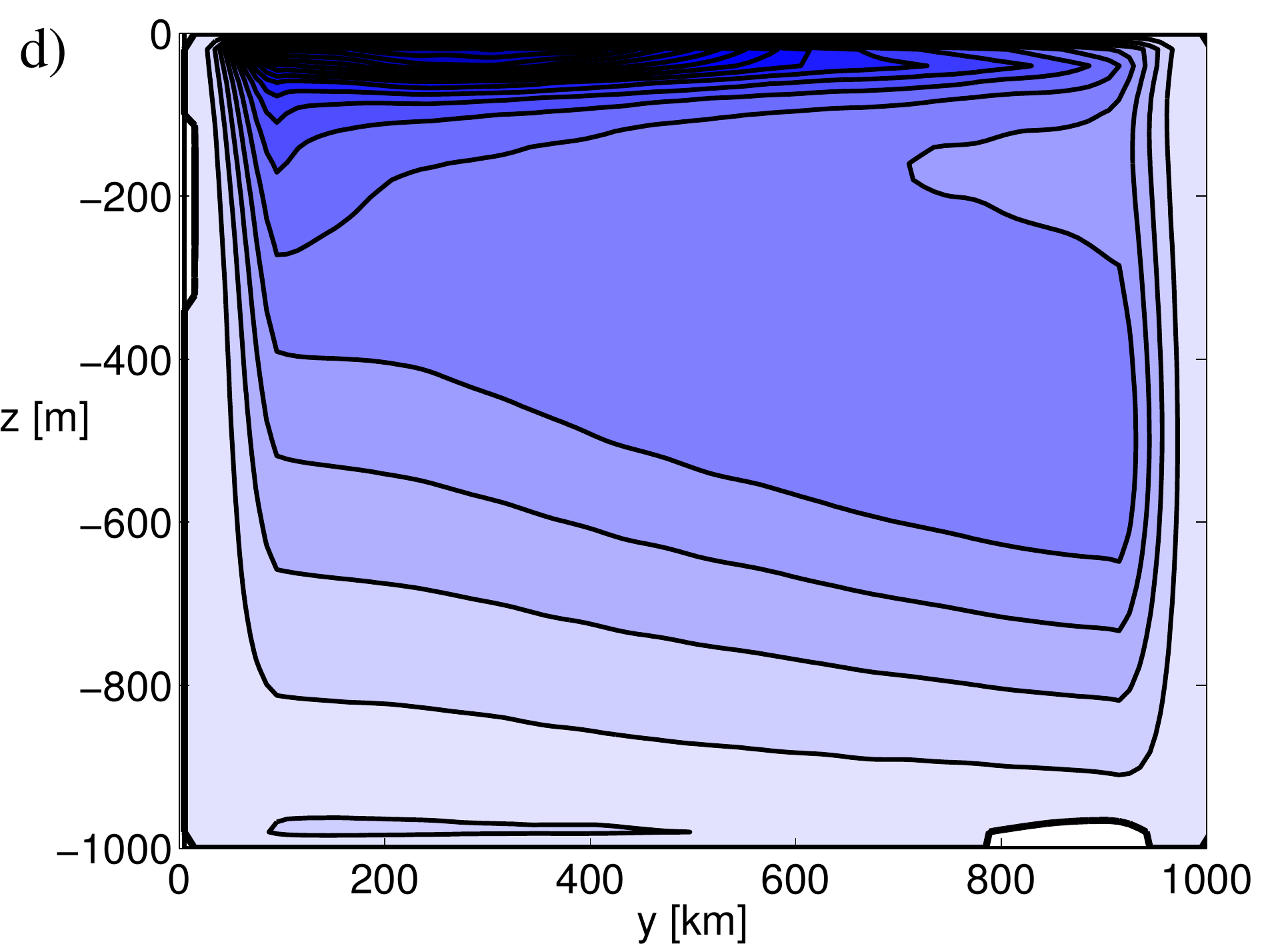}
  }
  \subfigure{
    \includegraphics[scale=0.269]{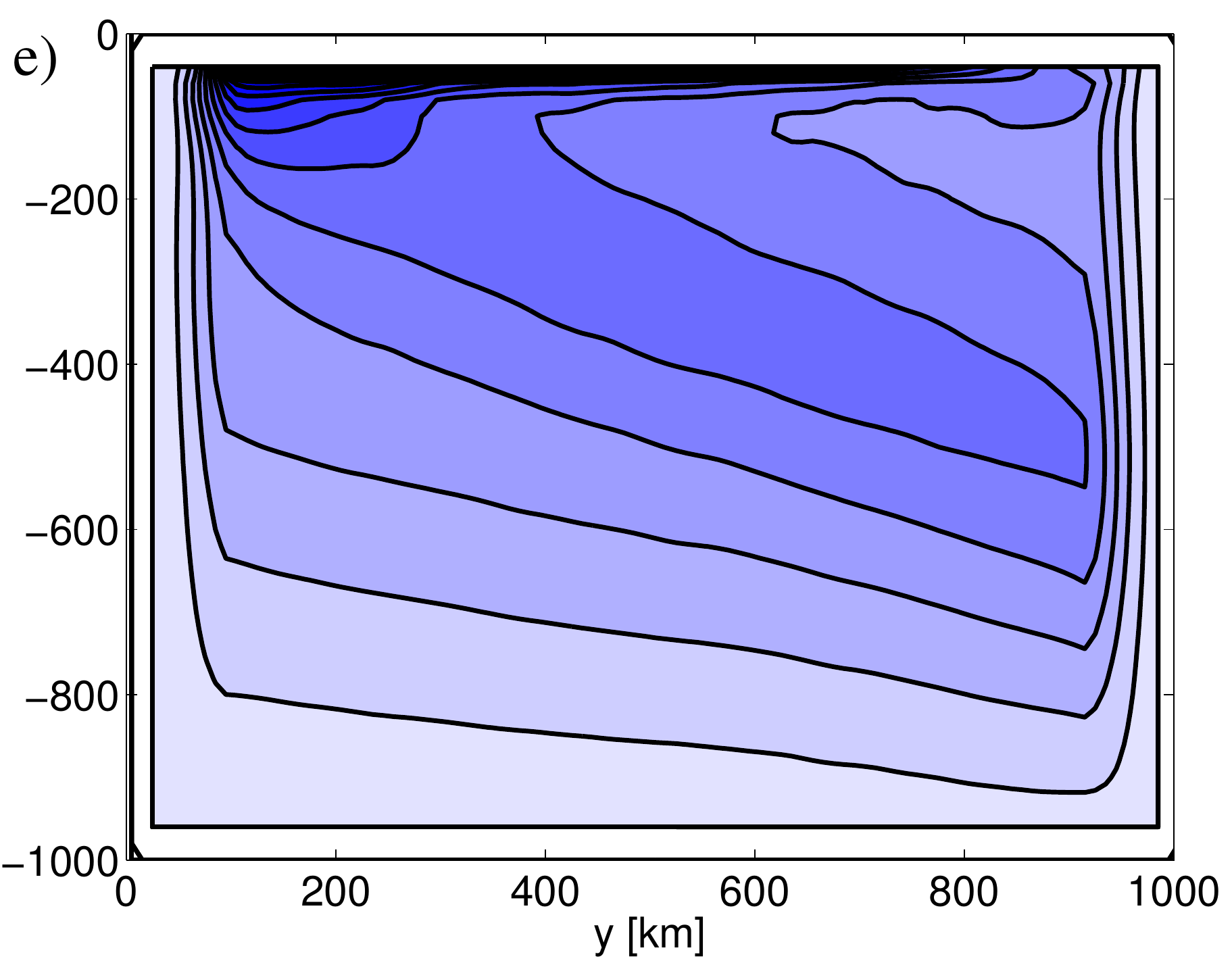}
  }
  \subfigure{
    \includegraphics[scale=0.269]{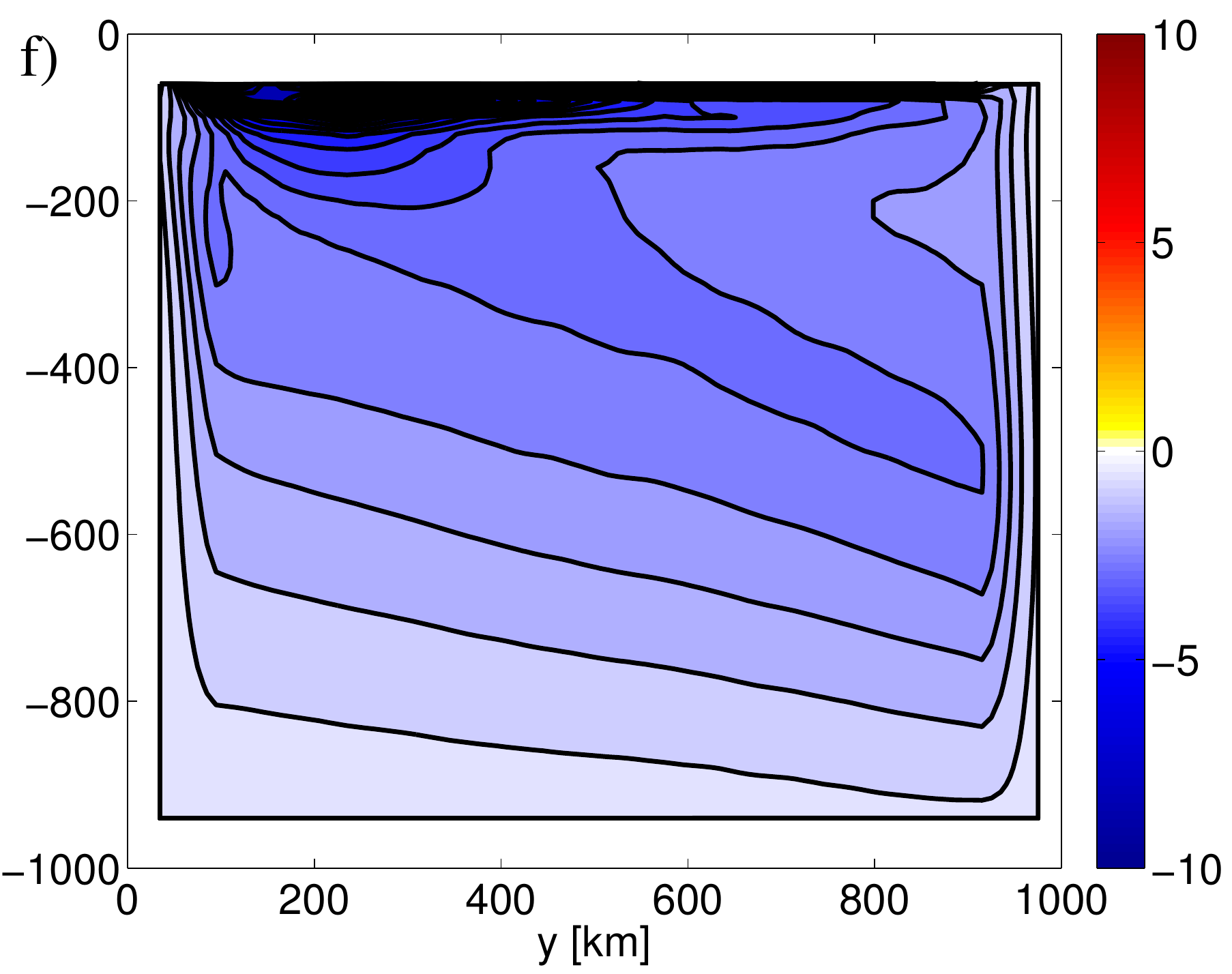}
  }
\caption{
The first three terms of the series expansion of $\psi^*$ for the NL case:
$\psi^*_{I}\equiv-J_1/|\nabla\ol{b}|$ (a), $\psi^*_{II}\equiv\partial_mJ_2/|\nabla\ol{b}|$ (b), $\psi^*_{III}\equiv-\frac{1}{2}\partial^2_mJ_3/|\nabla\ol{b}|$ (c)
and the corresponding residual streamfunctions including terms of $\psi^*$ up to the first (d), second (e), third (f) order.
The contour interval is $0.5$Sv and zero lines are thick.
}
\label{NLflat}
\end{figure}

Fig. \ref{NLflat} a) shows the first term $\psi^*_{I}\equiv-J_1/|\nabla\ol{b}|$ of the series expansion of $\psi^*$ (see Eq. (\ref{psiEdFULL})),
i.e. the TEM eddy streamfunction, and Fig. \ref{NLflat} d) shows ${\Lambda}+\psi^*_{I}$, i.e. the TEM residual MOC of the NL case.
Since ${\Lambda}\approx0$ (not shown) due to the lack of zonal wind stress, both are largely identical,
showing an anti-clockwise MOC with mainly along-isopycnal flow in the ocean interior and strong diapycnal flow in the three buoyancy restoring regions (but no bottom boundary layer).
Only at mid-depth around $y=750$km, the TEM residual MOC is slightly reduced in magnitude compared to $\psi^*_{I}$.
However, physically inadequate is the extremely strong recirculation cell in the surface layer,
which does not tend to zero at the surface (in Fig. \ref{NLflat} we simply set the surface values to zero).

\cite{Nurser:04b} find a similar circulation pattern (see their Fig. 1).
They discuss the unphysical surface circulation in the context of the classical TEM formalism,
but the approaches of solving the problems at the boundaries by merging different eddy streamfunctions into each other
are only partially successful and of unclear physical basis, hence remain unsatisfactory.
In the physically logical approach of \cite{Eden:07}, which we adopt in this study (see appendix A for a synopsis),
the problems at the boundaries are theoretically solved by incorporating the appropriate rotational eddy flux (given by Eq. (\ref{AthetaFULL})).
However, since the appropriate rotational eddy flux is only given by a series expansion,
the practical problem of including a sufficient number of terms of the series expansion
in order to obtain an adequate approximation emerges.

Fig. \ref{NLflat} b,c) show the second term $\psi^*_{II}$ and the third term $\psi^*_{III}$ of the series expansion of $\psi^*$.
In line with the expectation that $\psi^*_{I}$ represents a good approximation of $\psi^*$ in the nearly adiabatic ocean interior,
the dominant values of the next higher order terms $\psi^*_{II}$ and $\psi^*_{III}$ are found in the surface diabatic boundary layer (for $\psi^*_{IV}$ and $\psi^*_{V}$ as well, not shown),
although changes of the corresponding residual MOCs (Fig. \ref{NLflat} e,f)) are also visible in the ocean interior
(compare also Fig. \ref{NLflat} d-f) with Fig. \ref{isopycnalframework} a)).
More precisely, $\psi^*_{II}$ mainly opposes the surface layer circulation of $\psi^*_{I}$,
while $\psi^*_{III}$ amplifies it again.
Hence, the series expansion (\ref{psiEdFULL}) (or Eq. (\ref{AthetaFULL})) appears to be alternating,
which is obvious in Fig. \ref{NLflat} a-c) and continued for the next higher order terms $\psi^*_{IV}$ and $\psi^*_{V}$ (not shown).
This behaviour hampers the determination of an order at which the series expansion may be appropriately cut off,
since the next higher order always compensates a part of the previous order.
Moreover, it holds $|\psi^*_{II}|<|\psi^*_{III}|$ in the surface layer.
More generally, the ratio $|\psi^*_i|/|\psi^*_I|$ (not shown) increases with increasing order $i$ in the surface boundary layer,
while it decreases in the ocean interior.
That is, the magnitude of the terms of the series expansion (\ref{AthetaFULL}) appears to be increasing in the diabatic surface region
with increasing order,
while a decreasing behaviour, which is found in the ocean interior,
is necessary for an adequate approximation of $\psi^*$ by low order terms.

We conclude: In the NL case, the series expansion of $\psi^*$ may be adequately approximated by low order terms in the ocean interior,
since there they appear to be decreasing with higher order -
$\psi^*_{I}$ alone already may seem sufficient (but notice section \ref{IsoPsi}).
However, in the diabatic surface boundary layer the series expansion of $\psi^*$ is
alternating and initially increasing\footnote{
In this study,
we call a series expansion $s=\sum_{i=1}^{\infty}s_i$ \emph{increasing} (or \emph{decreasing}),
if the magnitude of the terms of $s$ is increasing (or decreasing) with higher order $i$, i.e. $|s_i|<|s_{i+1}|$ (or $|s_i|>|s_{i+1}|$).
},
which precludes an adequate approximate surface layer representation by low order terms,
i.e. leads to the ``break down'' of an approximation of $\psi^*$.
Hence, the series approach of \cite{Eden:07} seems to be unable to practically solve the problems at the boundaries appearing in the residual-mean framework.

\subsection{Flat case}\label{RMflatcase}
\begin{figure}[t]
  \centering
  \subfigure{
    \includegraphics[scale=0.270]{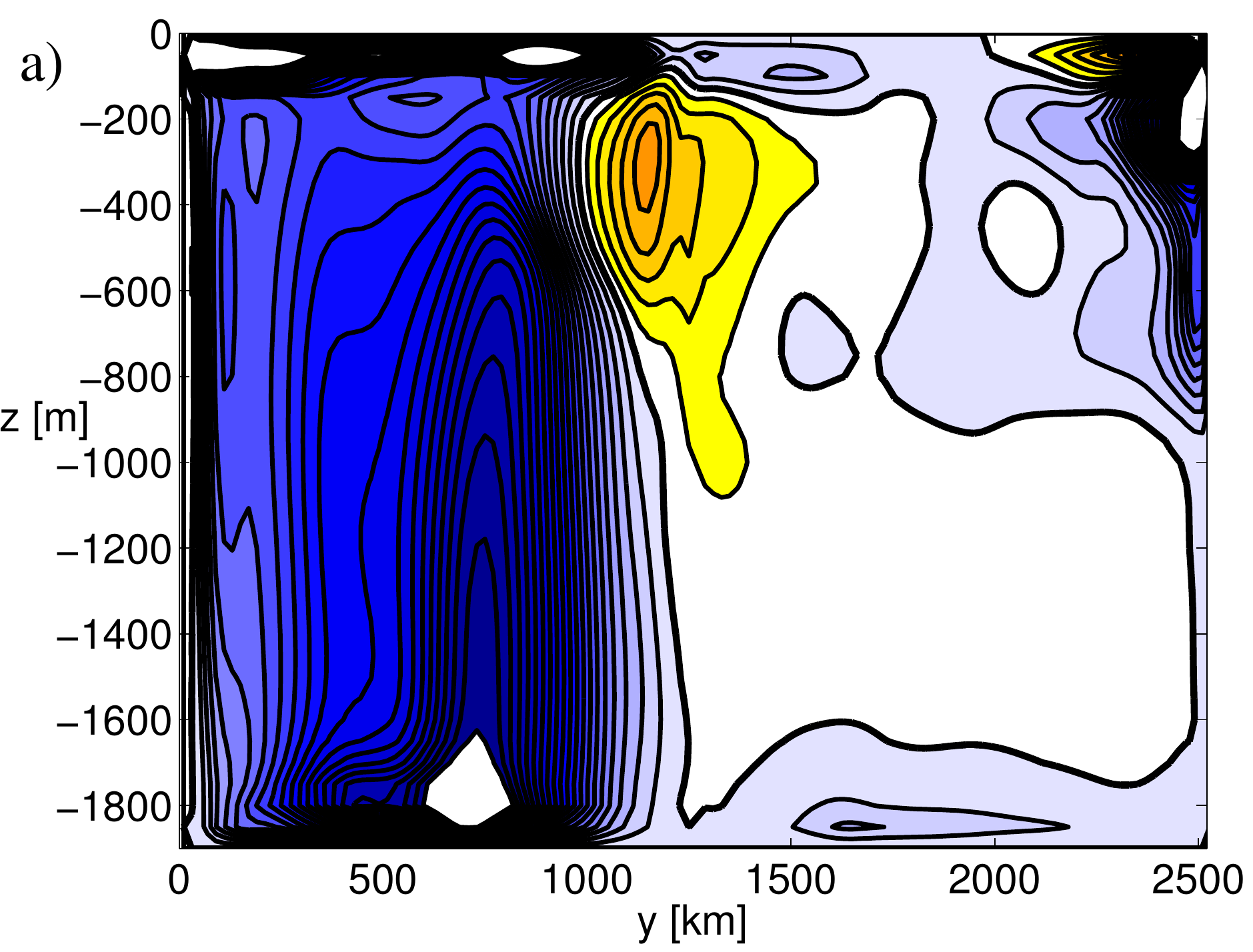}
  }
  \subfigure{
    \includegraphics[scale=0.270]{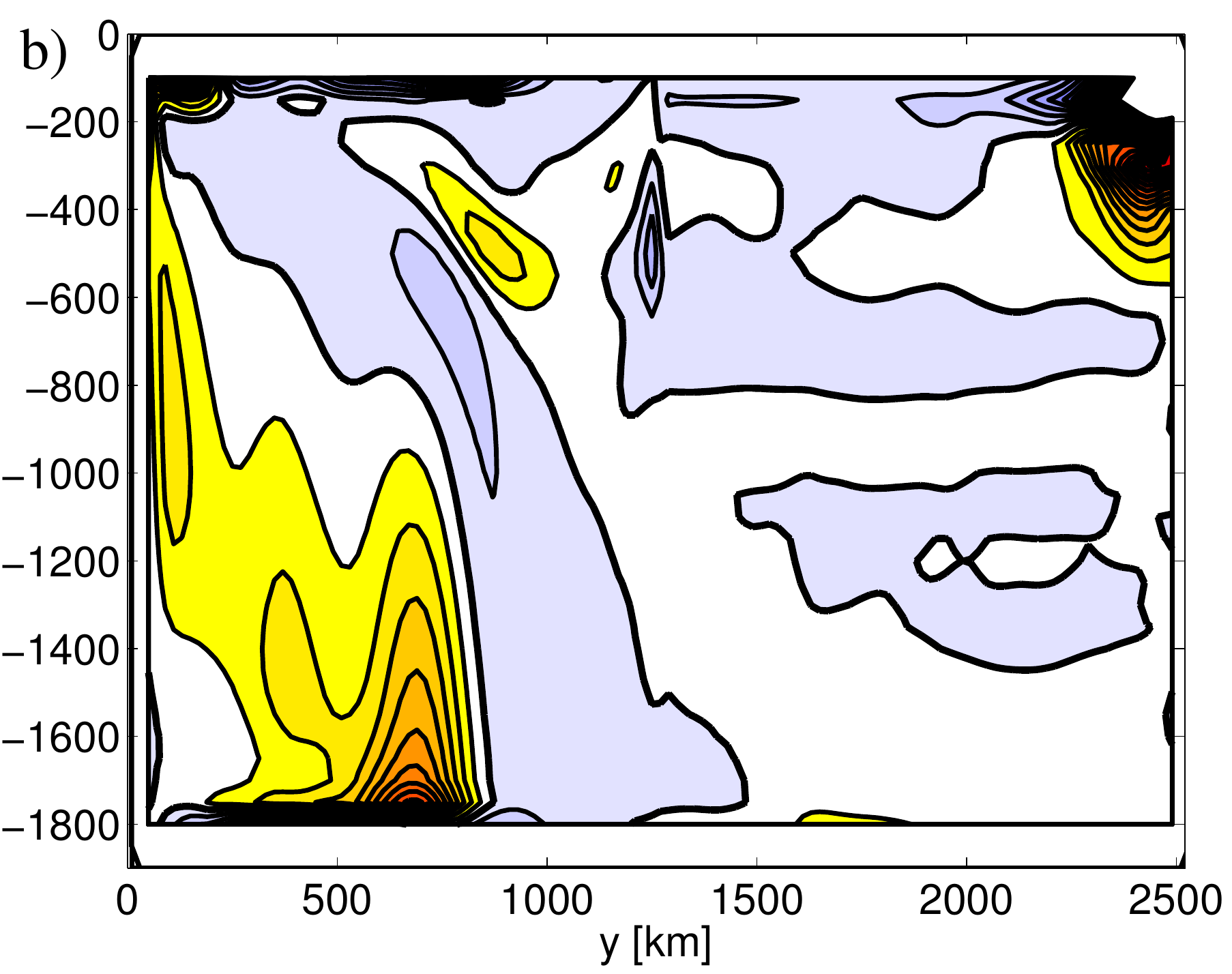}
  }
  \subfigure{
    \includegraphics[scale=0.270]{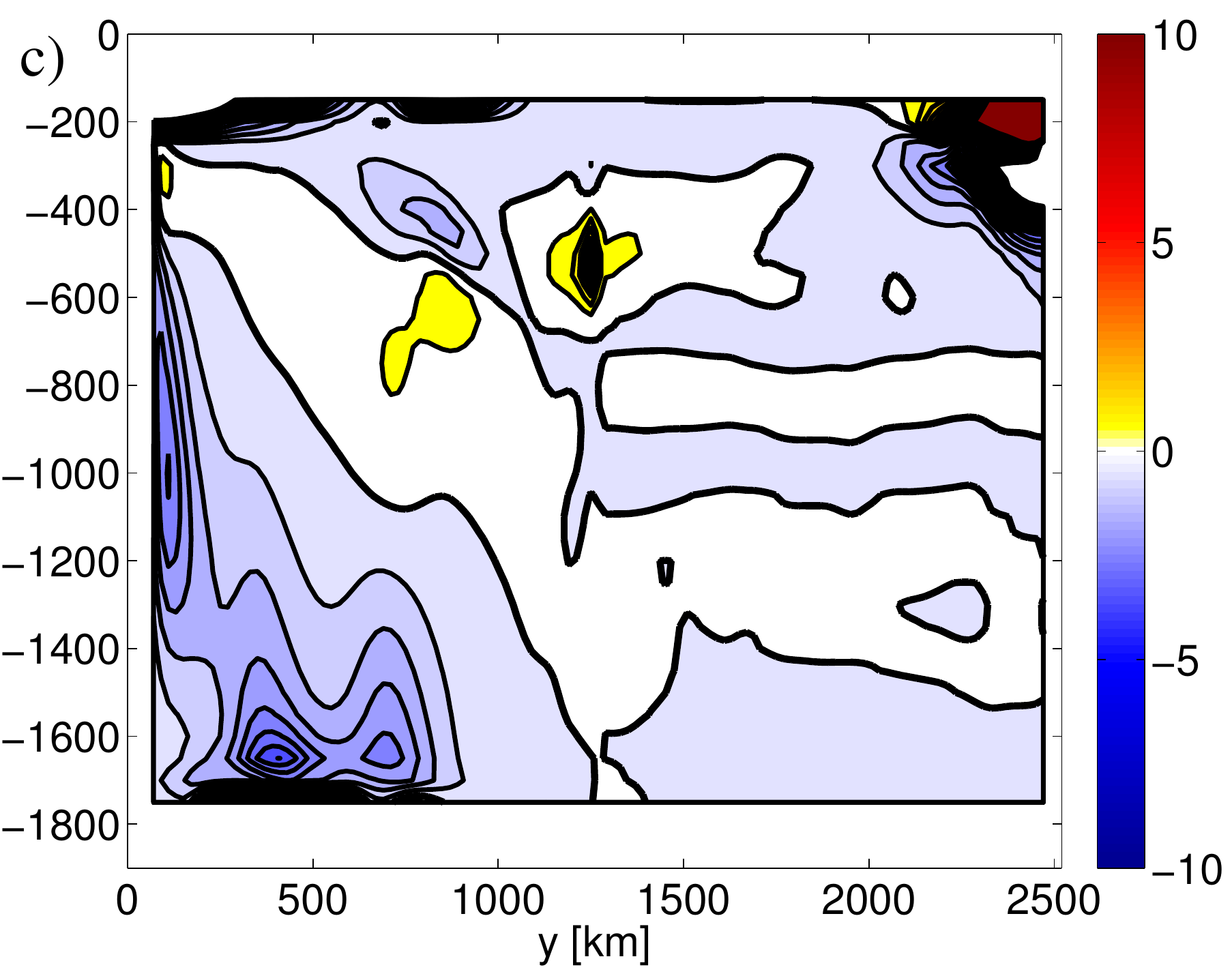}
  }
  \subfigure{
    \includegraphics[scale=0.270]{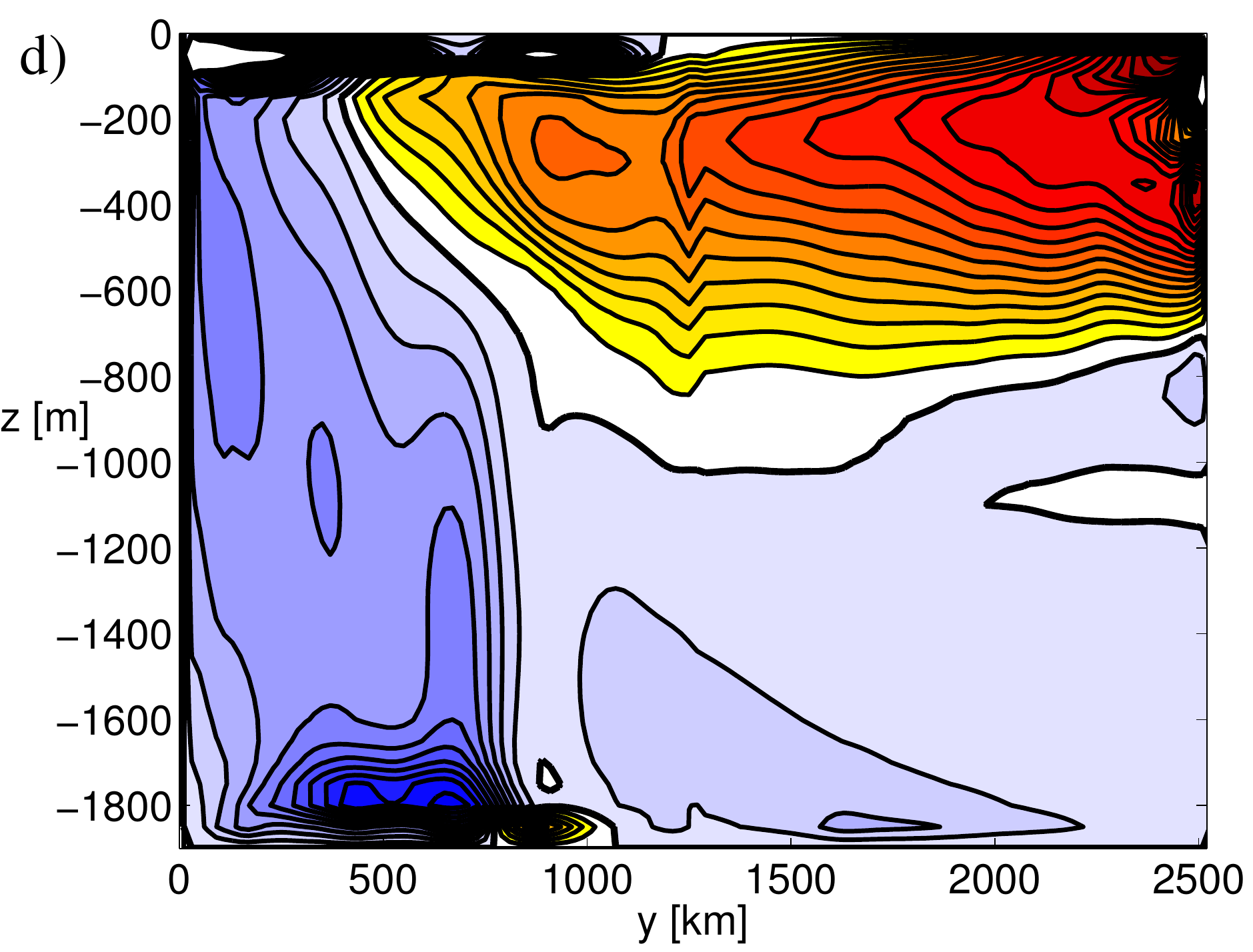}
  }
  \subfigure{
    \includegraphics[scale=0.270]{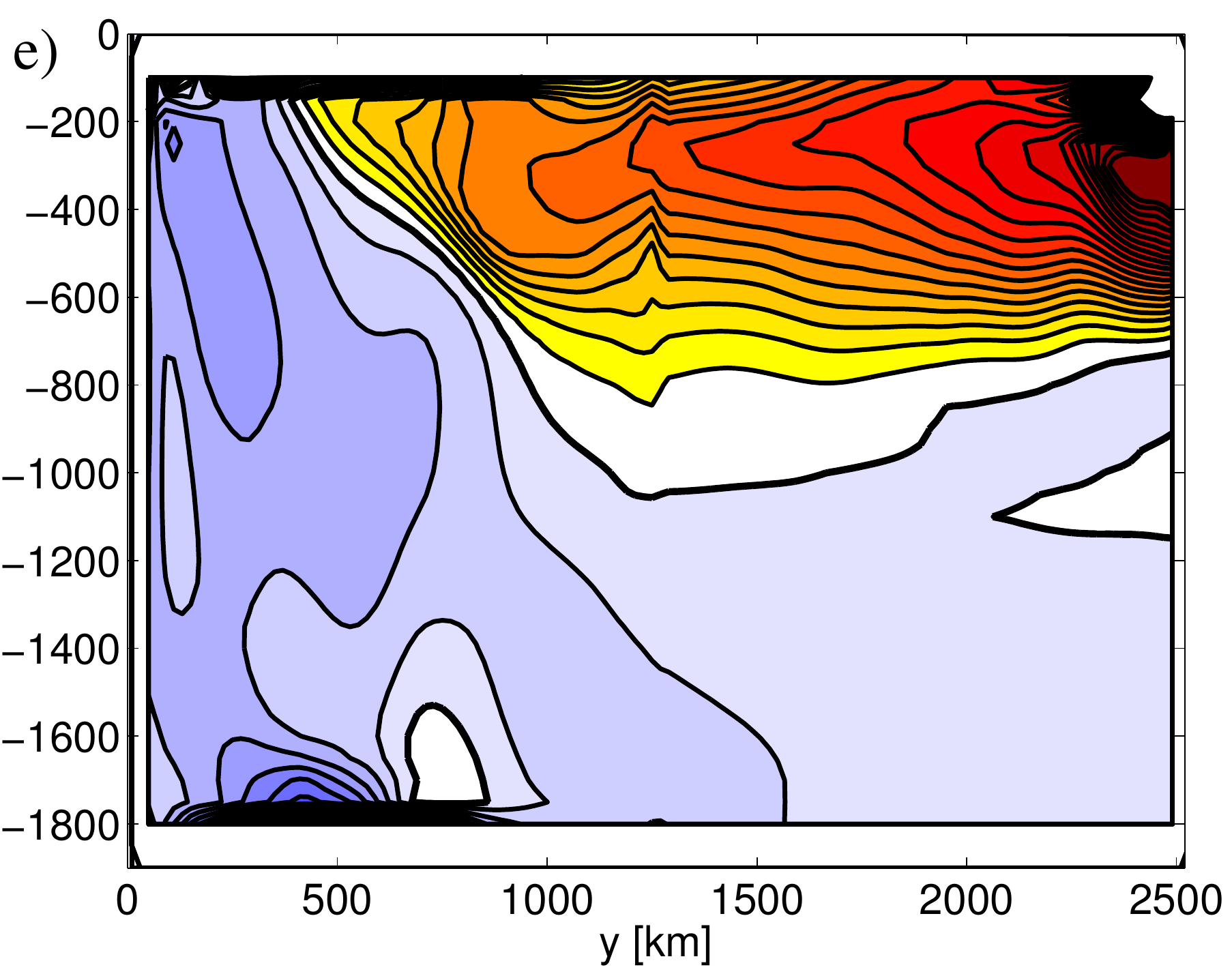}
  }
  \subfigure{
    \includegraphics[scale=0.270]{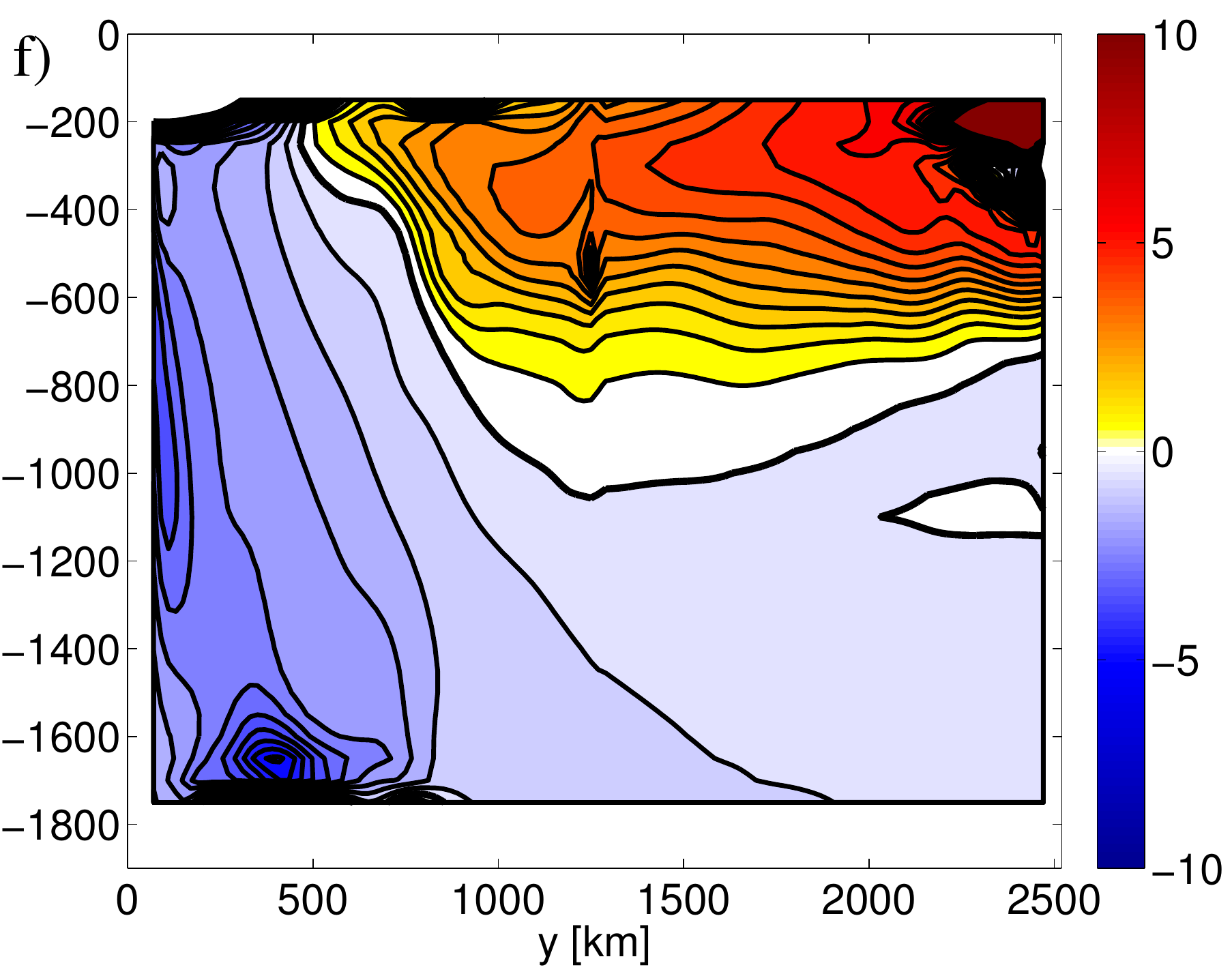}
  }
  \subfigure{
    \includegraphics[scale=0.270]{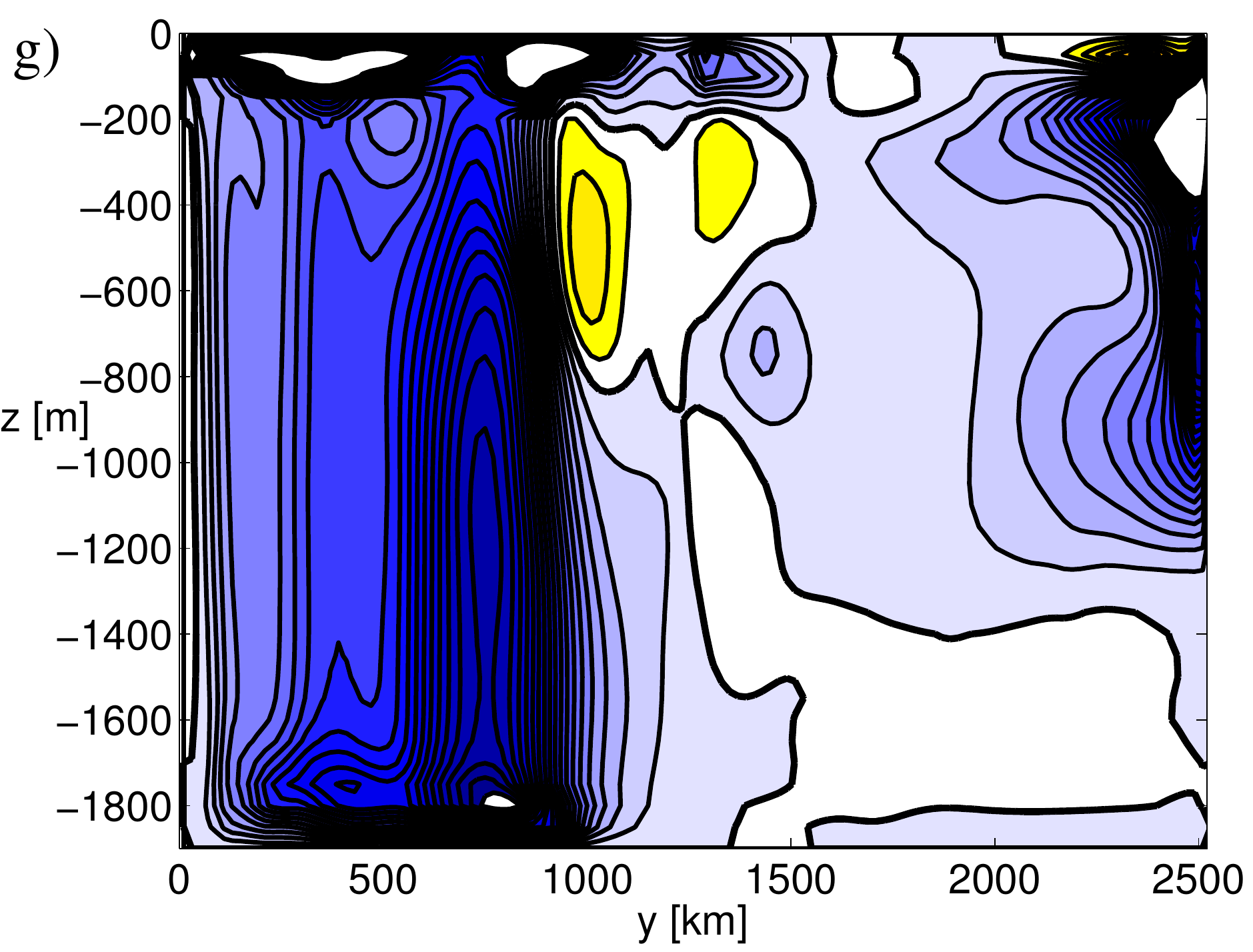}
  }
  \subfigure{
    \includegraphics[scale=0.270]{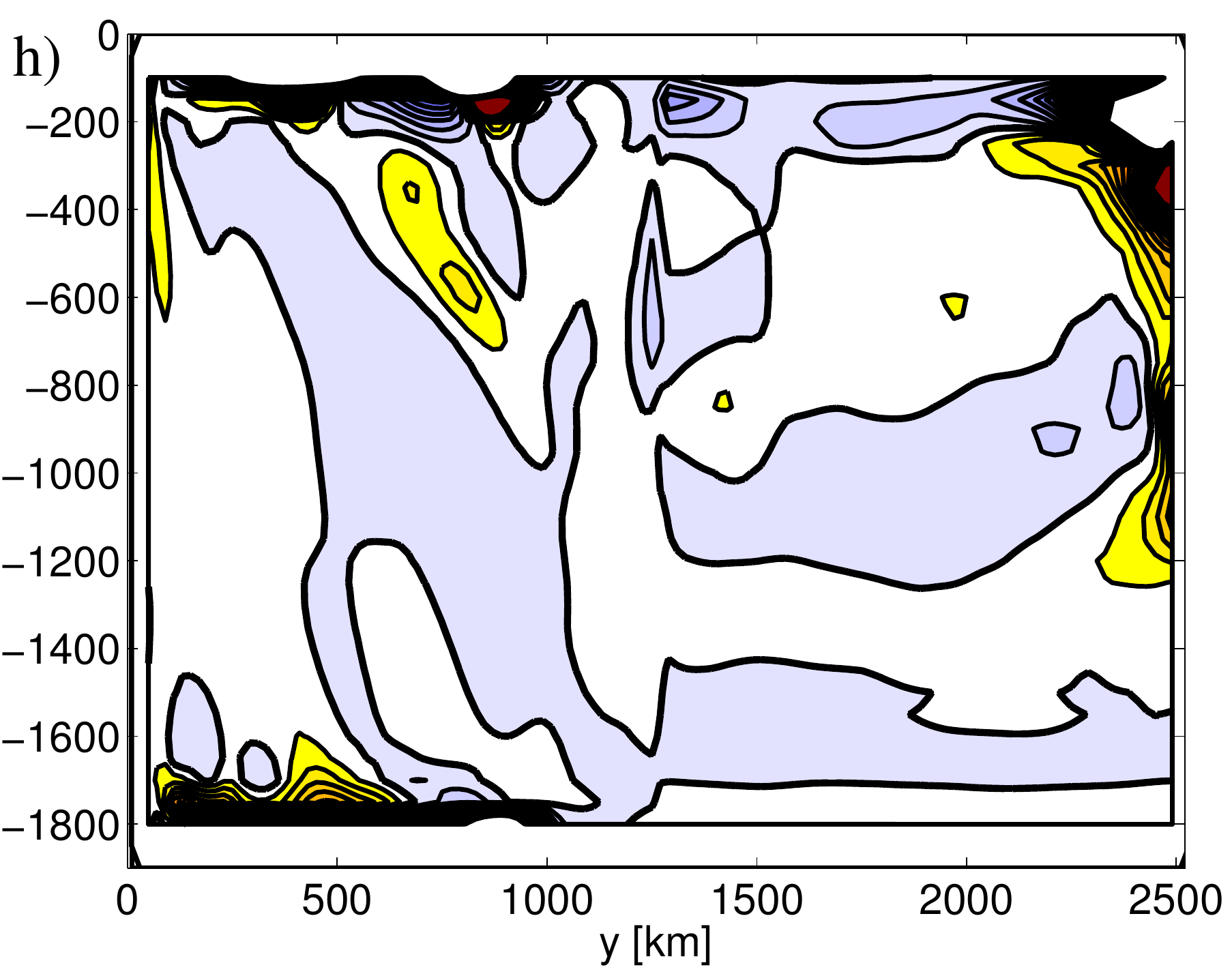}
  }
  \subfigure{
    \includegraphics[scale=0.270]{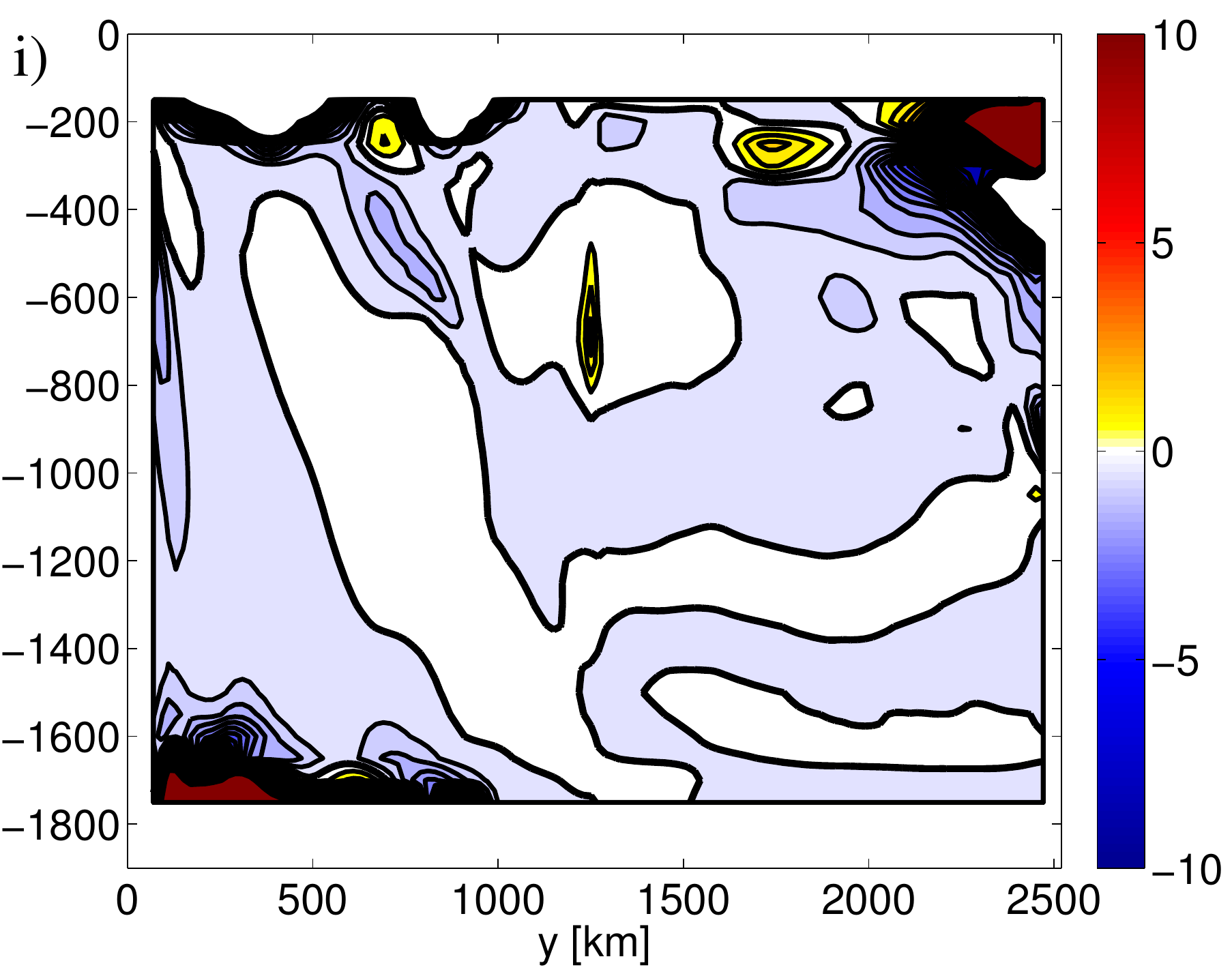}
  }
  \subfigure{
    \includegraphics[scale=0.270]{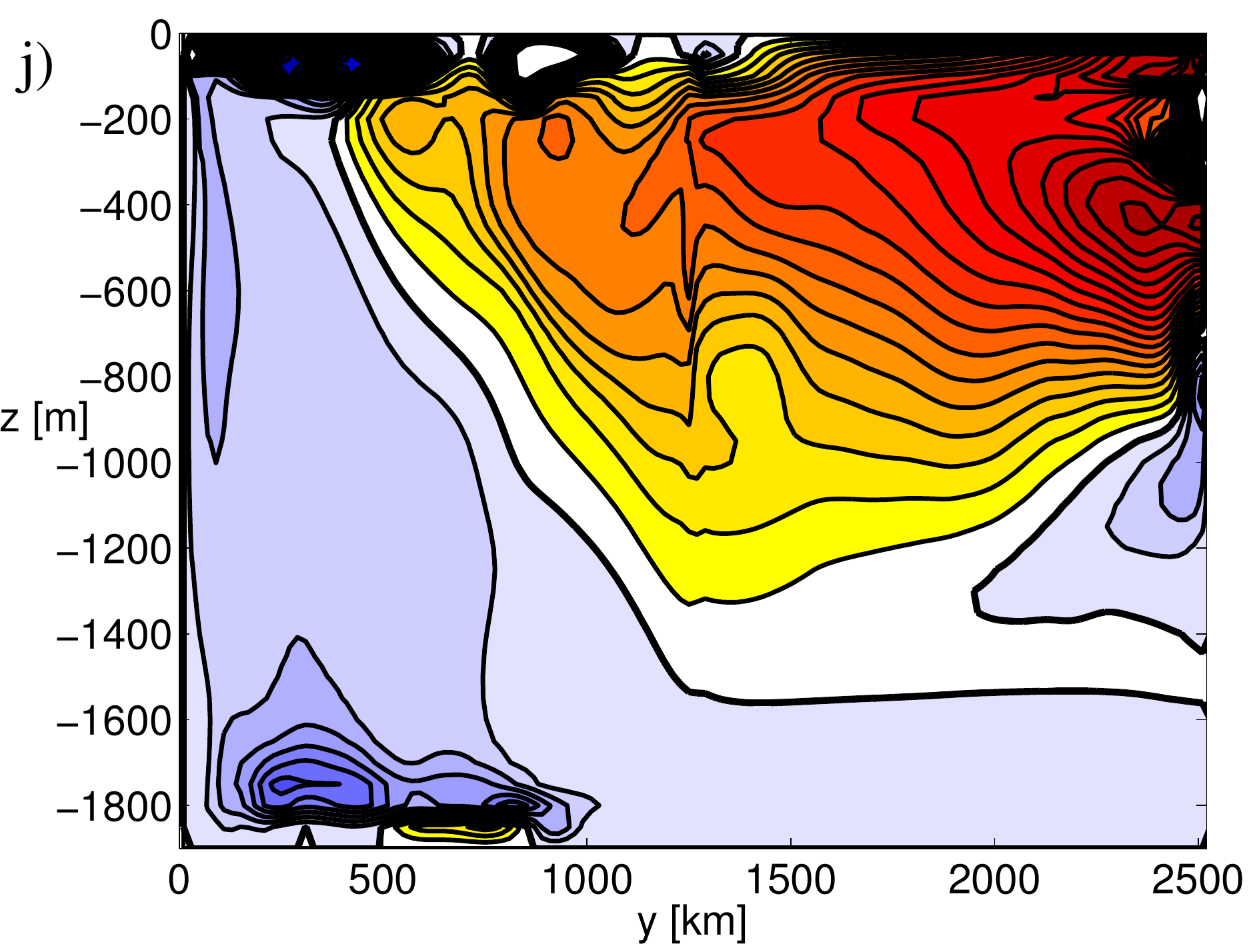}
  }
  \subfigure{
    \includegraphics[scale=0.270]{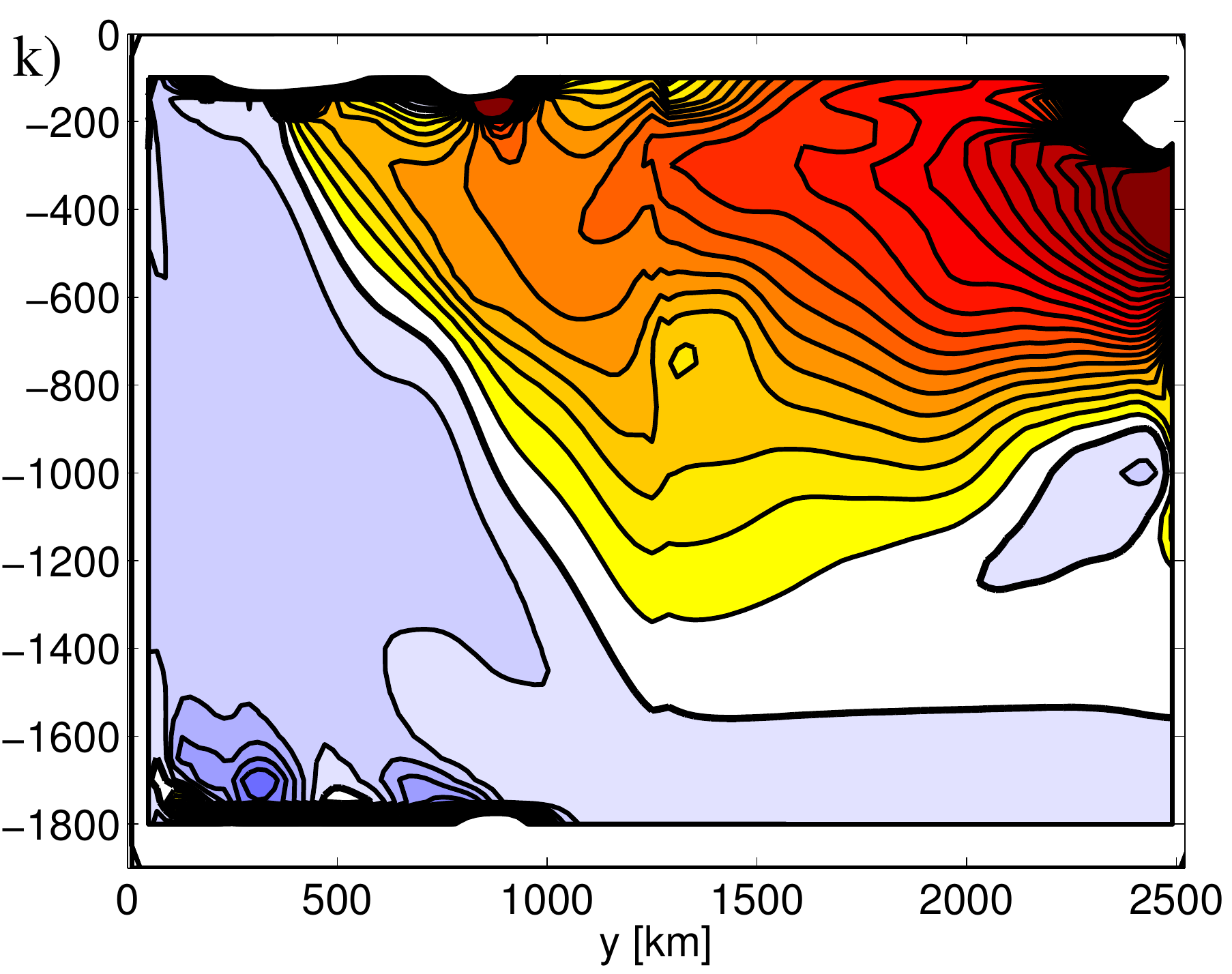}
  }
  \subfigure{
    \includegraphics[scale=0.270]{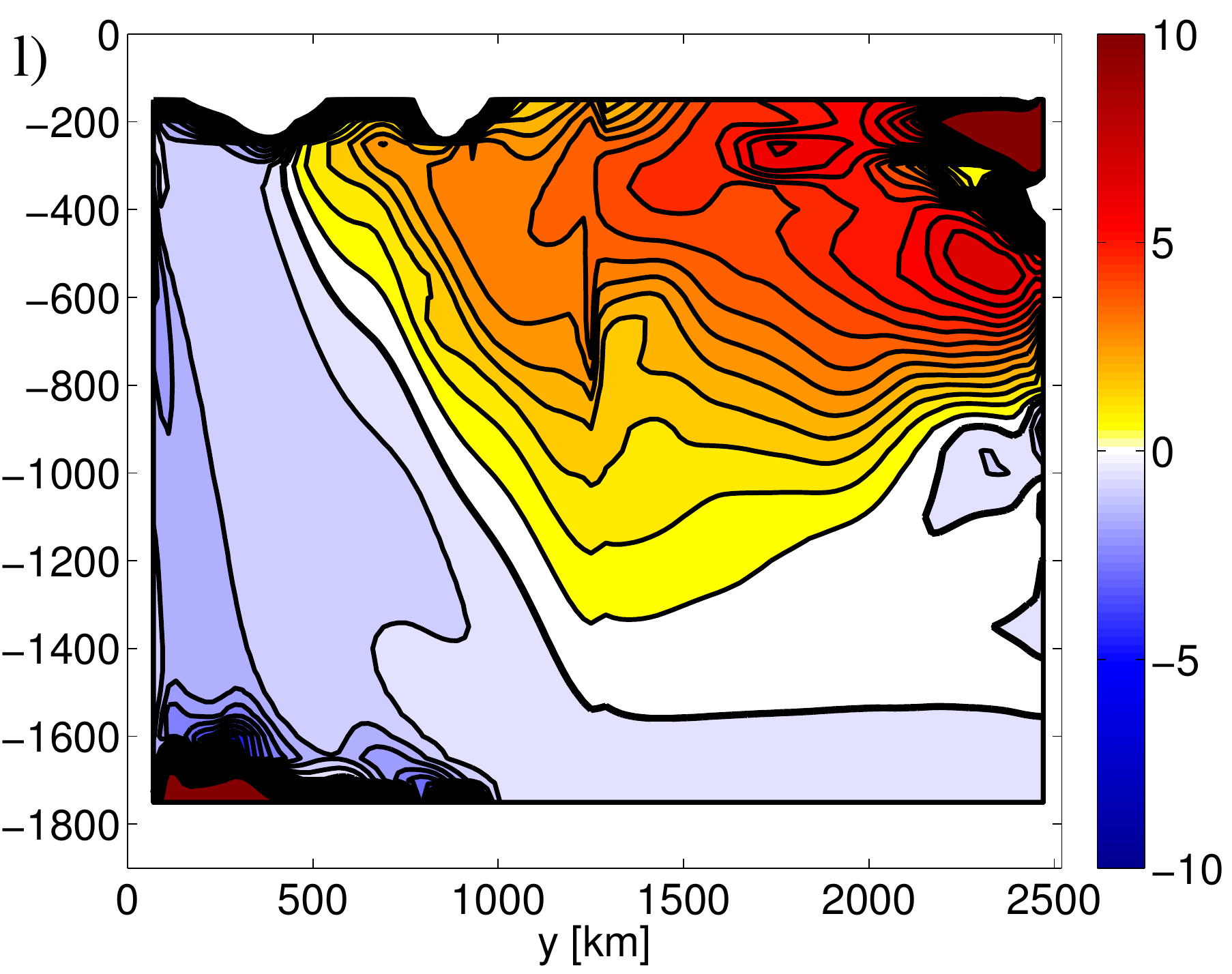}
  }
\caption{
The first three terms of the series expansion of $\psi^*$ for the flat case:
$\psi^*_{I}$ (a), $\psi^*_{II}$ (b), $\psi^*_{III}$ (c)
and the corresponding residual streamfunctions including terms of $\psi^*$ up to the first (d), second (e), third (f) order.
(g-i) and (j-l) show the same quantities, but for the flat case experiment including a harmonic viscosity of $A_h=2000\mathrm{m}^2\mathrm{s}^{-1}$.
The contour interval is $0.5$Sv and zero lines are thick.
}
\label{RMflat}
\end{figure}

Fig. \ref{RMflat} a) shows $\psi^*_{I}$, i.e. the TEM eddy streamfunction of the flat case,
which shows the well-known eddy-induced streamfunction behaviour in the SO (extending from $x=0$km to $x=L/2$):
A strong negative circulation pattern, which opposes the positive circulation pattern of ${\Lambda}$ (not shown, see e.g. \cite{Viebahn:12})
in the SO, such that the sum of both, i.e. the TEM residual streamfunction shown in Fig. \ref{RMflat} d), is given by two global overturning cells.
Namely, a positive circulation cell which connects the SO and the Atlantic and a bottom reaching negative circulation cell.
We notice that, on the one hand, the residual MOC in the SO is of equal magnitude as the residual MOC of the NL case\footnote{
That is, the water mass transformations are similar in both cases \citep{Walin:82,Marshall:03}.
},
but, on the other hand, the magnitude of $\psi^*_{I}$ is significantly larger, since ${\Lambda}$ is not small.
However, we find problems at the boundaries analog to the problems already encountered in the NL case.
$\psi^*_{I}$ now shows large and unphysical\footnote{
We note that ``unphysical'' primarily means that $\psi^*_{I}$ does not tend to zero at the horizontal boundaries,
while, for averaging along latitude circles, negative recirculation cells at the surface are also found in an isopycnal averaging framework - as shown for both the flat case and the hill case in Fig. \ref{isopycnalframework} b,c) and discussed by \cite{Viebahn:12}.
}
negative recirculation cells in the surface boundary layer and in the bottom boundary layer in the SO,
which are also present in the TEM residual MOC ${\Lambda}+\psi^*_{I}$.
Due to the coarser vertical resolution of our second model configuration, we can not discuss in detail the boundary layer behaviour of the next higher order terms of the series expansion of $\psi^*$.
Nevertheless, the few boundary layer values suggest the same unsolved practical problem as in the NL case, namely,
next higher order terms $\psi^*_{i}$ of increasing magnitude
such that an adequate approximate boundary layer circulation given by low order terms is impossible.

But we can consider the behaviour of the subsequent terms of the series expansion of $\psi^*$ in the ocean interior.
Fig. \ref{RMflat} b,c) show $\psi^*_{II}$ and $\psi^*_{III}$.
As expected, $\psi^*_{I}$ dominates over $\psi^*_{II}$ and $\psi^*_{III}$ in the interior of the SO.
However, both $\psi^*_{II}$ and $\psi^*_{III}$, although being smaller than $\psi^*_{I}$ everywhere in the interior of the SO,
show magnitudes of the same order as $\psi^*_{I}$ below $z=-950$m and at mid-depth around $y=1000$km.
Moreover, we notice the strong changes in the diabatic northern convective region.
Hence, both terms induce significant changes in the residual MOC (Fig. \ref{RMflat} e,f)):
Especially, the negative circulation cell changes both magnitude and circulation pattern by the inclusion of each term.
In case both $\psi^*_{II}$ and $\psi^*_{III}$ are included (Fig. \ref{RMflat} f)), the streamlines of the residual MOC in the SO are significantly more aligned along the time-zonal-mean isopycnals in the interior (compare Fig. \ref{RMflat} d-f) with Fig. \ref{isopycnalframework} b,e)).
Hence, in the flat case the inclusion of the next higher order terms $\psi^*_{II}$ and $\psi^*_{III}$ distinctly improves
the approximation of $\psi^*$ by $\psi^*_{I}$ alone.

Furthermore, $\psi^*_{II}$ mainly opposes the interior circulation of $\psi^*_{I}$, while $\psi^*_{III}$ largely amplifies it.
That is, the series expansion of $\psi^*$ is again alternating (similar to the NL case),
which is obvious in Fig. \ref{RMflat} a-c) and continued for the next higher order terms $\psi^*_{IV}$ and $\psi^*_{V}$ (not shown).
However, with increasing order $i$ the ratio $|\psi^*_i|/|\psi^*_I|$ (not shown) decreases in the interior of the SO,
which suggests that the desired behaviour of an in general decreasing series expansion essentially holds in the interior of the SO.
We can conclude: In the flat case, the series expansion of $\psi^*$ is alternating and may be adequately approximated by low order terms in the ocean interior, since there they appear to be decreasing.
However, the next higher order terms $\psi^*_{II}$ and $\psi^*_{III}$ significantly improve an approximation of $\psi^*$ by $\psi^*_{I}$ alone.
Hence, the series approach of \cite{Eden:07} represents an advancement of the description of the ocean interior circulation
within the residual-mean framework.

Now, in order to minimise the impact of the next higher order terms $\psi^*_{II}$ and $\psi^*_{III}$,
i.e. in order to obtain a faster convergence of the series expansion of $\psi^*$ in the interior of the SO,
we introduced a harmonic viscosity of $A_h=2000\mathrm{m}^2\mathrm{s}^{-1}$ in the flat case.
$A_h$ acts to damp EKE and hence
the corresponding TEM eddy streamfunction $\psi^*_I$, shown in Fig. \ref{RMflat} g),
has a weaker but still strong negative circulation cell in the SO
(while the negative circulation in the northern convective region is increased and extended).
The TEM residual MOC, i.e. ${\Lambda}+\psi^*_I$ (Fig. \ref{RMflat} j)), shows a significantly weaker bottom reaching negative circulation cell,
while the global positive circulation cell extends much deeper, but remains of the same magnitude in the SO (of course, also ${\Lambda}$ (not shown) changes).
The next higher order terms $\psi^*_{II}$ and $\psi^*_{III}$ shown in Fig. \ref{RMflat} h,i) (as well as $\psi^*_{IV}$ and $\psi^*_{V}$, not shown)
are now significantly reduced (except for an increase in the northern convective region),
such that in the interior of the SO, $\psi^*$ is essentially given by $\psi^*_I$.
Only in a small band at mid-depth around $y=1000$km the next higher order terms $\psi^*_{II}$ and $\psi^*_{III}$ show significant values,
which almost disappear for $\psi^*_{IV}$ and $\psi^*_{V}$ (not shown).
The reduced impact of the next higher order terms is more accurately expressed by the behaviour of the ratio $|\psi^*_i|/|\psi^*_I|$ (not shown),
which is also drastically reduced in the interior of the SO.
Consequently, in the flat case the impact on the TEM eddy streamfunction $\psi^*_I$ by the gauge potential introduced by \cite{Eden:07} (see Eq. (\ref{AthetaFULL})) related to rotational eddy fluxes depends directly on the ``strength'' of the eddy field, i.e. the magnitude of the EKE.
Hence, if the EKE is adequately reduced, it seems acceptable to approximate $\psi^*$ by $\psi^*_I$ in the nearly adiabatic interior of the SO in the flat case.
We notice that the series expansion remains alternating
and note that in the diabatic regions at the surface, at the bottom and at the northern and southern boundaries
convergence is far from being reached by including the first three orders (see Fig. \ref{RMflat} g-i)).

\subsection{Hill case}\label{RMhillcase}
\begin{figure}[t]
  \centering
  \subfigure{
    \includegraphics[scale=0.270]{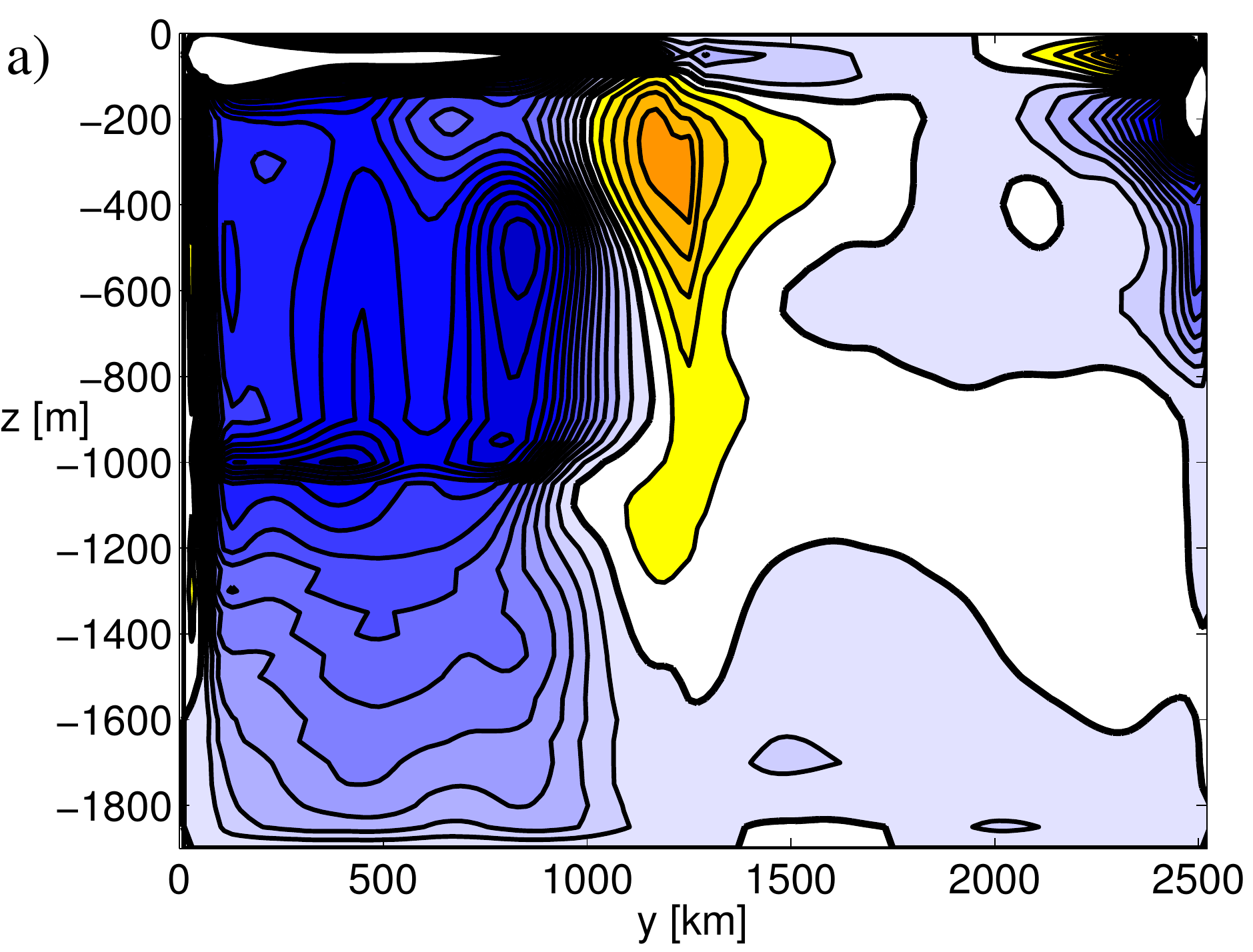}
  }
  \subfigure{
    \includegraphics[scale=0.270]{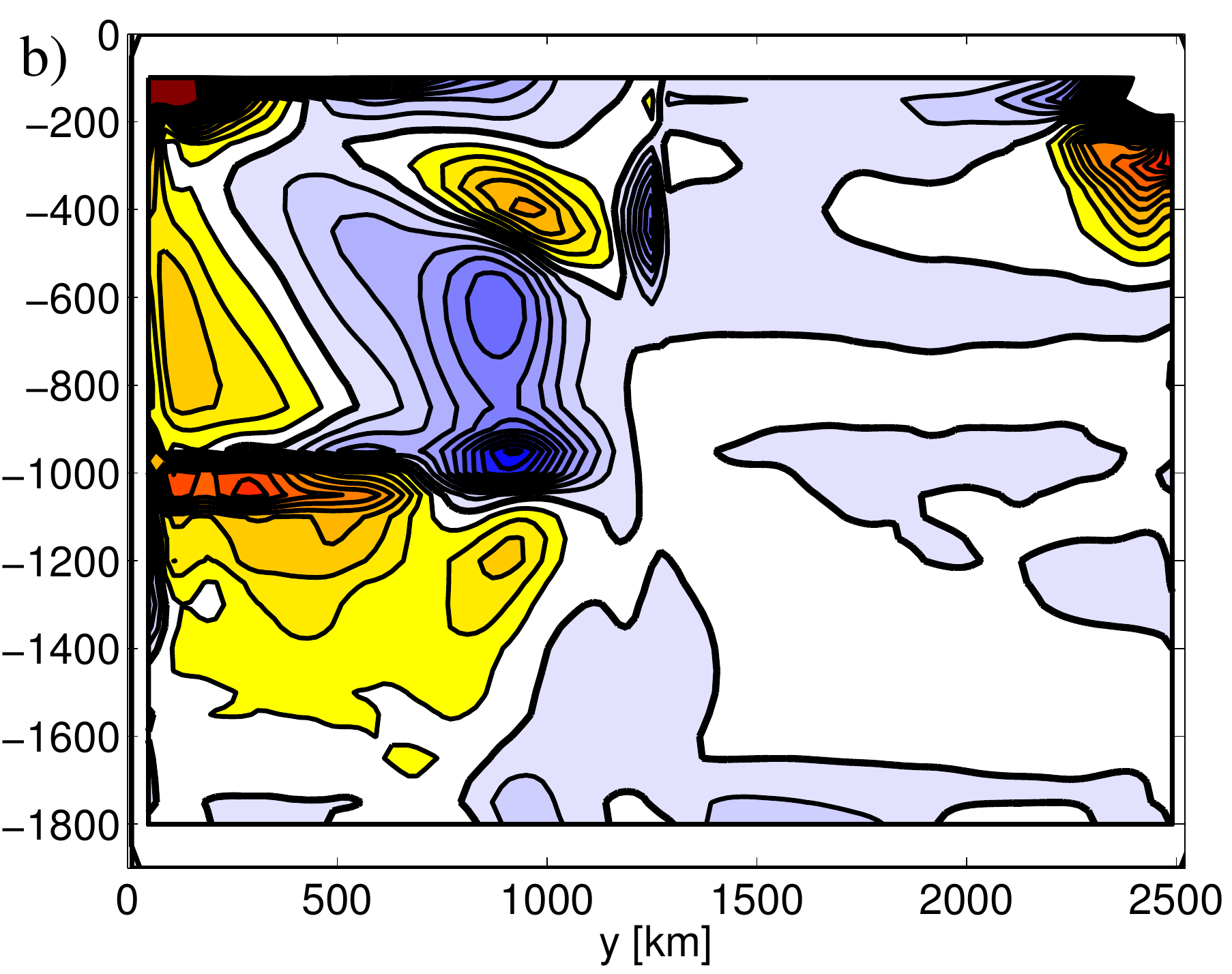}
  }
  \subfigure{
    \includegraphics[scale=0.270]{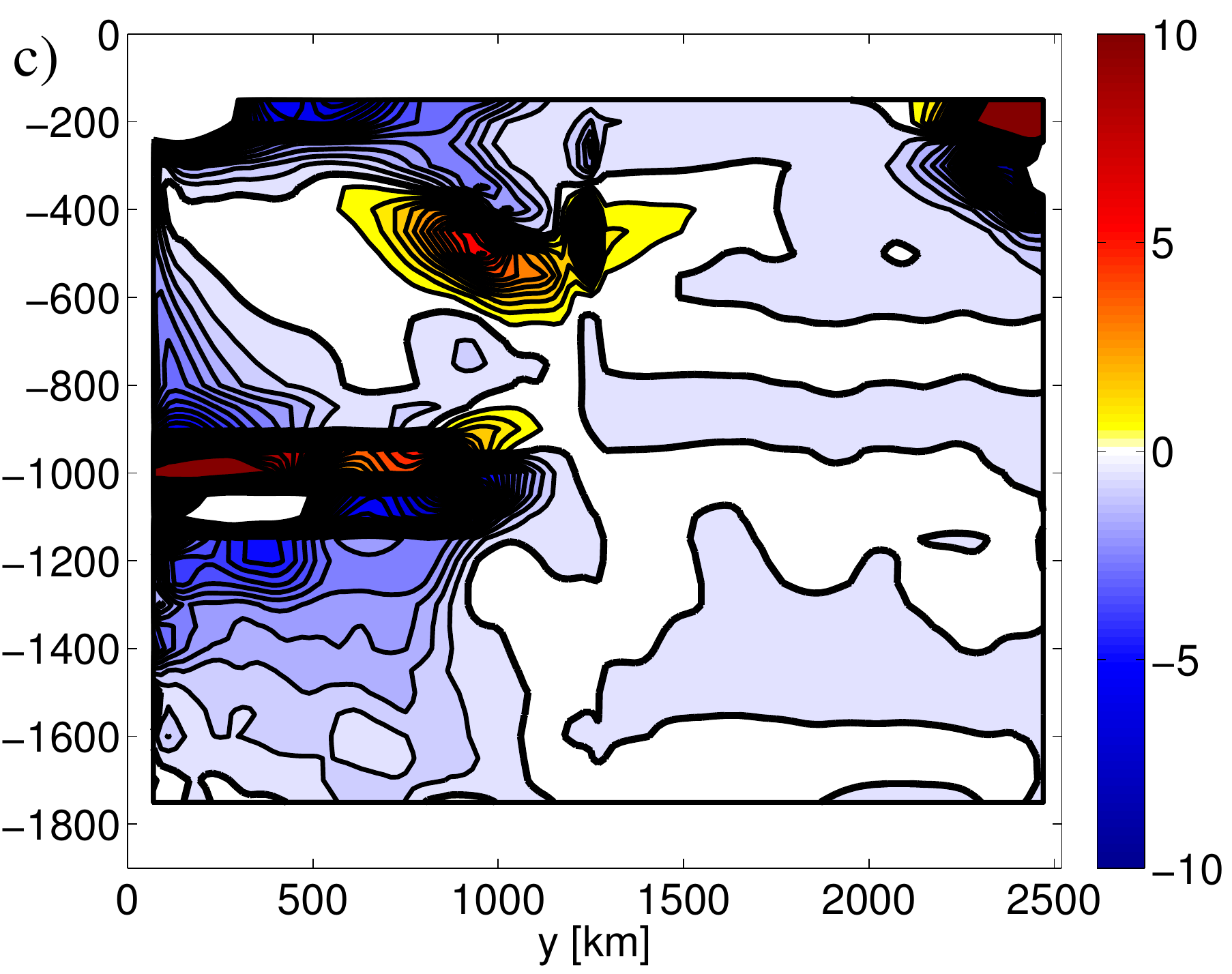}
  }
  \subfigure{
    \includegraphics[scale=0.270]{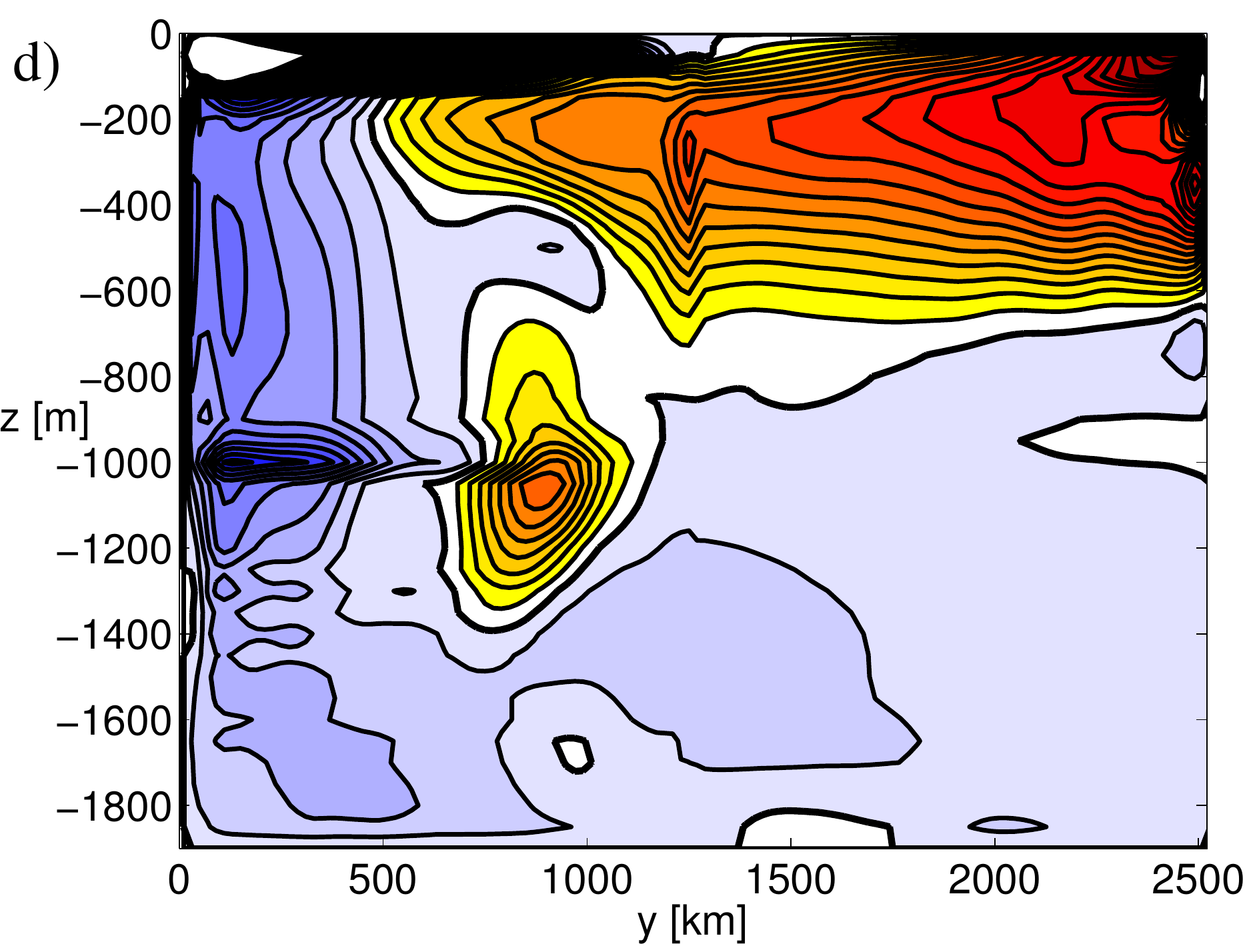}
  }
  \subfigure{
    \includegraphics[scale=0.270]{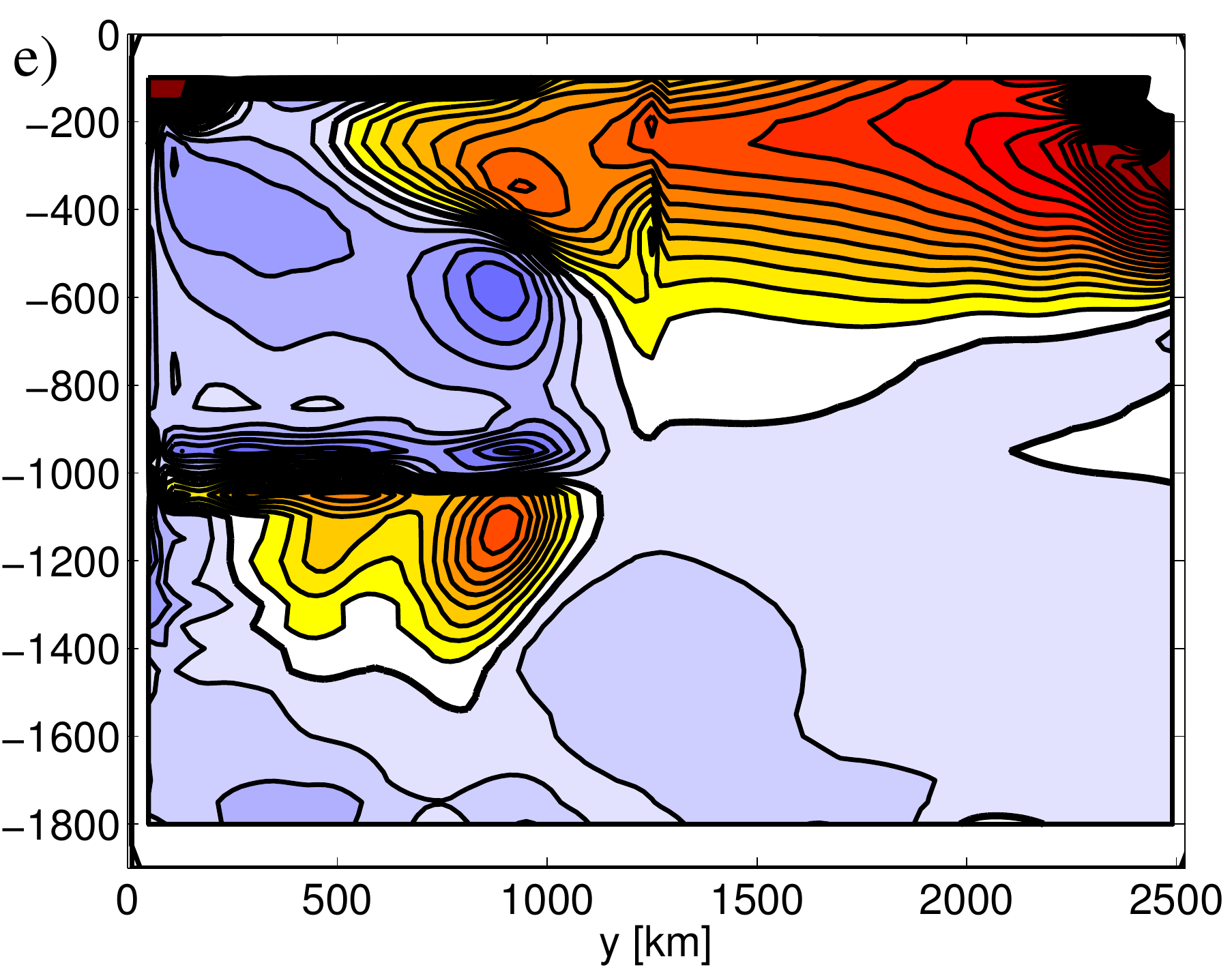}
  }
  \subfigure{
    \includegraphics[scale=0.270]{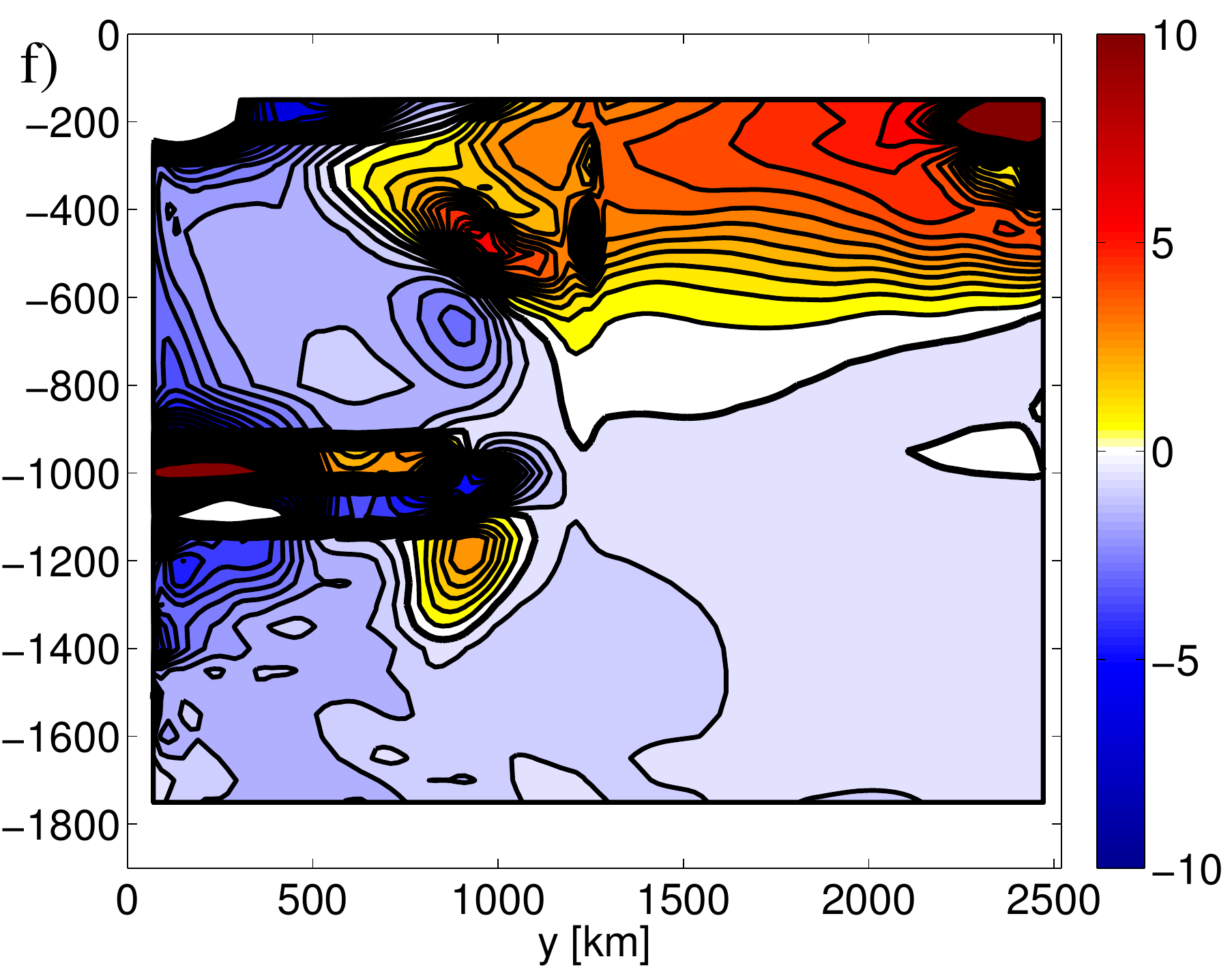}
  }
\caption{
The first three terms of the series expansion of $\psi^*$ for the hill case:
$\psi^*_{I}$ (a), $\psi^*_{II}$ (b), $\psi^*_{III}$ (c)
and the corresponding residual streamfunctions including terms of $\psi^*$ up to the first (d), second (e), third (f) order.
The contour interval is $0.5$Sv and zero lines are thick.
}
\label{hillsill}
\end{figure}

Fig. \ref{hillsill} a) shows the TEM eddy streamfunction $\psi^*_I$ of the hill case.
Compared to the flat case, the negative circulation is reduced at topographic depths,
in particular a bottom boundary layer is absent.
This is in accordance with the geostrophic return flow of ${\Lambda}$ (not shown, see \cite{Viebahn:12}) in the hill case,
which extends over the depth range below the hill depth 
and is not confined to a bottom boundary layer as in the flat case.
On the other hand, the number of local maxima in $\psi^*_I$ is increased:
One local maximum is found above topography around $y=800$km and $500$m depth.
Moreover, meridional recirculation cells appear around the 
hill depth\footnote{
Note that in the isopycnal eddy streamfunction meridional recirculation cells around the hill depth do not appear,
as shown by \cite{Viebahn:12}. See also Fig. \ref{isopycnalframework} c).
}
($z=-950$m),
which induce a negative and a positive recirculation cell around the hill depth in the TEM residual MOC ${\Lambda}+\psi^*_I$ (Fig. \ref{hillsill} d)).
These recirculation cells represent strong diapycnal flow and hence contradict the physical picture of a nearly adiabatic flow in the interior of the SO.
Consequently, we would expect the next higher order terms to reduce these cells in order to obtain a physically more consistent circulation pattern.

Fig. \ref{hillsill} b,c) show $\psi^*_{II}$ and $\psi^*_{III}$.
While $\psi^*_{II}$ and $\psi^*_{III}$ exhibit the same circulation pattern in the Atlantic part as in the flat case,
they are drastically increased in the SO part, in particular around the hill depth (along the entire meridional extension of the SO)
and above topography (around $y=800$km and $400$m depth).
The maximal values now lie around the hill depth and not near the bottom as in the flat case.
$\psi^*_{II}$ partially compensates for the spurious diabatic recirculation cells in the residual MOC ${\Lambda}+\psi^*_{I}$
with a negative circulation around $y=900$km and a positive circulation around $y=400$km at the hill depth.
Nevertheless, the inclusion of $\psi^*_{II}$ appears to overcompensate (Fig. \ref{hillsill} e)):
The number and the magnitude of the recirculation cells around the hill depth and above topography is increased such that the overall circulation pattern becomes more unphysical.
This tendency of intensifying the recirculation cells and complicating the circulation pattern continues, if $\psi^*_{III}$ is included (Fig. \ref{hillsill} f)).

Furthermore, the magnitude of $\psi^*_{III}$ is even larger than the magnitude of $\psi^*_{II}$ for most parts of the SO.
More precisely, we find that with increasing order $i$ the ratio $|\psi^*_i|/|\psi^*_I|$ (not shown) increases in the interior of the SO,
with values greater $1$ around the hill depth already for $i=II$.
As in the flat case, the next higher order terms $\psi^*_{II}$ and $\psi^*_{III}$ (also $\psi^*_{IV}$ and $\psi^*_V$, not shown) of the series expansion of $\psi^*$ still show a type of alternating behaviour,
but the behaviour of a decreasing series expansion seems to be completely lost in the hill case.

This drawback of a, at least initially, increasing series expansion does not disappear, if a harmonic viscosity $A_h$ is introduced in the hill case\footnote{
We do not show further hill case figures due to their physical disqualification.
}.
By increasing $A_h$, expectedly the overall magnitude of the TEM eddy streamfunction $\psi^*_I$ decreases.
Moreover, the circulation pattern of $\psi^*_I$ deforms with increasing $A_h$,
such that the negative recirculation cell at the hill depth of the residual MOC ${\Lambda}+\psi^*_I$ decreases.
For example, if $A_h=10000\mathrm{m}^2\mathrm{s}^{-1}$ is used, the negative circulation cell of the TEM residual MOC is nearly void of recirculation cells in the nearly adiabatic interior, but the accompanying positive recirculation cell is drastically increased.
While $A_h=2000\mathrm{m}^2\mathrm{s}^{-1}$ has a rather small impact on $\psi^*$ and the residual MOC in the hill case,
$A_h=5000\mathrm{m}^2\mathrm{s}^{-1}$ leads to the strongest reduction of the next higher order terms $\psi^*_{II}$ and $\psi^*_{III}$ (in particular above topography) for our set of $A_h$ values.
Nevertheless, for all three values of $A_h$ the next higher order terms $\psi^*_i$ (we considered terms up to $i=V$) conserve the previous features:
In an alternating manner, the magnitudes increase and the circulation patterns complicate with higher orders $i$.
In particular, the ratios $|\psi^*_i|/|\psi^*_I|$ (not shown) increase in the SO with more and more regions in the SO of values greater than $1$.

Consequently, in the hill case the series expansion of $\psi^*$ may not have a reasonable cut off,
since it seems to be, at least initially, an increasing series expansion in broad regions of the SO.
Now even in the interior the residual-mean approach, both in its classical TEM version and in its advancement by \cite{Eden:07}, is dissatisfying.

\section{Comparison with the quasi-Stokes streamfunction}\label{STOKES}
Assuming that the buoyancy field $b(x,y,z,t)$ is vertically strictly monotonic in the entire ocean,
the instantaneous isopycnal $b_a$ lies at an instantaneous height $z(x,y,b_a,t)=z_a(y,b_a)+z'_a(x,y,b_a,t)$,
where $z_a$ is the time-zonal-mean height of $b_a$ and $z_a'$ is the deviation from the time-zonal-mean, i.e. $\ol{z_a'}=0$.
The time-zonal-mean isopycnal streamfunction ${\psi_I}$ is then the temporally averaged and zonally integrated
meridional transport below the instantaneous isopycnal $b_a$.
We may write
\beq
{\psi_I}(y,b_a)=-L_x\ol{\int_{bottom}^{z_a+z_a'}v\ dz}\ .
\eeq
${\psi_I}$ may be transformed to Eulerian space by identifying each $b_a$ with its mean height $z_a$.
Therewith, the quasi-Stokes streamfunction $\Psi^*$ is defined via the decomposition
\beq
-L_x\ol{\int_{bottom}^{z_a+z_a'}v\ dz}={\Lambda}(y,z_a)+\Psi^*(y,z_a)\ ,
\eeq
that is,
\beq
\Psi^*(y,z_a)\equiv-L_x\ol{\int_{z_a}^{z_a+z_a'}v\ dz}\ ,
\eeq
and gives the transport of ${\psi_I}$ related to the perturbation $z'_a$.
$\Psi^*$ is the eddy-induced streamfunction of ${\psi_I}$ in Eulerian space\footnote{
In \cite{Viebahn:12} the corresponding decomposition is defined in an isopycnal framework.
}.
Expressions of both $z'_a$ and $\Psi^*$ by Eulerian mean quantities may be obtained by expanding $b$ and $v$ in Taylor series centred around $z_a$ \citep{McDougall:01,Nurser:04b} as outlined in appendix B.
If we define the orders of the series expansion by the perturbations of $b$ in order to obtain a form comparable to the residual-mean framework (see Eq. (\ref{psiEdFULL}) and Eq. (\ref{psiEdAPPROX})),
we find the following series expansion for $\Psi^*$ expressed by Eulerian mean quantities
(extending the approximations of previous studies, see appendix B)
\beq\label{qSexpan}
\Psi^*&=&\Psi^*_{I}+\Psi^*_{II}+\Psi^*_{III}+O(b'^4)\ ,
\eeq
where
\beq
\Psi^*_{I}&=&\frac{L_x\ol{b'v'}}{\partial_z\ol{b}}=\psi^*_I\\
\Psi^*_{II}&=&-\frac{1}{\partial_z\ol{b}}\partial_z\biggl(\frac{L_x\ol{\phi_2v}}{\partial_z\ol{b}}\biggr)
+\frac{\ol{v}}{\partial_z\ol{b}}\partial_z\biggl(\frac{L_x\ol{\phi_2}}{\partial_z\ol{b}}\biggr)=\psi^*_{II}+\psi^*_{\Delta II}\\
\Psi^*_{III}&=&\frac{1}{2}\frac{1}{\partial_z\ol{b}}\partial_z\biggl(\frac{1}{\partial_z\ol{b}}\partial_z\biggl(\frac{L_x\ol{\phi_3v}}{\partial_z\ol{b}}\biggr)\biggr)-\nonumber\\
&&-\frac{1}{2}\frac{\ol{v}}{\partial_z\ol{b}}\partial_z\biggl(\frac{1}{\partial_z\ol{b}}\partial_z\biggl(\frac{L_x\ol{\phi_3}}{\partial_z\ol{b}}\biggr)\biggr)-\frac{L_x^{-1}}{\partial_z\ol{b}}\partial_z\biggl(\frac{L_x\ol{\phi_2}}{\partial_z\ol{b}}\biggr)\partial_z\biggl(\frac{L_x\ol{v'b'}}{\partial_z\ol{b}}\biggr)\\
&=&\psi^*_{III}+\psi^*_{\Delta IIIa}+\psi^*_{\Delta IIIb}\nonumber
\eeq

\subsection{Formal comparison of $\psi^*$ and $\Psi^*$}\label{formalCOMP}
It is obvious that the complete series expansions of $\psi^*$ and $\Psi^*$ are essentially different,
since $\psi^*$ and $\Psi^*$ advect different time-zonal-mean buoyancy distributions.
However, the corresponding time-zonal-mean buoyancy distributions mainly differ in the boundary layers,
while in the nearly adiabatic interior of the ocean they are generally found to be similar \citep{Killworth:01,Nurser:04a,Viebahn:12}.
Hence, the two streamfunctions are expected to be similar there too.

By comparing Eq. (\ref{qSexpan}) with Eq. (\ref{psiEdAPPROX}),
the similarity of $\psi^*$ and $\Psi^*$ is suggested by the identity of the first order terms.
However, we find that
with increasing order the series expansions of $\psi^*$ and $\Psi^*$ deviate more and more from each other.
The difference between the second order terms, $\psi^*_{\Delta II}$, is given by a term including the time-zonal-mean meridional velocity,
which is generally small in a zonal channel.
The difference between the third order terms is constituted by a corresponding term, $\psi^*_{\Delta IIIa}$,
including the time-zonal-mean meridional velocity
and an additional term, $\psi^*_{\Delta IIIb}$, of a type not present in the residual-mean series expansion,
namely, a product of a first order term (the vertical derivative of $\Psi^*_{I}$) with a second order term (a variance term).
For higher order terms we expect even more complicated discrepancies,
especially further products between different orders, i.e. types of terms not present in the residual-mean series expansion.

Moreover, only in case of $\Psi^*$ we are able to compute the streamfunction directly in an isopycnal framework
without referring to the series expansion \citep{Nurser:04a,Viebahn:12}.
Hence, we know the result to which the series expansion of $\Psi^*$ must converge.
In case of $\psi^*$, we do not have another computational option besides the series expansion.
Especially in the diabatic regions, the residual-mean circulation may therefore not be properly determined so far,
as demonstrated in section \ref{RME}.
Furthermore, it is not even secure so far that the residual-mean series is a converging series expansion.
Up to now, the advantage of the residual-mean series over series expansion of $\Psi^*$
is that it is given in a compact and complete form,
while we have not found a corresponding expression for the series expansion of $\Psi^*$ yet.

Now we return to our three model experiments.
In each case, we discuss the three additional terms appearing in Eq. (\ref{qSexpan}) and not in Eq. (\ref{psiEdAPPROX}),
and we consider the streamfunctions to which the series expansion $\Psi^*$ has to converge.

\subsection{Differences between the approximations of $\psi^*$ and $\Psi^*$ in model experiments}
In the NL case, both $\psi^*_{\Delta II}$ and $\psi^*_{\Delta IIIa}$ (not shown), related to $\ol{v}\approx0$, essentially vanish,
even in the surface boundary layer (as far as it is resolved).
In contrast, the third order difference term $\psi^*_{\Delta IIIb}$ (not shown),
related to the product of a variance term and an eddy buoyancy flux term,
exhibits significant values in the surface boundary layer.
Consequently, the difference between the terms of the series expansions of $\Psi^*$ and $\psi^*$ appears to increase with higher order.
In accordance with our expectation, the differences $\psi^*_{\Delta II}$, $\psi^*_{\Delta IIIa}$ and $\psi^*_{\Delta IIIb}$ suggest
that $\psi^*$ and $\Psi^*$ mainly differ in the diabatic surface layer, while they are similar in the nearly adiabatic ocean interior.
More precisely, $\psi^*_{\Delta IIIb}$ is small and of the same sign as $\psi^*_{III}$ (Fig. \ref{NLflat} c)) in the northern part of the channel,
while $\psi^*_{\Delta IIIb}$ is significantly opposing $\psi^*_{III}$ in the southern part of the channel.
Hence, the maximal absolute values of $\Psi^*_{III}$ are slightly smaller than those of $\psi^*_{III}$.
This might indicate that the series expansion of $\Psi^*$ converges faster then that of $\psi^*$.
Nevertheless, the overall characteristics of the low order terms of the series expansion of $\Psi^*$ remain those described in section \ref{NLcase}.

In the flat case, 
the terms $\psi^*_{\Delta II}$, $\psi^*_{\Delta IIIa}$ and $\psi^*_{\Delta IIIb}$ (not shown) are small in the interior of the ocean,
such that also the second and third orders of the series expansions of $\psi^*$ and $\Psi^*$ coincide in the ocean interior for the flat case
(with characteristics described in section \ref{RMflatcase}).
In particular, the terms $\psi^*_{\Delta II}$ and $\psi^*_{\Delta IIIa}$ are small in the interior, although it holds $\ol{v}\neq0$.
Significant values of $\psi^*_{\Delta II}$, $\psi^*_{\Delta IIIa}$ and $\psi^*_{\Delta IIIb}$ are present in the northern convective region,
while in the surface and bottom boundary layers they are only obvious for $\psi^*_{\Delta IIIb}$
(and probably lost due to the few vertical grid points and the smaller extension for $\psi^*_{\Delta II}$ and $\psi^*_{\Delta IIIa}$).
The significant values tend to counteract the corresponding next higher order contributions shown in Fig. \ref{RMflat} b,c).
Furthermore and similar to the NL case,
the term of highest order in perturbation quantities, $\psi^*_{\Delta IIIb}$,
exhibits the highest values in the northern convective region and in the southern bottom and surface boundary layers.
This tendency of increasing compensation again might indicate that the series expansion of $\Psi^*$ converges faster than the one of $\psi^*$
in the diabatic regions in the flat case.

By including a harmonic viscosity of $A_h=2000\mathrm{m}^2\mathrm{s}^{-1}$ in the flat case configuration (not shown),
the situation is essentially unchanged:
The terms $\psi^*_{\Delta II}$, $\psi^*_{\Delta IIIa}$ and $\psi^*_{\Delta IIIb}$ remain small in the ocean interior.
In accordance with Fig. \ref{RMflat} h,i), the magnitudes of the significant values in the boundary regions are increased compared to the case of vanishing $A_h$,
such that significant values also appear in $\psi^*_{\Delta II}$ in the surface layer.
The term $\psi^*_{\Delta IIIb}$ still shows the highest values, in particular in the southern bottom and surface boundary layers of the SO,
so that a tendency of increasing compensation is furthermore present.

So far all considered cases are in line with expectations:
In the ocean interior $\psi^*$ and $\Psi^*$ essentially coincide,
while in the boundary regions significant differences between $\psi^*$ and $\Psi^*$ are given by $\psi^*_{\Delta II}$, $\psi^*_{\Delta IIIa}$ and $\psi^*_{\Delta IIIb}$,
such that the next higher order terms of the series expansion of $\Psi^*$ tend to be smaller than the corresponding terms of $\psi^*$.

\begin{figure}[t]
  \centering
  \subfigure{
    \includegraphics[scale=0.270]{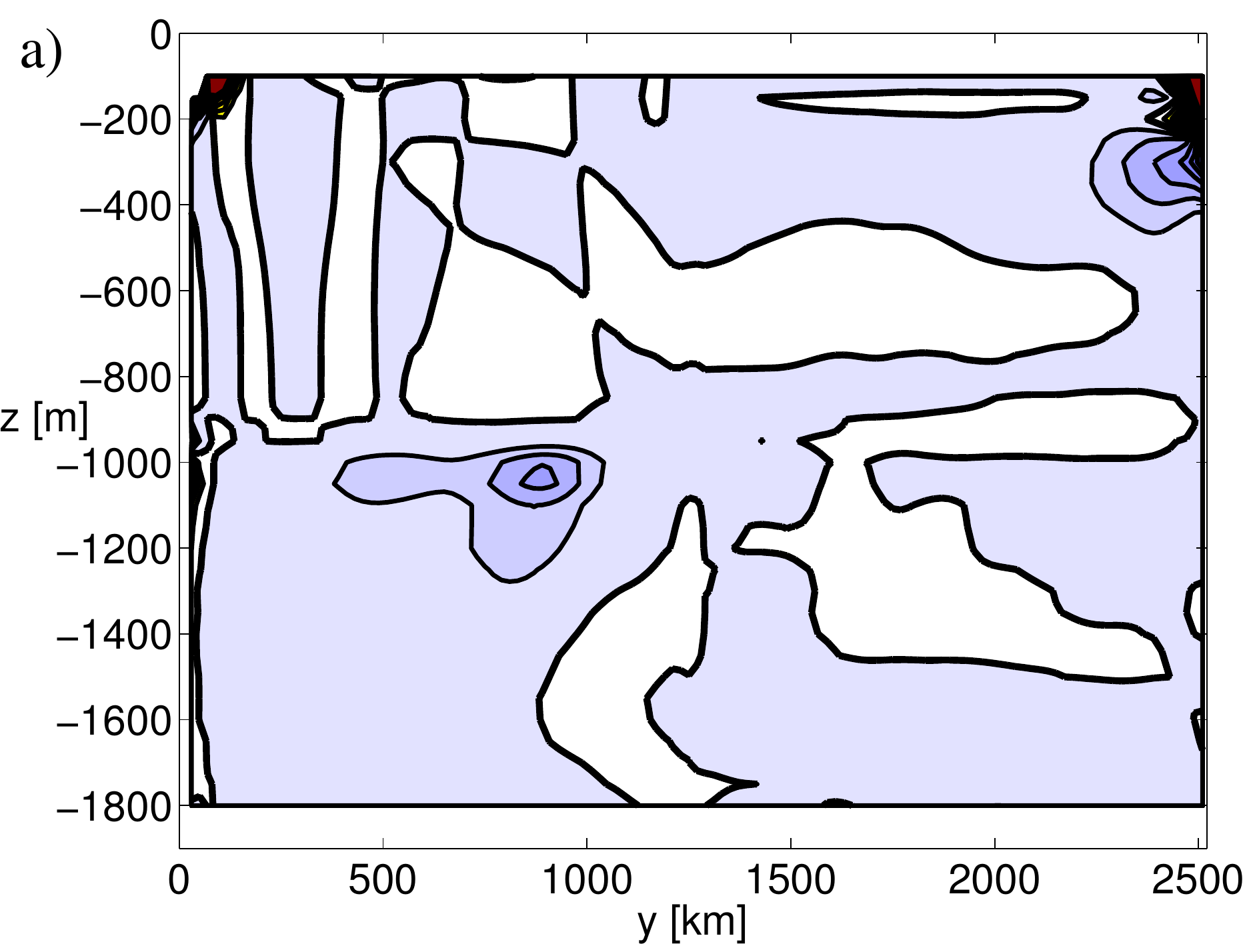}
  }
  \subfigure{
    \includegraphics[scale=0.270]{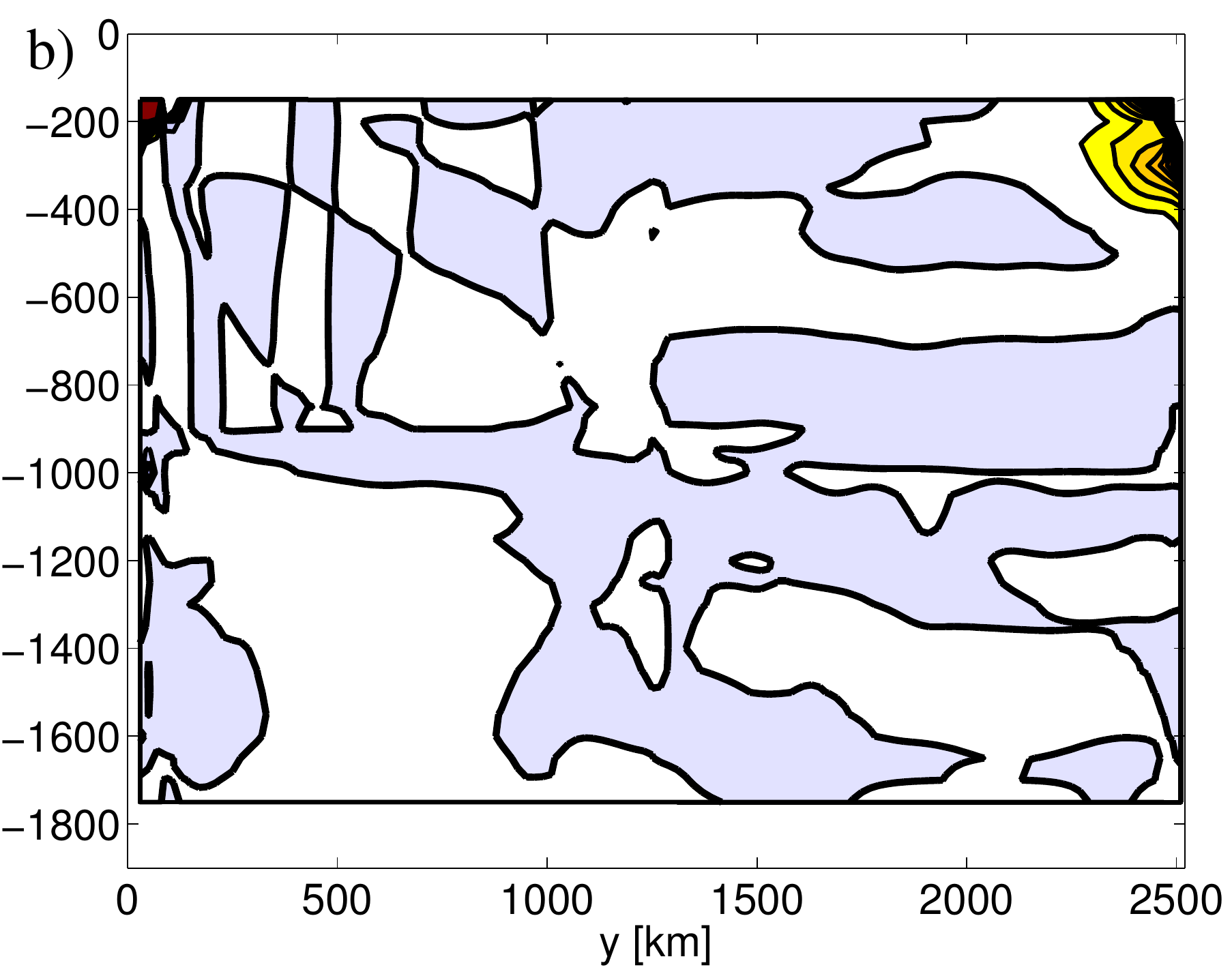}
  }
  \subfigure{
    \includegraphics[scale=0.270]{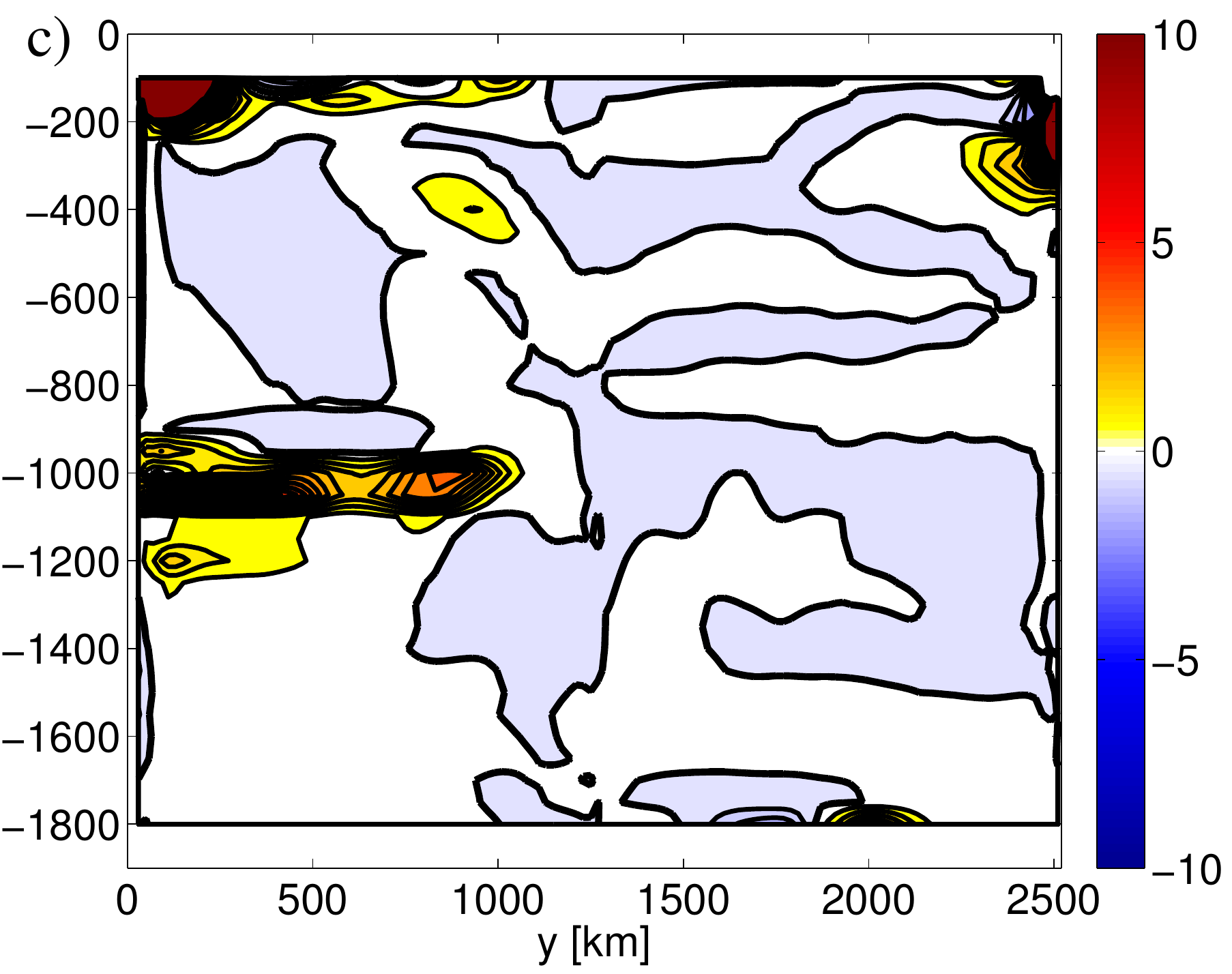}
  }
\caption{
$\psi^*_{\Delta II}$ (a), $\psi^*_{\Delta IIIa}$ (b) and $\psi^*_{\Delta IIIb}$ (c) for the hill case.
The contour interval is $0.5$Sv and zero lines are thick.
}
\label{qS}
\end{figure}

Turning to the hill case, the question is:
Do the additional terms $\psi^*_{\Delta II}$, $\psi^*_{\Delta IIIa}$ and $\psi^*_{\Delta IIIb}$ add to the residual-mean terms at each order
such that the series expansion of $\Psi^*$ is decreasing and that the recirculation cells in the ocean interior, encountered in section \ref{RMhillcase}, disappear?
This is not the case.
Fig. \ref{qS} a-c) show $\psi^*_{\Delta II}$, $\psi^*_{\Delta IIIa}$ and $\psi^*_{\Delta IIIb}$ for the hill case.
In the basin part ($x>1250$km), each term shows nearly the same pattern as in the flat case,
similar to the terms $\psi^*_{I}$, $\psi^*_{II}$ and $\psi^*_{III}$ encountered in section \ref{RMhillcase}.
However, in the SO now significant values appear in the ocean interior.
The second order term $\psi^*_{\Delta II}$, related to $\ol{v}$, exhibits a small negative recirculation cell around $y=800$km at $1100$m depth.
This leads to a small reduction of the corresponding positive recirculation cell in Fig. \ref{hillsill} e), but the overall pattern remains unchanged (not shown).
The same holds for the third order: Although $\psi^*_{\Delta IIIa}$, related to $\ol{v}$, vanishes,
$\psi^*_{\Delta IIIb}$ induces a strong positive recirculation cell at the hill depth
and a smaller one at mid-depth ($y=900$km and $z=-400$m),
which, however, only slightly change the circulation pattern of Fig. \ref{hillsill} f) (not shown).
Consequently, also in the hill case we find
that the significant values of the next higher order terms of the series expansion of $\Psi^*$ tend to be reduced compared to than those of $\psi^*$.
But the overall circulation pattern in not essentially changed by including the lower orders,
so that the unphysical recirculation cells in the ocean interior remain.

Including a harmonic viscosity $A_h$ in the hill case configuration does not change the situation.
In case of $A_h\equiv5000\mathrm{m}^2\mathrm{s}^{-1}$ (not shown),
the magnitudes of both $\psi^*_{\Delta II}$ and $\psi^*_{\Delta IIIb}$ are reduced in the ocean interior, but still significant,
while $\psi^*_{\Delta IIIa}$ now shows a negative recirculation cell around $y=800$km at $1100$m depth.
However, the unphysical circulation patterns 
are only slightly changed by the inclusion of the quasi-Stokes terms (not shown).
The analog situation is met if $A_h$ is set to $A_h\equiv10000\mathrm{m}^2\mathrm{s}^{-1}$ (not shown).
Each term exhibits recirculation cells around the top of topography with drastically increased magnitudes.
However, since the magnitudes of $\psi_{II}$ and $\psi_{III}$ are even more increased, the overall effect remains small (not shown).

\begin{figure}[t]
  \centering
  \subfigure{
    \includegraphics[scale=0.270]{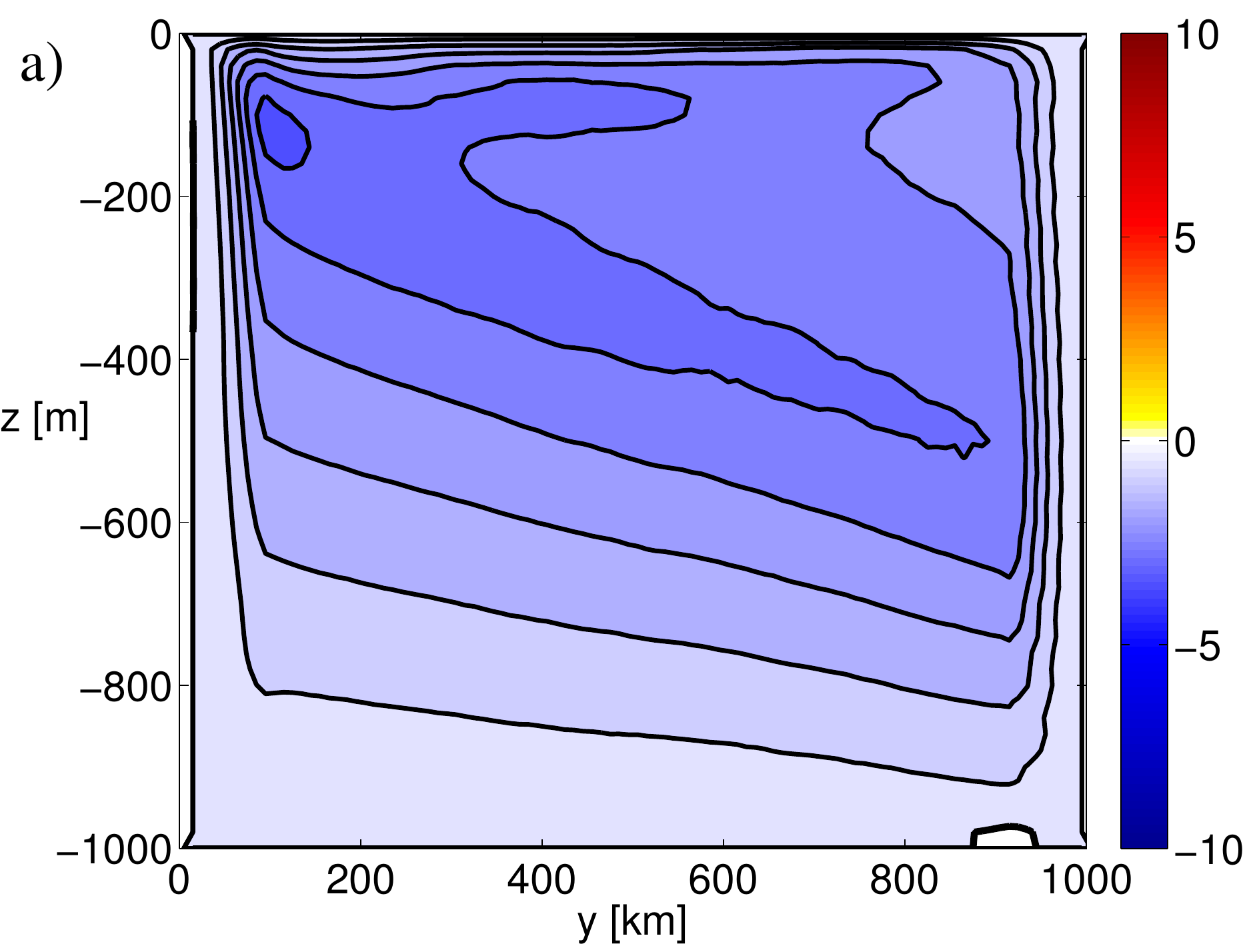}
  }
  \subfigure{
    \includegraphics[scale=0.270]{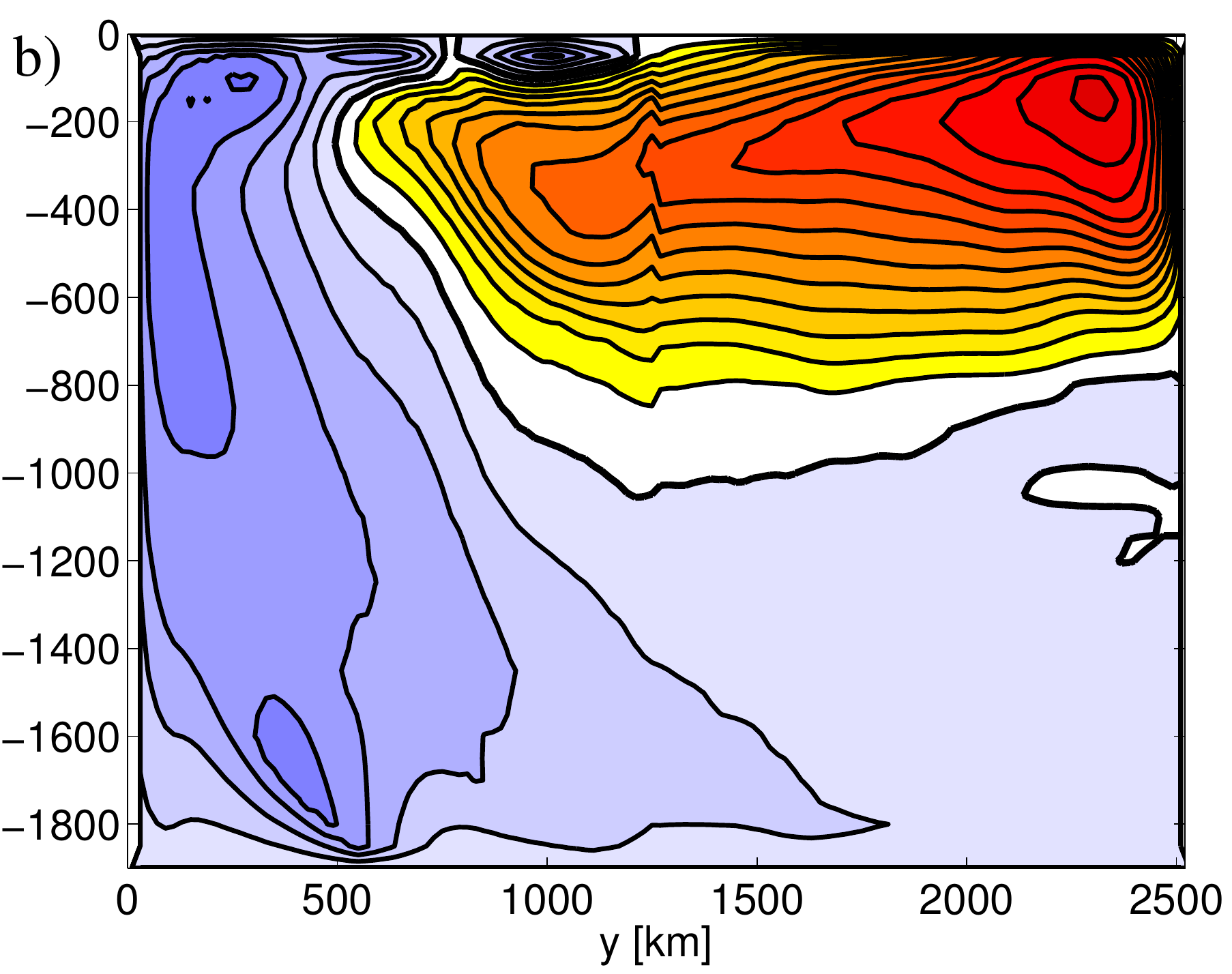}
  }
  \subfigure{
    \includegraphics[scale=0.270]{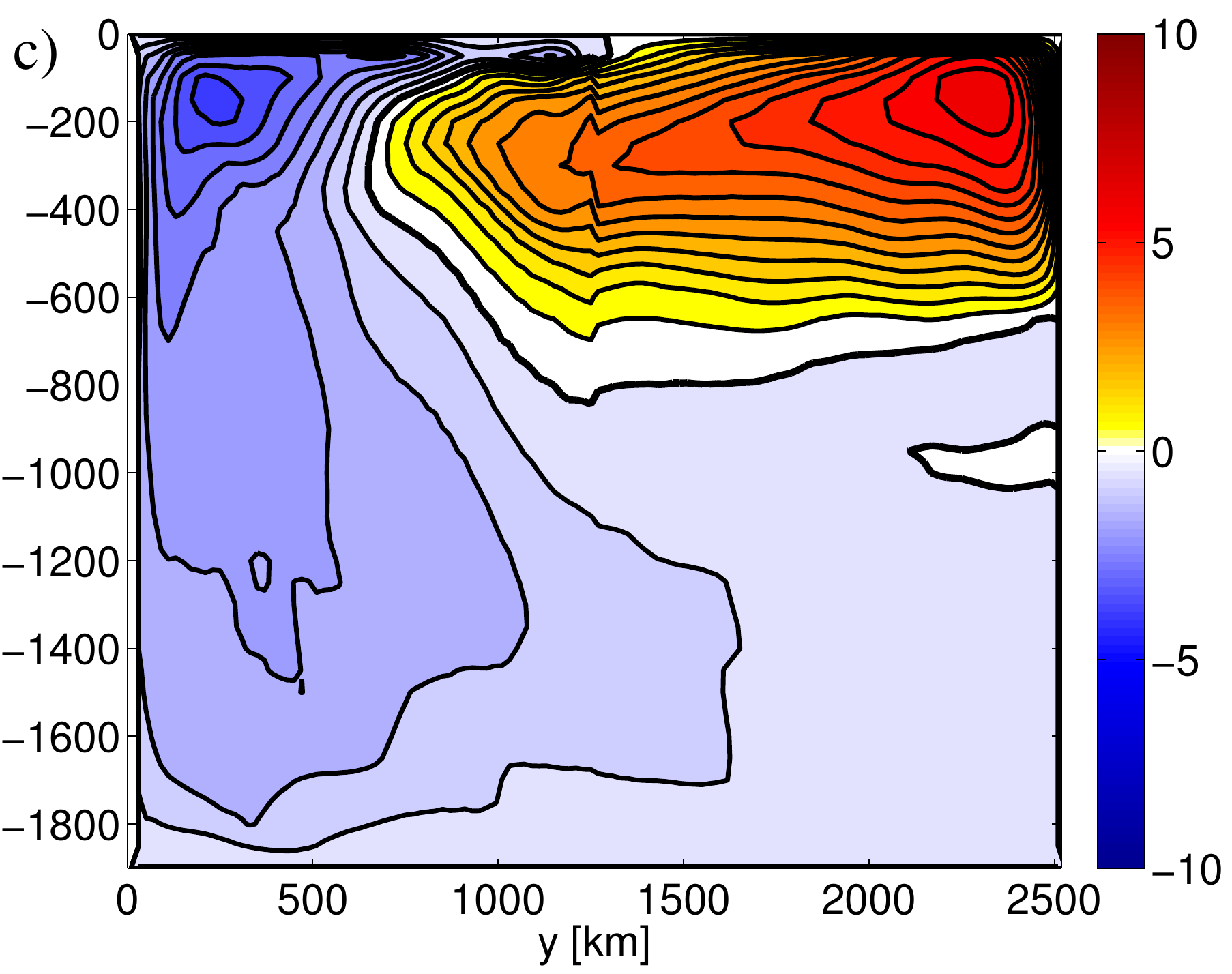}
  }
  \subfigure{
    \includegraphics[scale=0.265]{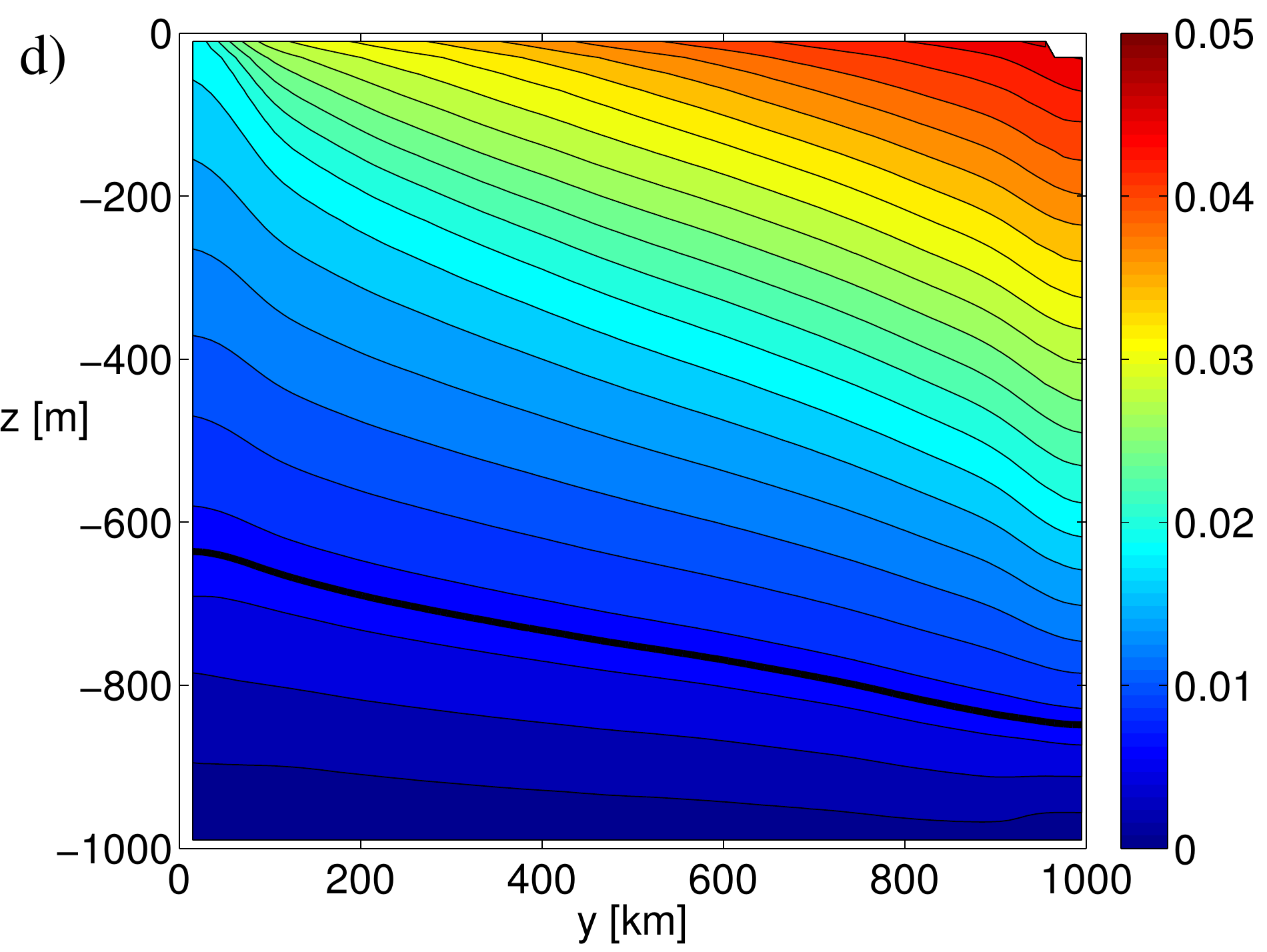}
  }
  \subfigure{
    \includegraphics[scale=0.265]{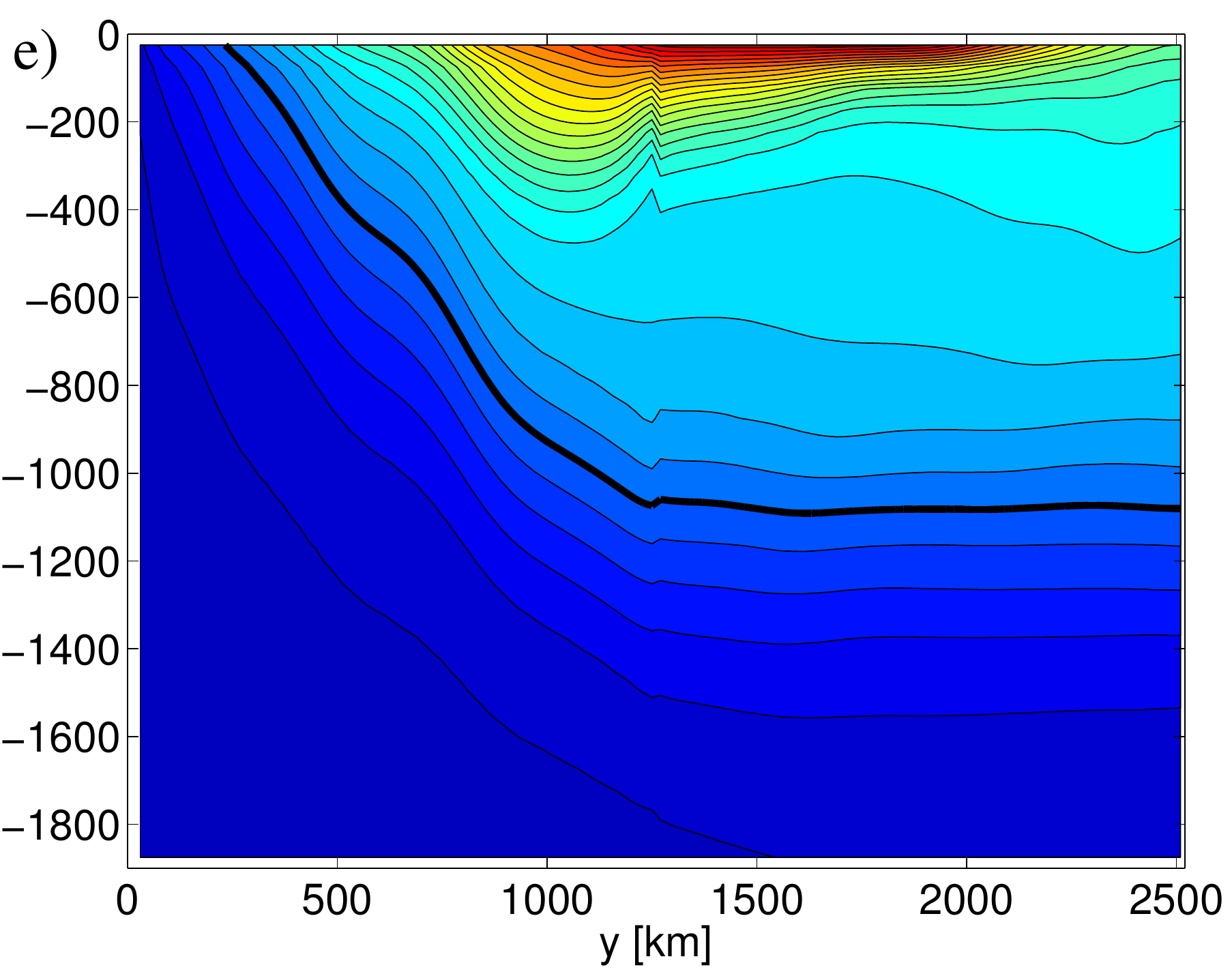}
  }
  \subfigure{
    \includegraphics[scale=0.265]{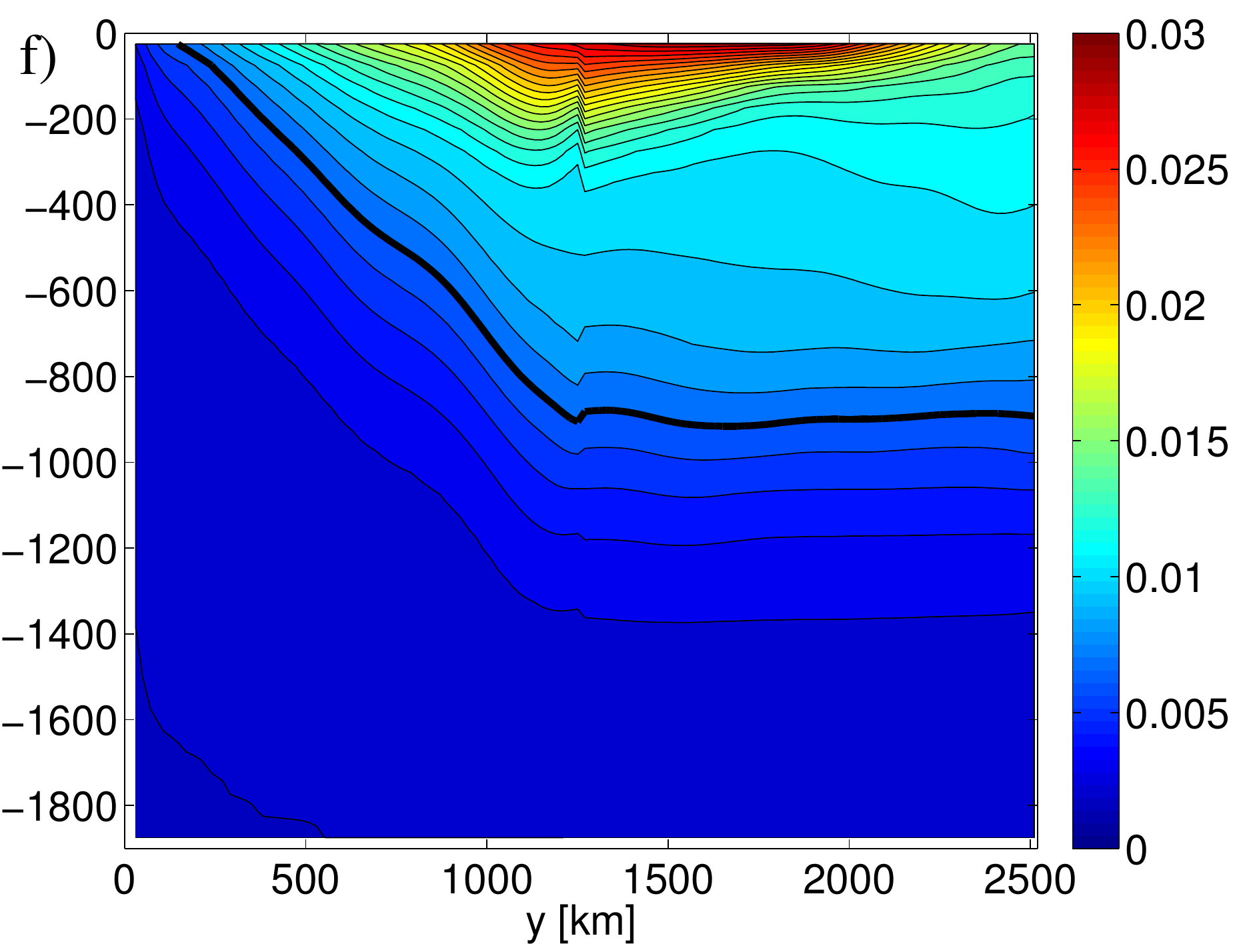}
  }
\caption{
Isopycnal streamfunction transformed to depth coordinates via the mean height of isopycnals \citep{Nurser:04a}
for the NL case (a), flat case (b) and hill case (c). The contour interval is $0.5$Sv and zero lines are thick.
Below are shown the corresponding mean isopycnals (i.e. the isopycnally averaged buoyancy distributions),
where in the NL case (d) the contour interval is $0.002$m/s$^2$, while in the flat case (e) and the hill case (f)
the contour interval is $0.001$m/s$^2$. In all three cases the $0.007$m/s$^2$ line is thick.
}
\label{isopycnalframework}
\end{figure}

For completeness we show in Fig. \ref{isopycnalframework} the isopycnal streamfunction of the NL case (a), flat case (b) and hill case (c),
and the corresponding mean isopycnals,
which largely have been discussed in \cite{Nurser:04a} and \cite{Viebahn:12}.
Comparing Fig. \ref{isopycnalframework} a) with Fig. \ref{NLflat} d-f) we again find that the interior circulation is significantly improved by
the incorporation of the second and third order terms, while in the upper $200$m convergence is far from being reached.
In the flat case, comparing Fig. \ref{isopycnalframework} b) and Fig. \ref{RMflat} d-f), the problems are slightly more severe,
since also, beside the surface layer, in the northern convective region, the southern boundary and the bottom boundary layer in the SO
convergence is far from being reached.
Finally, the worst scenario we find in the hill case (compare Fig. \ref{isopycnalframework} c) with Fig. \ref{hillsill} d-f)),
where even the interior circulation of the SO becomes completely unphysical by including next higher order terms.

\section{Series number}\label{seriesN}
In sections \ref{RME} and \ref{STOKES}, we demonstrated that an approximation of $\psi^*$ and $\Psi^*$ by low order terms of their series expansion
is impossible in certain regions of the ocean, since the series expansion of $\psi^*$ and $\Psi^*$ are increasing there.
These regions are mainly the horizontal boundary layers, which are generally characterised by diabatic processes,
i.e. by large diapycnal diffusivities.
A serious aggravation we found in the more realistic hill case, where the series expansion of $\psi^*$ and $\Psi^*$ are also initially increasing at mid-depth above topography.
We are able to give an indicator of whether the series expansion of $\psi^*$
is initially increasing or decreasing
by consulting the results of \cite{Eden:09b}.
Since the increasing behaviour of the series expansion of $\Psi^*$ is similar to the one of $\psi^*$ in our model experiments, $S$ may also apply to $\Psi^*$.

\cite{Eden:09b} were able to derive the generalised Osborn-Cox relation,
\beq\label{Osborn-Cox}
\kappa+\mu=\mu\ol{(1+C)\exp(-Db')}\ ,
\eeq
which relates the turbulent diapycnal diffusivity $\kappa$ to the molecular diffusivity $\mu$,
the Cox number $C=|\nabla b'|^2/|\nabla\ol{b}|^2$ and
the dimensionless ratio $Db'$ relating the buoyancy perturbation with the mean curvature scale $D=(\nabla^2\ol{b})/|\nabla\ol{b}|^2$.
The ratio $Db'$ appears in the argument of the exponential map,
which represents a standard example of a converging series which is initially increasing, if the argument is greater than $1$.
Hence, as a measure of whether the series expansion of $\kappa$, given by Eq. (\ref{Osborn-Cox}),
is initially increasing or decreasing,
we define the \emph{series number},
\beq
S\equiv|D|\sqrt{2\ol{\phi_2}}\ ,
\eeq
such that we expect for $S>1$ (or near $1$) an initially increasing series expansion.
Since $\kappa$ appears on the one side and $\psi^*$ on the other side of the Eulerian mean buoyancy budget (see Eq. (\ref{TXbbugetdecomp}) in appendix A),
we carry this criterion over to the series expansion of $\psi^*$.

\begin{figure}[t]
  \centering
  \subfigure{
    \includegraphics[scale=0.27]{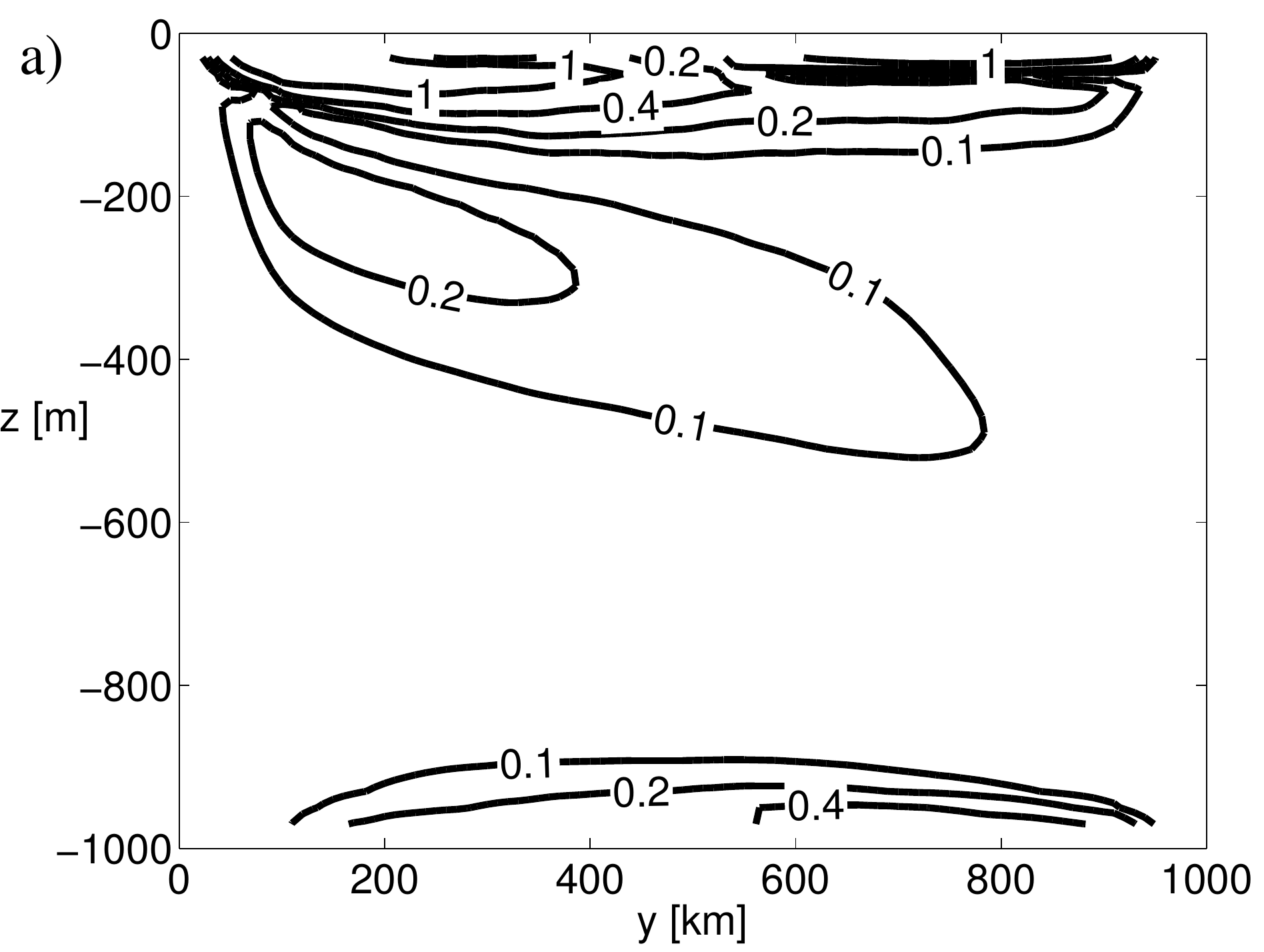}
  }
  \subfigure{
    \includegraphics[scale=0.27]{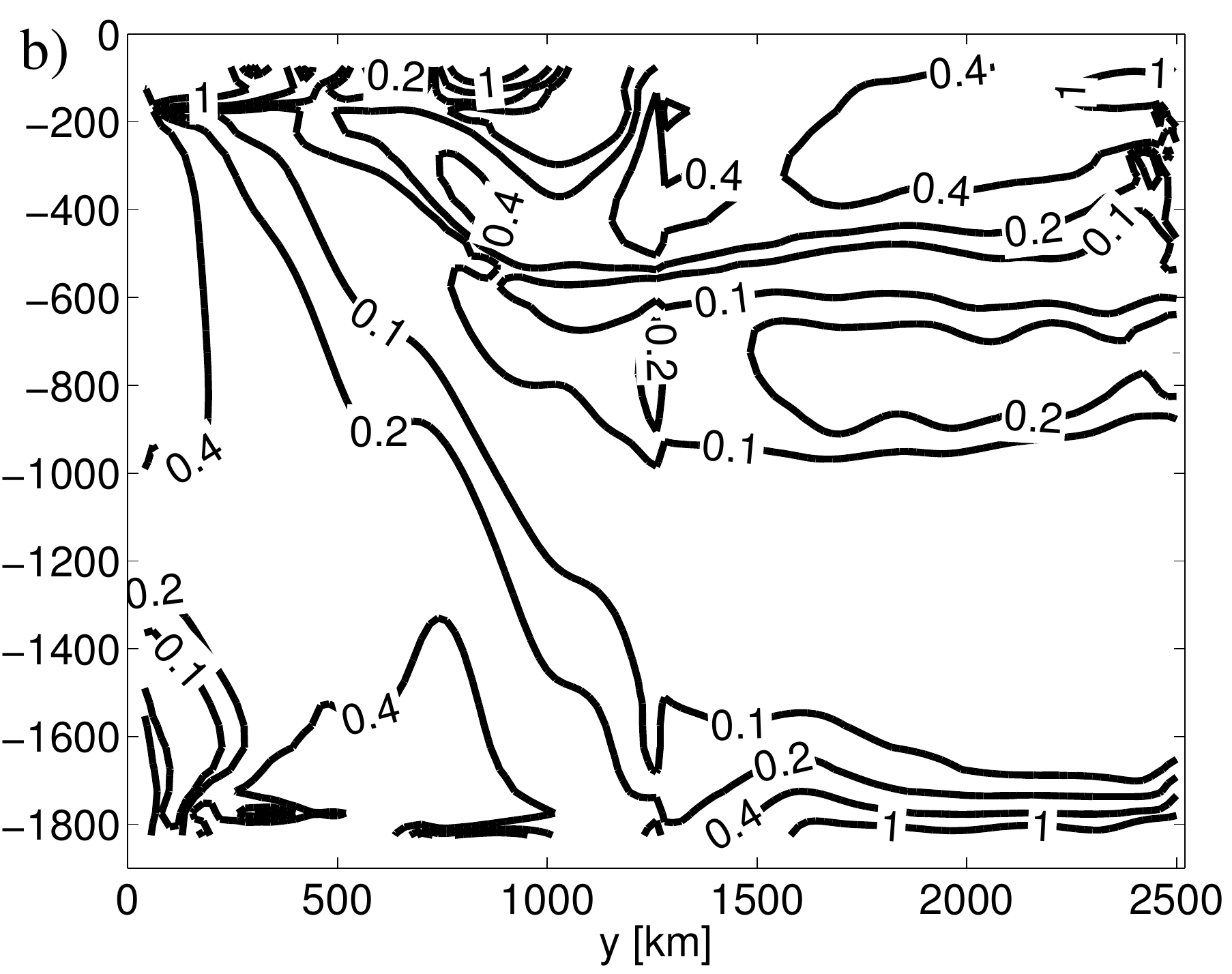}
  }
  \subfigure{
    \includegraphics[scale=0.27]{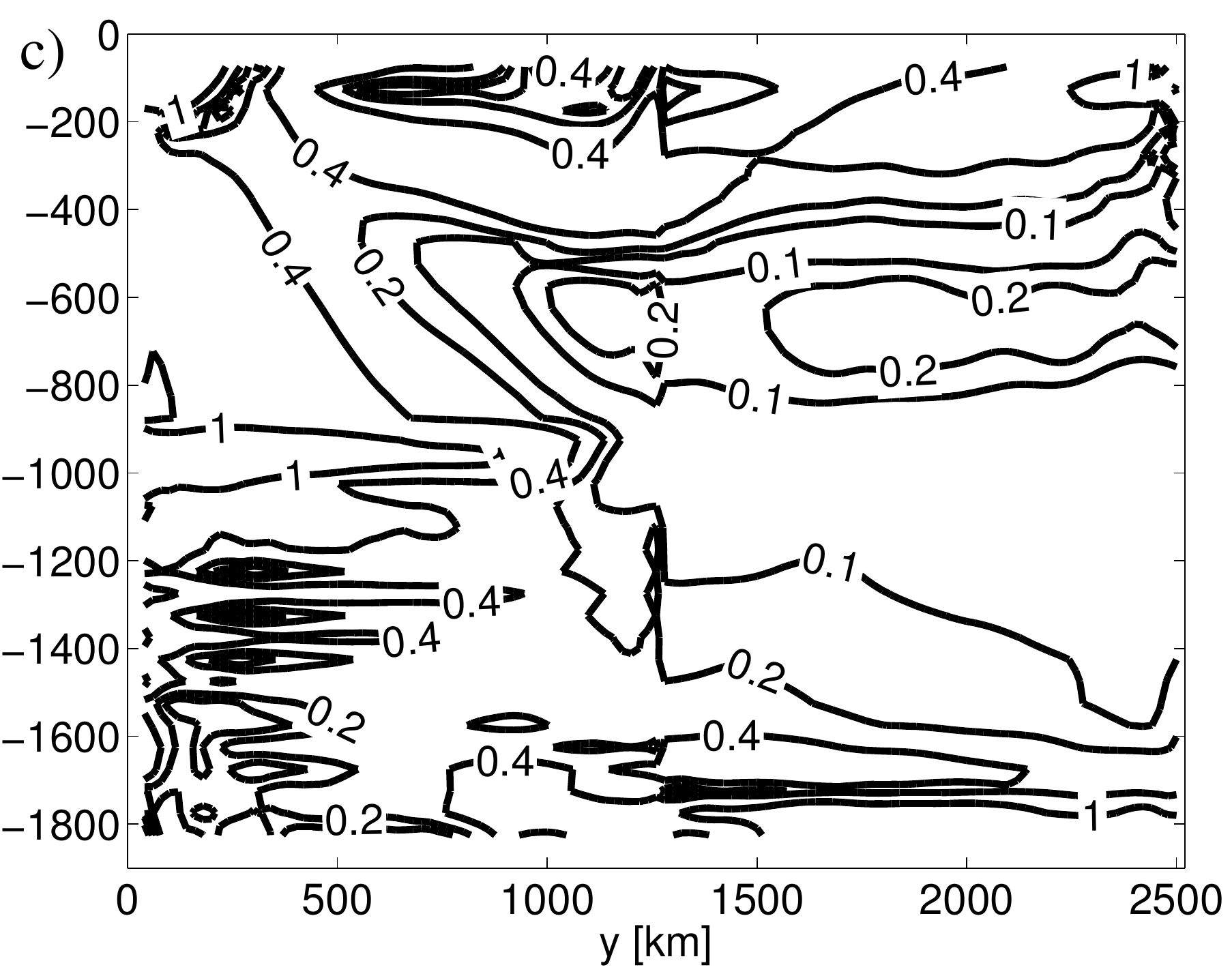}
  }
\caption{
Series number $S$ for the NL case (a), the flat case (b) and the hill case (c). Contour lines are $0.1$, $0.2$, $0.4$ and $1$.
}
\label{olbers}
\end{figure}

Fig. \ref{olbers} a) shows $S$ for the NL case.
As expected, we find $S>1$ only in the surface boundary layer.
Below the surface boundary layer, it holds $S<1$, with the highest values near the bottom.
Fig. \ref{olbers} b) shows $S$ for the flat case.
We find $S>1$ in the southern surface boundary layer, the northern convective region and in the bottom boundary layer\footnote{
In the boundary layer of the SO, the values of $S$ are around $0.96$ (and the lowest two grid points are missing due to the second order derivatives),
while in the cases with $A_h$, $S$ significantly exceeds $1$ in the bottom boundary layer of the SO.
}.
In the ocean interior of the SO, it holds $S<1$.
Introducing $A_h$ (not shown) generally decreases $S$ in the interior of the SO, but increases $S$ in the boundary layers and in the Atlantic part.
Finally, Fig. \ref{olbers} c) shows $S$ for the hill case.
As expected, we find $S>1$ now at the top of topography and not at the bottom in the SO,
while in the rest of the ocean interior it holds $S<1$.
In particular, at mid-depth above topography in the SO, where the series expansion of $\psi^*$ is initially increasing,
$S$ is increased compared to the flat case, but we still have $S<1$.
Hence, in regions of smaller diapycnal diffusivity the criterion is of reduced evidence.
For $A_h=5000\mathrm{m}^2\mathrm{s}^{-1}$ the series number $S$ decreases at the hill depth, but it still holds $S>1$ (not shown).
For $A_h=10000\mathrm{m}^2\mathrm{s}^{-1}$ the series number $S$ is again drastically increased in the entire domain (not shown).

Consequently, in the diabatic boundary regions and at topographic depths the series number $S$ represents a successful\footnote{
Notice that the ratios $|\psi^*_i|/|\psi^*_I|$ (not shown) generally are greater than $1$ in the bottom boundary layer of the Atlantic part,
so that $S>1$ is appropriate there.
}
measure in our model experiments of whether the series expansion of $\psi^*$ is, at least initially, increasing or not,
while in the nearly adiabatic interior above topography the criterion is of reduced evidence in the hill case.

\section{Summary and discussion}\label{Summary}
In this study, we have considered
the series expansion of the residual-mean eddy streamfunction $\psi^*$
and the Taylor expansion of the quasi-Stokes streamfunction $\Psi^*$
up to third order in buoyancy perturbation $b'$.
Beside a formal comparison,
we analysed the resulting MOCs at each order in
three different eddy-permitting numerical model experiments
which have been previously used to analyse eddy streamfunctions,
namely the NL case experiment, which is essentially a reproduction of the idealised zonal channel model setup considered by \cite{Nurser:04a,Nurser:04b},
and the flat case and hill case experiments of the idealised SO model setup introduced by \cite{Viebahn:10,Viebahn:12}.

Formally, the series expansions of $\psi^*$ and $\Psi^*$ increasingly differ from each other with increasing order.
While the first order terms are identical,
the difference between the second order terms is related to the time-zonal-mean meridional velocity $\ol{v}$.
Since $\ol{v}$ is generally small in a zonal channel, the second order difference may be expected to be small there as well.
The third order difference is constituted by a corresponding term related to $\ol{v}$
and an additional term, which is related to $v'$ and, hence, is of fourth order in perturbation quantities $b'$ and $v'$.
For orders higher than three, we expect the emergence of further types of terms related to $v'$.
Regarding a zonal channel, it is likely that the terms related to $v'$
primarily need to be considered in order to distinguish between $\psi^*$ and $\Psi^*$.

This expectation is confirmed in each of our three model experiments,
where the third order difference term related to $v'$ shows the largest magnitudes.
Hence, at least initially and in regions of significant values,
the magnitudes of the differences between $\psi^*$ and $\Psi^*$ tend to increase with higher order.
Significant differences between the terms of the series expansion of $\psi^*$ and the Taylor series of $\Psi^*$
are present in the diabatic boundary regions in the NL case and the flat case,
while in the hill case differences are also found in the ocean interior (around hill depth and above\footnote{
That is, differences primarily appear in the regions where both series expansions are initially increasing - see two paragraphs further down.
}).
In the NL case and the flat case, this is in accordance with the expectation that both streamfunctions largely coincide in the nearly adiabatic interior, since the corresponding mean buoyancy distributions largely coincide there \citep{Nurser:04a,Viebahn:12}.
Finally, we find that the terms of $\Psi^*$ generally tend to have smaller magnitudes than the corresponding terms of $\psi^*$,
which might indicate that the series expansion of $\Psi^*$ converges faster than that of $\psi^*$.

However, despite significant differences in certain regions,
the series expansion of $\psi^*$ and the Taylor series of $\Psi^*$, considered up to the third order in our model experiments,
show the same behaviour in several aspects:
Both series expansions generally tend to be of alternating character,
such that the next higher order always compensates a part of the previous order.
Furthermore, in the NL case and the flat case,
both series expansions may be adequately approximated by low order terms in the ocean interior,
since there they appear to be decreasing with higher order.
Nevertheless, including terms up to the third order still significantly improves the interior circulations in these two cases,
in the sense that they further approach the corresponding circulation patterns of the isopycnal streamfunction
and that streamlines become more aligned along the mean isopycnals in the ocean interior.
For the flat case, we showed that the impact of the next higher order terms in the ocean interior
may be reduced by the introduction of a harmonic viscosity $A_h$, which acts to damp EKE
and also changes the strength and depth of the circulation cells.

In contrast, in the typically diabatic boundary regions,
i.e. the surface boundary layer in the NL case and
the surface and bottom boundary layers as well as the northern convective region in the flat case,
both series expansions are alternating and increasing,
which rules an adequate approximation by low order terms out,
as previously discussed by \cite{Killworth:01,McDougall:01,Nurser:04b}.
This intractable behaviour becomes more pronounced and severe in the hill case.
There, physically inconsistent recirculation cells appear around the hill depth in the first order MOC,
which are not effectively reduced by the inclusion of next higher order terms.
On the contrary, the magnitude of the next higher order terms now even is increasing in the ocean interior
(around hill depth and above topography around $500$m depth),
which further intensifies the recirculation cells and complicates the circulation patterns.
This drawback of initially increasing series expansions does not disappear,
if a harmonic viscosity $A_h$ is introduced in the hill case.
Consequently, an approximation of the ocean interior circulation by low order terms seems not to be possible in the hill case.

The increasing behaviour of both series expansions in certain regions of the ocean
is the handicap which precludes a satisfying approximation of $\psi^*$ or $\Psi^*$ by low order terms.
As an indicator of whether the series expansion of $\psi^*$ is initially increasing or decreasing,
we proposed the series number $S$, i.e. a dimensionless ratio relating the buoyancy perturbation with the mean isopycnal curvature scale.
We find that in the diabatic boundary regions and at topographic depths, $S$ represents a successful
measure in our model experiments of whether the series expansion of $\psi^*$ is initially increasing or not,
while in the nearly adiabatic interior above topography, $S$ is of reduced evidence.
Since the increasing behaviour of the series expansion of $\Psi^*$ is similar to the one of $\psi^*$ in our model experiments,
$S$ applies in the same way to $\Psi^*$.

Consequently, in our model experiment which is equipped with a significant topographic feature
and which hence represents the most realistic model setup,
the approximations of the zonal-mean streamfunctions $\psi^*$ and $\Psi^*$ are most inappropriate.
In order to interpret this problematic behaviour in the ocean interior 
in the hill case,
we distinguish two regions, namely,
on the one hand, the region around the hill depth and below,
and, on the other hand, the interior region above topography.
We interpret the problematic behaviour of the low order terms of $\psi^*$ and $\Psi^*$
in the former region as a zonally integrated boundary layer effect.
As demonstrated in the NL case (section \ref{NLcase}) and the flat case (section \ref{RMflatcase}) and discussed in several previous studies \citep{Killworth:01,McDougall:01,Nurser:04b},
the approximations of $\psi^*$ and $\Psi^*$ typically break down in the horizontal boundary layers,
which are generally characterised by diabatic processes and a vertically non-monotonic buoyancy field.
While in the flat case these regions (surface, bottom) remain at fixed depth in the zonal dimension,
the bottom boundary layer extends zonally over the hill-like topography in the hill case.
More precisely, in the hill case we find significant values of the vertical diffusivity (not shown) indicating a small bottom boundary layer all along the bottom,
but most pronounced at the hill depth.
Hence, in the hill case, bottom boundary layer regions and interior parts are mixed up at topographic depths in the zonal integration carried out at fixed depth and along latitude circles.
This mixture of boundary and interior regions precludes appropriate approximations of $\psi^*$ and $\Psi^*$ at topographic depths in a zonal-mean framework.

The second region of significant and increasing contributions in the lower order terms $\psi^*_{II},\psi^*_{III},...$
and $\Psi^*_{II},\Psi^*_{III},...$
is found above topography and centred around $y=900$km.
We do not relate the pure appearance of these contributions to the presence of topography,
since they are also found, although weaker, in the flat case,
even if the EKE is reduced by the introduction of $A_h$ (see Fig. \ref{RMflat}).
But we ascribe the increasing behaviour of these contributions to the impact of topography on the zonal structure of the velocity and buoyancy fields:
Typically, undulations emerge horizontally in the physical fields as an effect of topography (so-called standing eddies, \cite{Viebahn:12}).
In a zonal-mean framework of zonal integration along latitude circles, these undulations induce the amplification of the significant contributions above topography.
However, the effect of standing eddies on the eddy streamfunctions vanishes,
if the zonal integration paths are redefined so that the topographic influence is taken into account,
or, more precisely, if the zonal integration is performed along time-mean isolines of buoyancy (which coincide with latitude circles in the flat case), as discussed by \cite{Viebahn:12}.
If the zonal integration paths are defined this way, the zonal-mean eddy circulations of the flat case and the hill case are more similar to each other.
Hence, we expect that in a framework of zonal integration along time-mean isolines of buoyancy,
an approximation of $\psi^*$ and $\Psi^*$ in the interior above topography would be possible again, just like in the flat case.
In other words, we interpret the increasing behaviour above topography in the hill case
not as a boundary layer effect in the zonal average,
but as a topographic effect which might be circumvented by the appropriate choice of zonal integration paths.
Moreover, also at topographic depths a reduction of the impact of the lower order terms might result from an appropriate redefinition of the zonal integration paths, but probably an effect of the boundary layer presence in the zonal average is inevitable.


\appendix
\section{Outline of the residual-mean framework}\label{ORMF}
The time-zonal-mean residual streamfunction $\psi_{res}(y,z)$ is defined as the meridional streamfunction which advects the Eulerian time-zonal-mean buoyancy $\ol{b}$.
The time-zonal-mean buoyancy budget under steady state conditions is given by
\beq\label{TXbbuget}
\ol{v}\partial_y\ol{b}+\ol{w}\partial_z\ol{b}+
L_x^{-1}\bigl(\partial_y\bigl(L_x\ol{v'b'}\bigr)+\partial_z\bigl(L_x\ol{w'b'}\bigr)\bigr)=\ol{Q}\ ,
\eeq
where we used the decompositions\footnote{
In a time-zonal-mean context, each quantity $q$ generally may be decomposed into  its temporal and zonal average $\ol{q}$
and its temporal and zonal deviation $q'\equiv q-\ol{q}$, i.e. $q=\ol{q}+q'$.
}
$b=\ol{b}+b'$, $v=\ol{v}+v'$ and $w=\ol{w}+w'$. 
The advection due to the time-zonal-mean velocity $(\ol{v},\ol{w})$
is described by the time-zonal-mean Eulerian streamfunction ${\Lambda}$,
\beq\label{EulerPsi}
\partial_z{\Lambda}=-L_x\ol{v}\ ,\qquad \partial_y{\Lambda}=L_x\ol{w}\ .
\eeq
The eddy buoyancy flux $\mathbf{F}_b\equiv L_x(\ol{v'b'},\ol{w'b'})$ may be decomposed into
an additional advective part\footnote{
The operator $\rnabla$ is defined as $\rnabla\equiv(-\partial_z,\partial_y)$.
}
$-\psi^*\rnabla\ol{b}$ (directed along the time-zonal-mean isopycnals)
and a diffusive part \citep{Andrews:76}.
The natural choice is to direct the diffusive part perpendicular to the advective part, i.e. along the buoyancy gradient $\nabla\ol{b}$ \citep{Andrews:78}.
This decomposition of the eddy buoyancy flux $\mathbf{F}_b$ is defined only up to an arbitrary rotational flux $\mathbf{F_*}=-\rnabla\theta$,
given by the gauge potential $\theta$,
since $\mathbf{F}_b$ appears in the mean buoyancy equation (\ref{TXbbuget}) inside the divergence operator.
In general, we have
\beq\label{decompG}
\mathbf{F}_b=\kappa\nabla\ol{b}-\psi^*\rnabla\ol{b}-\rnabla\theta\ .
\eeq
Using Eq. (\ref{decompG}), we obtain for the buoyancy budget (\ref{TXbbuget}),
\beq\label{TXbbugetdecomp}
v_{res}\partial_y\ol{b}+w_{res}\partial_z\ol{b}=\ol{Q}-L_x^{-1}\nabla\cdot\bigl(\kappa\nabla\ol{b}\bigr)\ ,
\eeq
where the residual velocities $v_{res}=\ol{v}-L_x^{-1}\partial_z\psi^*$ and $w_{res}=\ol{w}+L_x^{-1}\partial_y\psi^*$
represent the total advection velocities of $\ol{b}$.
Hence, the residual streamfunction $\psi_{res}$ is given by
\beq
\psi_{res}={\Lambda}+\psi^*\ .
\eeq
$\psi^*$ is the eddy streamfunction and defines the eddy-driven velocities
\beq
v^*=-L_x^{-1}\partial_z\psi^*\ ,\qquad w^*=L_x^{-1}\partial_y\psi^*\ ,
\eeq
while the flux component $\kappa\nabla\ol{b}$ corresponds to a diffusive flux,
and therefore the coefficient $\kappa$ represents the diapycnal diffusivity induced by meso-scale eddies.
Note that, since $\ol{b}$ does not retain the volumetric properties of the unaveraged buoyancy field $b$,
the effect of eddies on $\ol{b}$ is inevitably both advective and diffusive (in contrast to isopycnal averaging, \cite{Nurser:04b}).

The time-zonal-mean buoyancy $\ol{b}$ in Eq. (\ref{TXbbugetdecomp}) is forced by the small-scale diabatic forcing $\ol{Q}$
and the convergence of the meso-scale diffusive eddy flux $-L_x^{-1}\nabla\cdot\bigl(\kappa\nabla\ol{b}\bigr)$.
In order to ensure that, if there is no instantaneous diabatic buoyancy forcing $Q$, there is also no diabatic effects in the mean buoyancy budget,
we have to consider the rotational eddy fluxes.

While the choice of $\theta$ has no influence on the mean buoyancy equation\footnote{
More precisely, the sum of the additional eddy advection term $-\rnabla(\rnabla\theta\cdot\rnabla\ol{b}/|\nabla\ol{b}|^2)\cdot\nabla\ol{b}$ and the additional eddy diffusion term $\nabla\cdot(\rnabla\theta\cdot\nabla\ol{b}/|\nabla\ol{b}|^2\nabla\ol{b})$ identically vanishes.
},
it affects the eddy streamfunction $\psi^*$ and the diapycnal diffusivity $\kappa$,
\beq\label{psi_eddG}
\psi^*=-\frac{(\mathbf{F}_b+\rnabla\theta)\cdot\rnabla\ol{b}}{|\nabla\ol{b}|^2}\ ,\qquad
\kappa= \frac{(\mathbf{F}_b+\rnabla\theta)\cdot\nabla\ol{b}}{|\nabla\ol{b}|^2}\ .
\eeq
Further, the choice of the gauge potential $\theta$ affects the conservation equation of eddy variance $\ol{\phi_2}=\ol{b'b'}/2$, which is given by
\beq
\nabla\cdot\mathbf{F_2}=-\mathbf{F}_b\cdot\nabla\ol{b}+L_x\ol{b'Q'}\ ,
\eeq
where $\mathbf{F_2}=L_x(\ol{v}\ol{\phi_2}+\ol{v'\phi_2},\ol{w}\ol{\phi_2}+\ol{w'\phi_2})$ represents the total variance flux,
consisting of mean and turbulent variance advection.
The term $\ol{b'Q'}$ denotes dissipation of variance and
the term $-\mathbf{F}_b\cdot\nabla\ol{b}=-\kappa|\nabla\ol{b}|^2+\rnabla\theta\cdot\nabla\ol{b}$ is a variance production term.
The first term is positive for $\kappa>0$ and hence a source of variance, while the second term can have both signs.

By considering the analog budgets of the higher order buoyancy moments, defined as $\ol{\phi_n}=\ol{b'^n}/n$ for order $n$,
and applying decompositions of the corresponding fluxes $\mathbf{F}_n=L_x(\ol{v}\ol{\phi_n}+\ol{v'\phi_n},\ol{w}\ol{\phi_n}+\ol{w'\phi_n})$
analog to Eq. (\ref{decompG}), i.e. $\mathbf{F}_n=\kappa_n\nabla\ol{b}-\psi_n^*\rnabla\ol{b}-\rnabla\theta_n$,
\cite{Eden:07} are able to show that,
if the rotational flux potentials are specified as $n\theta_n=\psi^*_{n+1}$,
then the turbulent diffusivity $\kappa$ of Eq. (\ref{psi_eddG}) is given by the series
\beq\label{kappa}
\kappa|\nabla\ol{b}|^2=L_x\ol{b'Q'}-\mathcal{D}(L_x\ol{\phi_2Q})+\frac{1}{2}\mathcal{D}^2(L_x\ol{\phi_3Q})-\frac{1}{3!}\mathcal{D}^3(L_x\ol{\phi_4Q})+...\ ,
\eeq
where $\mathcal{D}()\equiv\nabla\cdot\nabla\ol{b}|\nabla\ol{b}|^{-2}()$.
In Eq. (\ref{kappa}) $\kappa$ is related to covariances between the small-scale forcing or mixing and buoyancy fluctuations.
Hence, by specifying the gauge potentials as $n\theta_n=\psi^*_{n+1}$,
there is no diapycnal turbulent mixing if there is no molecular mixing.
The gauge condition $\theta=\psi^*_{2}$ states that the rotational flux potential is given by the flux of variance circulating along the contours of $\ol{b}$ (where $\psi^*_{2}$ is affected by the rotational flux potential of eddy variance $\theta_2$).

Using $\psi^*_n|\nabla\ol{b}|^2=-(\mathbf{F}_n+\rnabla\theta_n)\cdot\rnabla\ol{b}$ and the gauge condition $n\theta_n=\psi^*_{n+1}$,
we obtain for the eddy streamfunction $\psi^*$,
\beq\label{ApsiEdFULL}
\psi^*|\nabla\ol{b}|=
-J_1+\partial_mJ_2-\frac{1}{2}\partial^2_mJ_3+\frac{1}{3!}\partial_m^3J_4-...\ ,
\eeq
where $\partial_m()\equiv|\nabla\ol{b}|^{-1}\nabla\ol{b}\cdot\nabla|\nabla\ol{b}|^{-1}()$
and the $J_n\equiv\mathbf{F}_n\cdot\rnabla\ol{b}|\nabla\ol{b}|^{-1}$ represent the along-isopycnal fluxes of the eddy buoyancy moments.
The first order term in the expansion for $\psi^*$ is identical to an eddy streamfunction of the transformed Eulerian mean (TEM) framework \citep{Andrews:76,Andrews:78},
i.e. the decomposition of $\mathbf{F}_b$ with $\rnabla\theta\equiv0$.
The remainder of the expansion is due to the introduction of the rotational flux potential $\theta$ given by
\beq\label{AthetaFULL}
\theta|\nabla\ol{b}|=-J_2+\frac{1}{2}\partial_mJ_3-\frac{1}{3!}\partial_m^2J_4+...\ .
\eeq
In the ocean interior, it typically holds $|\partial_y\ol{b}|\ll|\partial_z\ol{b}|$ and $|\ol{w}|\ll|\ol{v}|$ and we obtain
\beq
\psi^*&\approx&\label{ApsiEdAPPROX}
\frac{L_x\ol{v'b'}}{\partial_z\ol{b}}
-\frac{1}{\partial_z\ol{b}}\partial_z\biggl(\frac{L_x\ol{\phi_2v}}{\partial_z\ol{b}}\biggr)
+\frac{1}{2}\frac{1}{\partial_z\ol{b}}\partial_z\biggl(\frac{1}{\partial_z\ol{b}}\partial_z\biggl(\frac{L_x\ol{\phi_3v}}{\partial_z\ol{b}}\biggr)\biggr)-...\\
\theta&\approx&
\frac{L_x\ol{\phi_2v}}{\partial_z\ol{b}}
-\frac{1}{2}\frac{1}{\partial_z\ol{b}}\partial_z\biggl(\frac{L_x\ol{\phi_3v}}{\partial_z\ol{b}}\biggr)
+\frac{1}{3!}\frac{1}{\partial_z\ol{b}}\partial_z\biggl(\frac{1}{\partial_z\ol{b}}\partial_z\biggl(\frac{L_x\ol{\phi_4v}}{\partial_z\ol{b}}\biggr)\biggr)-...\\
\kappa&\approx&
\frac{L_x\ol{b'Q'}}{(\partial_z\ol{b})^2}
-\frac{1}{(\partial_z\ol{b})^2}\partial_z\biggl(\frac{L_x\ol{\phi_2Q}}{\partial_z\ol{b}}\biggr)
+\frac{1}{2}\frac{1}{(\partial_z\ol{b})^2}\partial_z\biggl(\frac{1}{\partial_z\ol{b}}\partial_z\biggl(\frac{L_x\ol{\phi_3Q}}{\partial_z\ol{b}}\biggr)\biggr)-...
\eeq

\section{Third-order approximation of the quasi-Stokes streamfunction}\label{ASTOKES}
We assume that the buoyancy field $b(x,y,z,t)$ is vertically strictly monotonic in the interior of the ocean.
The instantaneous isopycnal $b_a$ lies at an instantaneous height $z(x,y,b_a,t)=z_a(y,b_a)+z'_a(x,y,b_a,t)$,
where $z_a$ is the time-zonal-mean height of $b_a$ and $z_a'$ is the deviation from the time-zonal-mean, i.e. $\ol{z_a'}=0$.
The time-zonal-mean isopycnal streamfunction ${\psi_I}$ is the temporally averaged zonal and depth integral of the velocity $v$,
integrated below the isopycnal $b_a$.
We may write
\beq
{\psi_I}(y,b_a)=-L_x\ol{\int_{bottom}^{z_a+z_a'}v\ dz}\ .
\eeq
${\psi_I}$ may be transformed to Eulerian space by identifying each $b_a$ with its mean height $z_a$.
Therewith, the quasi-Stokes streamfunction $\Psi^*$ is defined via the decomposition
\beq
-L_x\ol{\int_{bottom}^{z_a+z_a'}v\ dz}={\Lambda}(y,z_a)+\Psi^*(y,z_a)\ ,
\eeq
that is,
\beq
\Psi^*(y,z_a)\equiv-L_x\ol{\int_{z_a}^{z_a+z_a'}v\ dz}\ ,
\eeq
and gives the transport of ${\psi_I}$ related to the perturbation $z'_a$.
$\Psi^*$ is the eddy-induced streamfunction of ${\psi_I}$ in Eulerian space\footnote{
In \cite{Viebahn:12} the corresponding decomposition is defined in an isopycnal framework.
}.
Approximations of both $z'$ and $\Psi^*$ by Eulerian mean quantities may be obtained by expanding $b$ and $v$ in Taylor series \citep{McDougall:01,Nurser:04b}.

A vertical Taylor series of $b$ centred at $z=z_a$ gives
\beq
b_a=b(z_a+z_a')=b(z_a)+z_a'\partial_zb|_{z=z_a}+\frac{1}{2}(z_a')^2\partial_z^2b|_{z=z_a}+\frac{1}{6}(z_a')^3\partial_z^3b|_{z=z_a}+...\nonumber
\eeq
Using the decomposition $b=\ol{b}+b'$, 
we obtain as terms up to third order in perturbation quantities (denoted by $a$)
\beq
b(z_a+z_a')=\ol{b}(z_a)+b'(z_a)+z_a'\partial_z\ol{b}|_{z=z_a}+z_a'\partial_zb'|_{z=z_a}+\frac{1}{2}(z_a')^2\partial_z^2\ol{b}|_{z=z_a}+\nonumber\\
+\frac{1}{2}(z_a')^2\partial_z^2b'|_{z=z_a}+\frac{1}{6}(z_a')^3\partial_z^3\ol{b}|_{z=z_a}+O(a^4)\nonumber
\eeq
Taking the temporal and zonal average of this equation yields
\beq
b_a=\ol{b}(z_a)+\ol{z_a'\partial_zb'|_{z=z_a}}+\frac{1}{2}\ol{(z_a')^2}\partial_z^2\ol{b}|_{z=z_a}
+\frac{1}{2}\ol{(z_a')^2\partial_z^2b'|_{z=z_a}}+\frac{1}{6}\ol{(z_a')^3}\partial_z^3\ol{b}|_{z=z_a}+O(a^4)\nonumber
\eeq
The difference of both equations gives
\beq
-z_a'\partial_z\ol{b}|_{z=z_a}&=&b'(z_a)+z_a'\partial_zb'|_{z=z_a}-\ol{z_a'\partial_zb'|_{z=z_a}}
+\frac{1}{2}\bigl((z_a')^2-\ol{(z_a')^2}\bigr)\partial_z^2\ol{b}|_{z=z_a}\nonumber\\
&&+\frac{1}{2}(z_a')^2\partial_z^2b'|_{z=z_a}-\frac{1}{2}\ol{(z_a')^2\partial_z^2b'|_{z=z_a}}+
\frac{1}{6}\bigl((z_a')^3-\ol{(z_a')^3}\bigr)\partial_z^3\ol{b}|_{z=z_a}+O(a^4)\nonumber
\eeq
From this last equation we obtain\footnote{
Note that if the topography varies vertically, then the zonal average and the vertical derivative do not commute due to the depth-dependent factor $L_x$.
}
a series expansion of $z'$ (extending the approximation of \cite{McDougall:01} about two orders),
\beq\label{Azexpan}
z'&=&-\frac{b'}{\partial_z\ol{b}}
+\frac{1}{\partial_z\ol{b}}\partial_z\biggl(\frac{\phi_2}{\partial_z\ol{b}}\biggr)
-\frac{L_x^{-1}}{\partial_z\ol{b}}\partial_z\biggl(\frac{L_x\ol{\phi_2}}{\partial_z\ol{b}}\biggr)
+\frac{L_x^{-1}}{\partial_z\ol{b}}\partial_z\biggl(\frac{b'}{\partial_z\ol{b}}\biggr)\partial_z\biggl(\frac{L_x\ol{\phi_2}}{\partial_z\ol{b}}\biggr)-\nonumber\\
&&-\frac{1}{2}\frac{1}{\partial_z\ol{b}}\partial_z\biggl(\frac{1}{\partial_z\ol{b}}\partial_z\biggl(\frac{\phi_3}{\partial_z\ol{b}}\biggr)\biggr)
+\frac{1}{2}\frac{L_x^{-1}}{\partial_z\ol{b}}\partial_z\biggl(\frac{1}{\partial_z\ol{b}}\partial_z\biggl(\frac{L_x\ol{\phi_3}}{\partial_z\ol{b}}\biggr)\biggr)+O(b'^4)
\eeq
By expanding $v$ in a vertical Taylor series, we obtain for the quasi-Stokes streamfunction $\Psi^*$,
\beq\label{AQSexpan}
\Psi^*&=&-L_x\Bigl(\ol{v(z_a)z'_a}+\frac{1}{2}\ol{(z_a')^2\partial_zv|_{z=z_a}}+\frac{1}{6}\ol{(z_a')^3\partial_z^2v|_{z=z_a}}+...\Bigr)\nonumber\\
&=&-L_x\Bigl(\ol{v'(z_a)z'_a}+\frac{1}{2}\ol{(z_a')^2}\partial_z\ol{v}|_{z=z_a}+\frac{1}{2}\ol{(z_a')^2\partial_zv'|_{z=z_a}}+\\
&&\qquad\qquad\qquad+\frac{1}{6}\ol{(z_a')^3}\partial_z^2\ol{v}|_{z=z_a}+\frac{1}{6}\ol{(z_a')^3\partial_z^2v'|_{z=z_a}}+...\Bigr)\ ,\nonumber
\eeq
where we used the decomposition $v=\ol{v}+v'$. 
Using Eq. (\ref{Azexpan}) in Eq. (\ref{AQSexpan}), we obtain up to third order in perturbation quantities
(extending the approximation of \cite{McDougall:01} about one order)
\beq\label{AQSexpan1}
\Psi^*&=&\Psi^*_{1}+\Psi^*_{2}+\Psi^*_{3}+O(a^4)\ ,
\eeq
where
\beq
\Psi^*_{1}&\equiv&0\\
\Psi^*_{2}&=&\frac{L_x\ol{b'v'}}{\partial_z\ol{b}}-\frac{L_x\ol{\phi_2}\partial_z\ol{v}}{(\partial_z\ol{b})^2}\\ 
\Psi^*_{3}&=&-\frac{1}{\partial_z\ol{b}}\partial_z\biggl(\frac{L_x\ol{\phi_2v'}}{\partial_z\ol{b}}\biggr)
+\frac{1}{2}\frac{\partial_z\ol{v}}{(\partial_z\ol{b})^2}\partial_z\biggl(\frac{L_x\ol{\phi_3}}{\partial_z\ol{b}}\biggr)+\frac{1}{2}\frac{1}{\partial_z\ol{b}}\partial_z\biggl(\frac{L_x\ol{\phi_3}\partial_z\ol{v}}{(\partial_z\ol{b})^2}\biggr)
\eeq
In order to make the relation between $\Psi^*$ and the residual-mean eddy streamfunction $\psi^*$ (see (\ref{ApsiEdAPPROX})) more obvious,
we arrange the expansion of $\Psi^*$ in orders of buoyancy perturbations $b'$
(extending the approximation of \cite{Nurser:04b} and of Eq. (\ref{AQSexpan1})),
\beq
\Psi^*&=&\Psi^*_{I}+\Psi^*_{II}+\Psi^*_{III}+O(b'^4)\ ,
\eeq
where
\beq
\Psi^*_{I}&=&\frac{L_x\ol{b'v'}}{\partial_z\ol{b}}=\psi^*_I\\
\Psi^*_{II}&=&-\frac{1}{\partial_z\ol{b}}\partial_z\biggl(\frac{L_x\ol{\phi_2v}}{\partial_z\ol{b}}\biggr)
+\frac{\ol{v}}{\partial_z\ol{b}}\partial_z\biggl(\frac{L_x\ol{\phi_2}}{\partial_z\ol{b}}\biggr)=\psi^*_{II}+\psi^*_{\Delta II}\\
\Psi^*_{III}&=&\frac{1}{2}\frac{1}{\partial_z\ol{b}}\partial_z\biggl(\frac{1}{\partial_z\ol{b}}\partial_z\biggl(\frac{L_x\ol{\phi_3v}}{\partial_z\ol{b}}\biggr)\biggr)-\nonumber\\
&&-\frac{1}{2}\frac{\ol{v}}{\partial_z\ol{b}}\partial_z\biggl(\frac{1}{\partial_z\ol{b}}\partial_z\biggl(\frac{L_x\ol{\phi_3}}{\partial_z\ol{b}}\biggr)\biggr)-\frac{L_x^{-1}}{\partial_z\ol{b}}\partial_z\biggl(\frac{L_x\ol{\phi_2}}{\partial_z\ol{b}}\biggr)\partial_z\biggl(\frac{L_x\ol{v'b'}}{\partial_z\ol{b}}\biggr)\\
&=&\psi^*_{III}+\psi^*_{\Delta IIIa}+\psi^*_{\Delta IIIb}\nonumber
\eeq

{\clearpage}
\bibliographystyle{amseng}
\bibliography{references}

\newpage

\newpage

\newpage

\newpage

\newpage

\newpage

\end{document}